\documentclass[PRX,twocolumn,nofootinbibs]{revtex4-1}
\usepackage{graphicx}% Include figure files
\usepackage{dcolumn}% Align table columns on decimal point
\usepackage{bm}% bold math
\usepackage{tabularx}
\usepackage{multirow}% http://ctan.org/pkg/multirow
\usepackage{hhline}% http://ctan.org/pkg/hhline
\newcommand{\di}{$MX_2$}
\newcommand{\tri}{$MX_3$}
\newcommand{\two}{$_2$}
\newcommand{\thr}{$_3$}

\usepackage{booktabs}
\usepackage{multirow}
\usepackage{soul}
\usepackage{microtype}
\usepackage{float}

\usepackage{upgreek}

\begin{document}

\title{Crystal and Magnetic Structures in Layered, Transition Metal Dihalides and Trihalides}

\author{Michael A. McGuire}
\email{McGuireMA@ornl.gov}
\affiliation{Materials Science and Technology Division, Oak Ridge National Laboratory, Oak Ridge, TN 37831 USA}

%=================================================================

\begin{abstract}
Materials composed of two dimensional layers bonded to one another through weak van der Waals interactions often exhibit strongly anisotropic behaviors and can be cleaved into very thin specimens and sometimes into monolayer crystals. Interest in such materials is driven by the study of low dimensional physics and the design of functional heterostructures. Binary~compounds with the compositions \di\ and \tri\ where $M$ is a metal cation and $X$ is a halogen anion often form such structures. Magnetism can be incorporated by choosing a transition metal with a partially filled $d$-shell for $M$, enabling ferroic responses for enhanced functionality. Here a brief overview of binary transition metal dihalides and trihalides is given, summarizing their crystallographic properties and long-range-ordered magnetic structures, focusing on those materials with layered crystal structures and partially filled \textit{d}-shells required for combining low dimensionality and cleavability with magnetism.
\end{abstract}

\maketitle

\section{Introduction}

Binary transition metal halides $MX_y$ ($M=$ metal cation, $X=$ halogen anion) provide a rich family of materials in which low dimensional magnetism can be examined, and such studies were carried out through much of the last century \cite{layered-TM-book}. The dihalides contain triangular nets of transition metal cations, and geometrical frustration is expected when the magnetic interactions are antiferromagnetic (AFM) \cite{Kadowaki-1987, Ramirez-1994, TriangularAFM}. Several of the \di\ compounds form helimagnetic structures and display multiferroic behavior \cite{Tokunaga-2011, Kurumaji-2011, Wu-2012, Kurumaji-2013}. In the trihalides, on the other hand, the transition metal cations form honeycomb nets. This lattice is not frustrated for simple AFM nearest neighbor interactions, but in the case of RuCl\thr\ more complex magnetic interactions and spin-orbit coupling are expected to result in a spin-liquid ground state that is currently of much interest \cite{Plumb-2014, Kim-2015, Sears-2015, Johnson-2015, Banerjee-2016}. For many of the materials considered here, the in plane interactions are ferromagnetic (FM). Chromium trihalides were identified as some of the earliest ferromagnetic semiconductors, and Cr$X_3$ compounds in general have received recent attention as candidate materials for the study of magnetic monolayers and for use in van der Waals heterostructures, in which their magnetism can be coupled to electronic and optical materials via proximity effects~\cite{Geim-2013, Wang-2011, CrI3-str, Zhang-2015, Liu-2016, Wang-2016, Zhong-2017, Huang-2017}. Developing cleavable ferroic materials, both magnetic and electric, is key to expanding the toolbox available for designing and creating custom, functional heterostructures and devices \cite{Lebegue-2013, PhysToday, Park-2016}. Although FM and ferroelectric materials play the most clear role in such applications, the development of spintronics employing antiferromagnetic materials may open the door to a much larger set of layered transition metal halides \cite{MacDonald-2011, Gomonay-2014, Jungwirth-2016}.

It is generally observed that binary halides often form low dimensional crystal structures, comprising either molecular units, one dimensional chains, or two dimensional layers. This holds true especially for the chlorides, bromides, and iodides, and can be attributed to the low ionic charge $X^{-}$ and relatively large ionic radii (1.8--2.2 {\AA}) of these anions. This results in multiple large anions for each metal cation assuming oxidation states typical for transition metals. Thus the cations are usually found in six-fold coordination and are well separated into structural units that are joined to one another in the crystal by van der Waals bonding between halogen anions. Fluoride has a much smaller ionic radius (1.3 {\AA}) and forms three dimensional crystal structures with the divalent and trivalent cations that are the focus of most of this work. Note that this simple ionic picture is not appropriate when metal-metal bonding is present, which is often the case in the early, heavy transition metals. Typically larger anion to cation ratios result in lower dimensional structures, but often polymorphs of different dimensionality occur for a single composition. For example, TiCl\thr\ forms with 1D chains of face sharing octahedra or 2D layers of edge sharing octahedra \cite{TiCl3-VCl3-str-BiI3, TiCl3-1D}. Dihalides (\di) and trihalides (\tri) represent the majority of layered binary transition metal halides. There are, however, other examples, including Nb$_3$Cl$_8$ \cite{von-Schnering-1961}, in which interesting magnetic behavior has been noted recently~\cite{Sheckelton-2017, Jiang-2017}.

Here, a brief review of \di\ and \tri\ compounds with partially filled $d$-shells is presented, with a focus on crystal and magnetic structures. A general overview of the crystallographic properties and magnetic behavior including static magnetic order in these two families is given. This short survey is not meant to be exhaustive, but rather to give a general introduction to these materials and a broad overview of the trends observed in their crystallographic and magnetic properties, with references to the literature where more detailed discussion can be found.

\begin{table*}
\caption{\label{tab:MX2} Summary of % call out is below the table, please confrim. -the call out is not below the table. It is on the page before the table, in the middle of Page 2. (MAM)
structural and magnetic data for layered \di\ compounds with partially filled \textit{d}-shells. A * indicates that the compound is known to undergo a crystallographic phase transition above or below room temperature. ``Magnetic order'' refers to long range 3D magnetic order, and is given as antiferromagnetic (AFM), ferromagnetic (FM), or helimagnetic (HM). ``Moments in layer'' refers to the arrangement of the magnetic moments within a single layer, with $||$ and $\perp$ indicating whether the moments are directed parallel to the plane of the layer (in plane) or perpendicular to it (out of plane), respectively. Ordering temperatures ($T_N$) and Weiss temperatures ($\theta$) are given. References for the magnetic data can be found in the associated text.
}
\resizebox{\textwidth}{!}{
\begin{tabular}{ccccccccc}					
\hline					
\multirow{2}{*}{{\bf Compound}}   &   \multirow{2}{*}{{\bf Structure Type}} &   \multirow{2}{*}{{\bf Ref.}} &   {\bf in Plane} {\boldmath{$M-M$}}  &   {\bf Layer}      &   \multirow{2}{*}{{\bf Magnetic Order}} &   {\bf Moments} &   {\boldmath{$T_N$}}   &   {\boldmath{$\theta$}}   \\
    &   &   &  {\bf Distance ({\AA})}    &   {\bf Spacing ({\AA})} &    &   {\bf in Layer}    &   {\bf{(K)}} &  {\bf{(K)}}  \\	% -I moved the ref column to the left in both tables so that it is now clearly associated with the structural data and not the magnetic data.
\hline
TiCl$_2$	&	CdI$_2$ ($P\overline{3}m1$)	&	\cite{TiCl2-str}	&	3.56	&	5.88	&	AFM	&	--	&	85	 &	 $-$702	\\
TiBr$_2$	&	CdI$_2$ ($P\overline{3}m1$)	&	\cite{TiBr2-str}	&	3.63	&	6.49	&	--	&	--	&	--	 &	--	 \\
TiI$_2$	&	CdI$_2$ ($P\overline{3}m1$)	&	\cite{TiI2-VBr2-str}	&	4.11	&	6.82	&	--	&	--	&	--	&	--	 \\
VCl$_2$	&	CdI$_2$ ($P\overline{3}m1$)	&	\cite{VCl2-str}	&	3.6	&	5.83	&	AFM	&	120$^{\circ}$	&	36	&	 $-$565, $-$437	 \\
VBr$_2$	&	CdI$_2$ ($P\overline{3}m1$)	&	\cite{TiI2-VBr2-str}	&	3.77	&	6.18	&	AFM	&	120$^{\circ}$	&	 30	&	$-$335	 \\
VI$_2$	&	CdI$_2$ ($P\overline{3}m1$)	&	\cite{VI2-str}	&	4.06	&	6.76	&	AFM	&	--	&	16.3, 15	&	 $-$143	 \\
MnCl$_2$	&	CdCl$_2$ ($R\overline{3}m$)	&	\cite{MnCl2-str}	&	3.71	&	5.86	&	AFM or HM	&	stripe or HM	&	2.0, 1.8	 &	$-$3.3	\\
MnBr$_2$ *	&	CdI$_2$ ($P\overline{3}m1$)	&	\cite{MnBr2-str}	&	3.89	&	6.27	&	AFM	&	stripe $||$	 &	 2.3, 2.16	&	--	\\  % changed from ; to comma, please confirm. -confirmed. I changed all ; to , in this table, and removed some , that were unnecessary. (MAM)
MnI$_2$	&	CdI$_2$ ($P\overline{3}m1$)	&	\cite{MnI2-CoI2-str}	&	4.16	&	6.82	&	HM	&	HM	&	 3.95, 3.8, 3.45	&	--	\\
FeCl$_2$	&	CdCl$_2$ ($R\overline{3}m$)	&	\cite{FeCl2-str}	&	3.6	&	5.83	&	AFM	&	FM $\perp$	 &	24	&	 9 ($||$), 21 ($\perp$)	\\
FeBr$_2$	&	CdI$_2$ ($P\overline{3}m1$)	&	\cite{FeBr2-str}	&	3.78	&	6.23	&	AFM	&	FM $\perp$	 &	14	 &	$-$3.0 ($||$), 3.5 ($\perp$)	\\
FeI$_2$	&	CdI$_2$ ($P\overline{3}m1$)	&	\cite{FeI2-str}	&	4.03	&	6.75	&	AFM	&	stripe $\perp$	&	9	&	 24 ($||$), 21.5 ($\perp$)	\\
CoCl$_2$	&	CdCl$_2$ ($R\overline{3}m$)	&	\cite{CoCl2-str}	&	3.54	&	5.81	&	AFM	&	FM $||$	 &	25	&	 38	\\
CoBr$_2$	&	CdI$_2$ ($P\overline{3}m1$)	&	\cite{CoBr2-str}	&	3.69	&	6.12	&	AFM	&	FM $||$	 &	19	&	 --	\\
CoI$_2$	&	CdI$_2$ ($P\overline{3}m1$)	&	\cite{MnI2-CoI2-str}	&	3.96	&	6.65	&	HM	&	HM	&	11	 &	--	 \\
NiCl$_2$	&	CdCl$_2$ ($R\overline{3}m$)	&	\cite{NiCl2-str}	&	3.48	&	5.8	&	AFM	&	FM $||$	 &	52	&	68	 \\
NiBr$_2$	&	CdCl$_2$ ($R\overline{3}m$)	&	\cite{NiBr2-str}	&	3.7	&	6.09	&	AFM, HM	&	FM $||$, HM	 &	 52, 23	&	--	\\
NiI$_2$*	&	CdCl$_2$ ($R\overline{3}m$)	&	\cite{NiI2-str}	&	3.9	&	6.54	&	HM	&	HM	&	75	&	--	 \\
ZrCl$_2$	&	MoS$_2$ ($R3m$)	&	\cite{ZrCl2-str}	&	3.38	&	6.45	&	--	&	--	&	--	&	--	\\
ZrI$_2$	&	MoTe$_2$ ($P2_1/m$)	&	\cite{ZrI2-str-MoTe2}	&	3.18, 3.74, 4.65	&	7.43	&	\multirow{2}{*}{--}	&	 \multirow{2}{*}{--}	&	\multirow{2}{*}{--}	&	\multirow{2}{*}{--}	\\
ZrI$_2$	&	WTe2 ($Pmn2_1$)	&	\cite{ZrI2-str-WTe2}	&	3.19, 3.74,  4.65	&	7.44	&		&		 &		&		 \\
\hline			
\end{tabular}	
}				
\end{table*}
%%%%%%%%%%%%%%%%%%%%%%%%%%%%%%%%%%%%%%%%%%%%%%%%

\section{Crystal Structures of Layered, Binary, Transition Metal Halides}

\subsection{\textit{MX}$_2$ Compounds}

\begin{figure*}
\begin{center}
\includegraphics[width=5.4in]{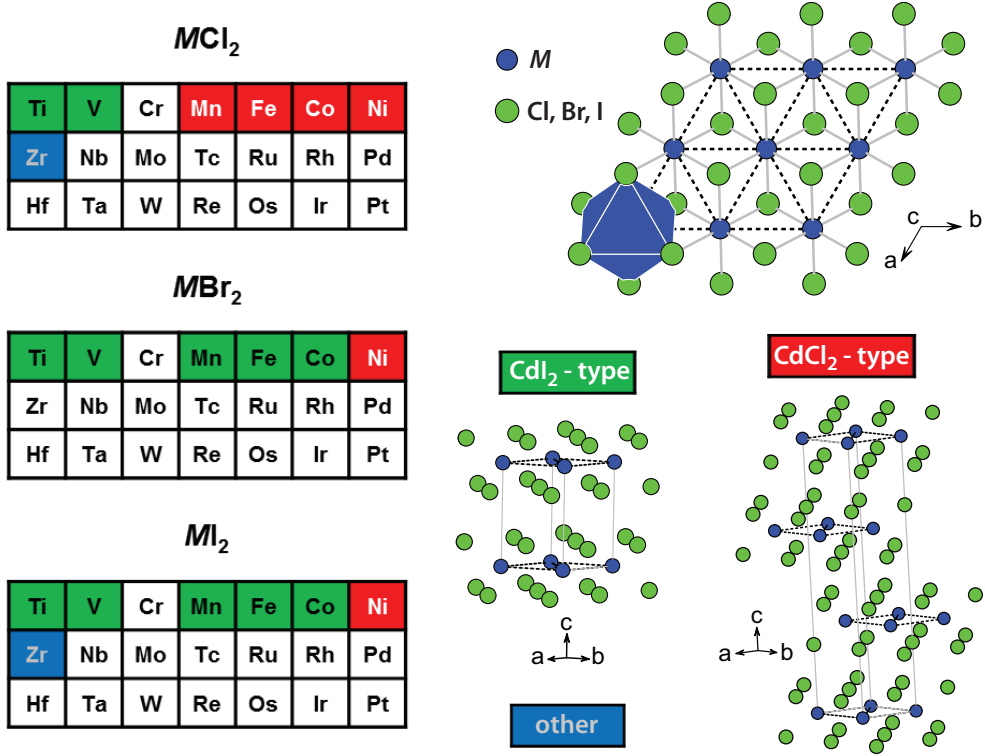}
\caption{\label{fig:dihalide-PT}
A section of the periodic table showing the transition metals for which layered \di\ compounds listed in Table \ref{tab:MX2} form. The metals are highlighted with colors that correspond to the structure types shown on the lower right. A plan view of a single layer common to both the CdI\two\ and CdCl\two\ structure types is shown on the upper right.
}
\end{center}
\end{figure*}
\begin{figure*}
\begin{center}
\includegraphics[width=5.4in]{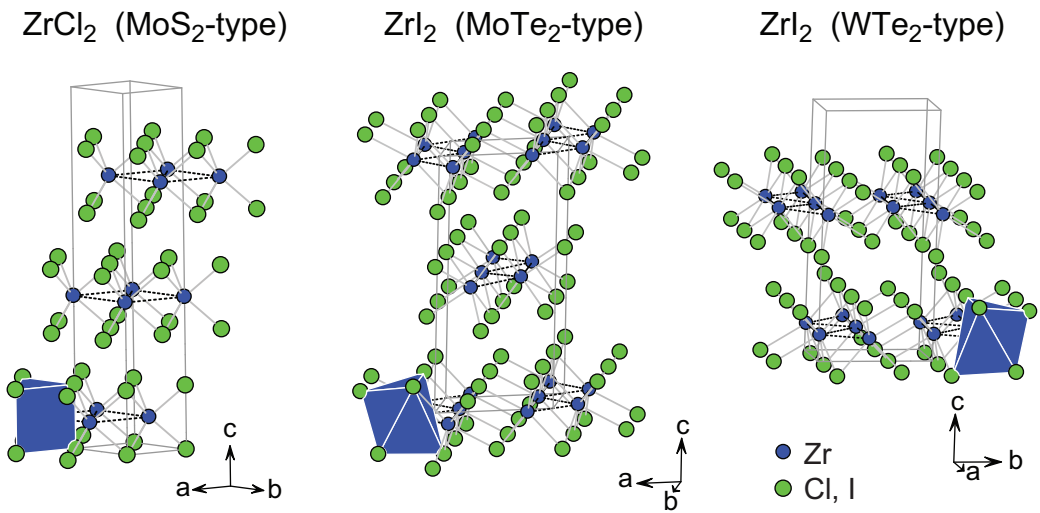}
\caption{\label{fig:ZrX2-structures}
Crystal structures of layered Zr$X_2$ compounds. A single Zr coordination polyhedron is shown for each structure.
}
\end{center}
\end{figure*}

\begin{table*}
\caption{\label{tab:MX3} Summary of structural and magnetic data for layered \tri\ compounds with partially filled \textit{d}-shells. A * indicates that the compound is known to undergo a crystallographic phase transition above or below room temperature. Multiple reported structure types are listed for some compounds. ``Magnetic order'' refers to long range 3D magnetic order, and is given as antiferromagnetic (AFM), ferromagnetic (FM), or helimagnetic (HM). ``Moments in layer'' refers to the arrangement of the magnetic moments within a single layer, with $||$ and $\perp$ indicating whether the moments are directed parallel to the plane of the layer (in plane) or perpendicular to it (out of plane), respectively, and ``canted'' indicating a canting away from either of these directions. Ordering temperatures ($T_N$, $T_C$) and Weiss temperatures ($\theta$) are given. References for the magnetic data can be found in the associated text.
}
\resizebox{\textwidth}{!}{
\begin{tabular}{ccccccccc}					
\hline			
\multirow{2}{*}{{\bf Compound}}   &   \multirow{2}{*}{{\bf Structure Type}} &   \multirow{2}{*}{{\bf Ref.}} &   {\bf in Plane} {\boldmath{$M-M$}}  &   {\bf Layer}      &   \multirow{2}{*}{{\bf Magnetic Order}} &   {\bf Moments} &   {\boldmath{$T_N$}{\bf or} {\boldmath{$T_C$}}}   &   {\boldmath{$\theta$}}   \\
    &   &   &  {\bf Distance ({\AA})}    &   {\bf Spacing ({\AA})} &    &   {\bf in Layer}    &   {\bf{(K)}} &  {\bf{(K)}}  \\	
%	&		&	in plane $M-M$ 	&	layer	&		&		&	moments	&	$T_N$ or $T_C$ 	&	$\theta$	\\
%compound	&	structure type 	&	distance ({\AA})	&	spacing ({\AA})	&	Ref.	&	magnetic order	&	in layer	 &	(K)	&	(K)	\\
\hline
TiCl$_3$ * 	&	BiI$_3$ ($R\overline{3}$)	&	\cite{TiCl3-VCl3-str-BiI3}	&	3.53	&	5.83	&	 \multirow{2}{*}{--}	&	\multirow{2}{*}{--}	 &	\multirow{2}{*}{--}	&   	\multirow{2}{*}{--}	 \\
TiCl$_3$ * 	&	Ti$_3$O ($P\overline{3}1c$)	&	\cite{Troyanov-1991}	&	3.55	&	5.86	&		&		&	 	&		 \\
TiBr$_3$ *	&	BiI$_3$ ($R\overline{3}$)	&	\cite{TiBr3-str}	&	3.74	&	6.21	&	--	&	--	&	--	 &	--	 \\
VCl$_3$	&	BiI$_3$ ($R\overline{3}$)	&	\cite{TiCl3-VCl3-str-BiI3}	&	3.47	&	5.78	&	AFM	&	--	&	 $\sim$20	&	$-$30	\\
VBr$_3$	&	BiI$_3$ ($R\overline{3}$)	&	\cite{VBr3-str}	&	3.7	&	6.21	&	--	&	--	&	--	&	--	 \\
CrCl$_3$ *	&	AlCl$_3$ ($C2/m$)	&	\cite{CrCl3-str}	&	3.44, 3.44	&	5.80	&	AFM	&	FM $||$	&	 15.5, 16.8	&	27	\\
CrBr$_3$ *	&	BiI$_3$ ($R\overline{3}$)	&	\cite{CrBr3-str}	&	3.64	&	6.11	&	FM	&	FM $\perp$	 &	37	 &	47	\\
CrI$_3$ *	&	AlCl$_3$ ($C2/m$)	&	\cite{CrI3-str}	&	3.96, 3.97	&	6.62	&	FM	&	FM $\perp$	&	61	&	70	 \\
FeCl$_3$	&	BiI$_3$ ($R\overline{3}$)	&	\cite{FeCl3-str}	&	3.50	&	5.80	&	\multirow{3}{*}{HM}	&	 \multirow{3}{*}{HM}	&	\multirow{3}{*}{9--10}	 &	 \multirow{3}{*}{$-$11.5}	\\ % changed from minus to endash, please confirm. -this change and subsequent changes of - to -- in this table are correct (MAM)
FeCl$_3$	&	Ti$_3$O ($P312$)	&	\cite{FeCl3-str-polymorphs}	&	3.50	&	5.80	&		&		 &		&		 \\
FeCl$_3$	&	FeCl$_3$ ($P\overline{3}$)	&	\cite{FeCl3-str-polymorphs}	&	3.50	&	5.81	&		&		 &		 &		\\
FeBr$_3$	&	BiI$_3$ ($R\overline{3}$)	&	\cite{FeBr3-str}	&	3.69	&	6.13	&	AFM	&	--	&	15.7	 &	 --	\\
MoCl$_3$	&	AlCl$_3$ ($C2/m$)	&	\cite{Schafer-1967}	&	2.76, 3.71	&	5.99	&	--	&	--	&	--	&	--	 \\
TcCl$_3$	&	AlCl$_3$ ($C2/m$)	&	\cite{TcCl3-str}	&	2.86, 3.60	&	5.86	&	--	&	--	&	--	&	--	 \\
RuCl$_3$ *	&	AlCl$_3$ ($C2/m$)	&	\cite{RuCl3-Cao-2016}	&	3.45, 3.45	&	5.69	&	\multirow{3}{*}{AFM}	&	 \multirow{3}{*}{zig-zag canted}	&	 \multirow{3}{*}{7--8, 13--14}	&	  \multirow{3}{*}{37 ($||$), $-$150($\perp$)}	\\
RuCl$_3$ *	&	Ti$_3$O ($P312$)	&	\cite{Fletcher-1967}	&	3.45	&	5.72	&		&		 &		&		 \\
RuCl$_3$ *	&	CrCl$_3$ ($P3_112$)	&	\cite{RuCl3-Stroganov1957}	&	3.44, 3.45	&	5.73	&		&		&		 &		 \\
RhCl$_3$	&	AlCl$_3$ ($C2/m$)	&	\cite{RhCl3-str}	&	3.44, 3.43	&	5.70	&	--	&	--	&	--	&	--	 \\
RhBr$_3$	&	AlCl$_3$ ($C2/m$)	&	\cite{RhBr3-RhI3-str}	&	3.62, 3.62	&	6.00	&	--	&	--	&	--	 &	--	 \\
RhI$_3$	&	AlCl$_3$ ($C2/m$)	&	\cite{RhBr3-RhI3-str}	&	3.91, 3.90	&	6.45	&	--	&	--	&	--	&	--	 \\
IrCl$_3$	&	AlCl$_3$ ($C2/m$)	&	\cite{RuCl3-IrCl3-Brodersen1965}	&	3.46, 3.45	&	5.64	&	--	&	--	&	 --	&	--	 \\
IrBr$_3$	&	AlCl$_3$ ($C2/m$)	&	\cite{IrBr3-IrI3-RuCl3-str}	&	3.67, 3.64	&	6.01	&	--	&	--	&	 --	&	 --	\\
IrI$_3$	&	AlCl$_3$ ($C2/m$)	&	\cite{IrBr3-IrI3-RuCl3-str}	&	--	&	6.54	&	--	&	--	&	--	&	--	 \\
\hline
\end{tabular}
}					
\end{table*}
Crystal structure information for \di\ compounds with partially filled $d$-shells is collected in Table \ref{tab:MX2}. Non-magnetic, layered dihalides of Zn, Cd, and Hg, with valence electronic configurations 3$d^{10}$, 4$d^{10}$, and 5$d^{10}$, respectively, are also known \cite{ZnCl2, ZnBr2, ZnI2, CdCl2, CdBr2, CdI2, HgI2}, but these are not considered here. It can be seen that most of the compounds in Table \ref{tab:MX2} adopt either the trigonal CdI\two\ structure type or the rhombohedral CdCl\two\ structure type. These structures are shown in Figure \ref{fig:dihalide-PT}. Both contain triangular nets of cations in edge sharing octahedral coordination forming layers of composition \di\ separated by van der Waals gaps between the $X$ anions. The structures differ in how the layers are stacked. The CdI\two\ structure type has AA stacking with one layer per unit cell, and the $X$ anions adopt a hexagonal close packed arrangement. The CdCl\two\ structure has ABC stacking with three layers per unit cell, and the $X$ anions adopt a cubic close packed arrangement.

Figure \ref{fig:dihalide-PT} also shows sections of the periodic table highlighting the transition metals for which the \di\ compounds listed in Table \ref{tab:MX2} form. Note that compounds with stoichiometry \di\ are known for other $M$, for example Cr, Mo, and Pd, but they form molecular (cluster) compounds or 1D chain structures. Among the dichlorides, the CdCl\two\ structure is found only for the later transition metals, and only for Ni in the dibromides and diiodides. However, Schneider et al. have shown that MnBr\two\ may undergo a crystallographic phase transition from the CdI\two\ structure type to the CdCl\two\ structure type at high temperature \cite{MnBr2-CdCl2}. NiI\two\ undergoes a crystallographic phase transition at 60\,K \cite{Kuindersma-1981}. It is monoclinic below this temperature, resulting from a slight distortion ($\beta = 90.2^{\circ}$) from the C-centered orthorhombic description of the hexagonal lattice. In addition, diffraction measurements on FeCl\two\ under pressure have shown a transition from the CdCl\two\ structure to the the CdI\two\ structure near 0.6\,GPa~\cite{Narath-1966, FeCl2-pressure}.

Interatomic distances between $M$ cations within the layers and the spacing of the layers are shown in Table \ref{tab:MX2}. For the CdCl\two\ and CdI\two\ structure types the in-plane $M-M$ distance is equal to the length of the crystallographic $a$ axis (hexagonal settings). The layer spacing, defined as the distance between the midpoints of neighboring layers measured along the stacking direction, is equal to the length of the $c$ axis in the CdI\two\ structure and $c$/3 in the CdCl\two\ structure. Moving across the series from Mn to Ni, both of these distances generally decrease, while less systematic behavior is seen for Ti$X_2$ and V$X_2$.

The layered phases are restricted to the first row of the transition metals, with the exception of the 4$d$ element Zr. Both ZrCl\two\ and ZrI\two\ are reported, but not the dibromide. The zirconium compounds are found to have different structures than the layered 3$d$ transition metal dihalides. As shown in Figure \ref{fig:ZrX2-structures}, ZrCl\two\ adopts the MoS\two\ structure type \cite{ZrCl2-str}, which has the same triangular nets of metal cations and ABC stacking found in CdCl\two. However, in ZrCl\two\ the Zr atoms are in trigonal prismatic coordination rather than octahedral coordination. As a result the Cl anions do not form a cubic close packed arrangement in ZrCl\two\, but instead an AABBCC stacking sequence. ZrI\two\ is reported to adopt both the MoTe\two\ and WTe\two\ structure types \cite{ZrI2-str-MoTe2, ZrI2-str-WTe2}. The closely related structures are shown in Figure \ref{fig:ZrX2-structures}. The regular triangular net of $M$ cations found in the compounds described previously is disrupted in ZrI\two, which has zigzag chains of Zr atoms (see $M-M$ in-plane distances in Table \ref{fig:dihalide-PT}). This~points to the tendency of heavier (4$d$ and 5$d$) transition metals to form metal-metal bonds. Indeed,~in addition to the layered MoS\two\ structure described above for ZrCl\two, a molecular crystal structure with Zr$_6$ clusters is also known \cite{ZrCl2-cluster}. Further examples of this tendency will be noted later in discussion of \tri\ compounds.

\subsection{\textit{MX}$_3$ Compounds}

\begin{figure*}
\begin{center}
\includegraphics[width=5.4in]{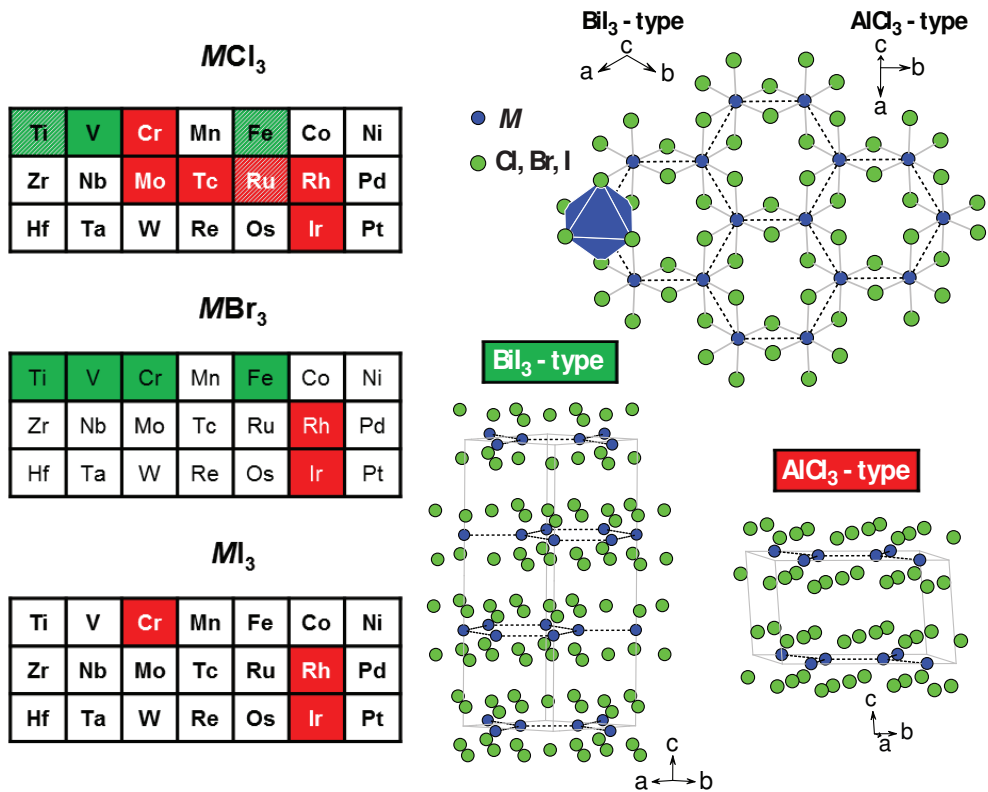}
\caption{\label{fig:trihalide-PT}
A section of the periodic table showing the transition metals for which layered \tri\ compounds listed in Table \ref{tab:MX3} form. The metals are highlighted with colors that correspond to the structure types shown on the lower right. Crosshatching indicates multiple structures have been reported (see Table \ref{tab:MX3}). A plan view of a single layer common to both the BiI\thr\ and AlCl\thr\ structure types is shown on the upper right, with coordinate systems corresponding to each structure type.
}
\end{center}
\end{figure*}

Crystal structure information for layered \tri\ compounds is collected in Table \ref{tab:MX3}. Only~compounds with transition metals containing partially filled \textit{d}-shells are included, since those are the materials in which magnetism may be expected. Non-magnetic, layered trihalides of Sc and Y, with valence electronic configurations 3$d^0$ and 4$d^0$, respectively, are also known \cite{TiCl3-VCl3-str-BiI3, ScBr3, ScI3, YCl3, YBr3, YI3}.  Note~that several of these materials also form in 1D chain structures, but those polymorphs are outside of the scope of the present work. All of the layered compounds have been reported to adopt either the monoclinic AlCl\thr\ structure type or the rhombohedral BiI\thr\ structure type, and this is indicated for each element on the periodic table sections shown in Figure \ref{fig:trihalide-PT}. In these materials the common structural motif is a honeycomb net of $M$ cations that are in edge sharing octahedral coordination, as shown in Figure \ref{fig:trihalide-PT}. In the BiI\thr\ structure the layer stacking sequence is strictly ABC, and the stacking in AlCl\thr\ is approximately ABC. In the former case subsequent layers are shifted along one of the $M-M$ ``bonds'' (to the right in the BiI\thr\ structure shown in Figure \ref{fig:trihalide-PT}), while in the latter case the layers are shifted perpendicular to this direction (into the page in the AlCl\thr\ structure shown in Figure \ref{fig:trihalide-PT}).

In the BiI\thr\ structure the honeycomb net is regular due to the three-fold symmetry. In the AlCl\thr\ structure the honeycomb net can be distorted, and the y-coordinate of the $M$ site determines the degree of distortion. This results in two unique in-plane $M-M$ distances (Table \ref{tab:MX3}). In most of the compounds these two distances are seen to be quite similar, that is the honeycomb nets are nearly undistorted. The two exceptions are the heavier transition metal compounds MoCl\thr\ and TcCl\thr, in which the net is broken into dimers that are separated from one another by a distance about one Angstrom longer than their intradimer distance. Metal-metal bonding in Tc halides including layered ($\beta$) TcCl\thr\ is discussed in [\citenum{Poineau-2013}]. % delete? -deleting the text "Ref." before the reference number is OK with me. (MAM)

Three elements, Ti, Fe, and Ru, are reported to form multiple layered crystal structures with stoichiometry $M$Cl$_3$ (Table \ref{tab:MX3}). This is indicated by the crosshatching on the table in Figure \ref{fig:trihalide-PT}. As noted in Table \ref{tab:MX3}, % there is no Table 3 in the text. -I have corrected this (MAM)
TiCl\thr\ is also reported to form in the trigonal Ti$_3$O structure type. This is similar to the BiI\thr\ structure shown in Figure \ref{fig:trihalide-PT}, but with an ABB stacking sequence. This same structure type is also found for one of the FeCl\thr\ polymorphs, which also forms in a third structure type (trigonal, $P\overline{3}$) with twelve honeycomb layers per unit cell and a $c$ axis length of 70\,{\AA}.

TiCl\thr\ undergoes a structural phase transition at low temperature \cite{Ogawa-1960}. Troyanov et al. demonstrated that the distortion upon cooling corresponds to a dimerization similar to that noted above in MoCl\thr\ and TcCl\thr\ \cite{Troyanov-1991}. Below 220\,K a monoclinic structure was reported. The space group, $C2/m$ is the same as the AlCl\thr\ structure type, but the structure is different, with three layers per unit cell. The dimerization is not as extreme in TiCl\thr\ as it is in MoCl\thr\ and TcCl\thr. At 160\,K the Ti-Ti distances within the distorted honeycomb net are 3.36 and 3.59\,{\AA} \cite{Troyanov-1991}, so the dimerization is not as strong at this temperature, 60\,K below the transition, as it is in MoCl\thr\ and TcCl\thr\ (Table \ref{tab:MX3}) at room temperature. A~structural phase transition is also reported for TiBr\thr, with a triclinic low temperature structure ($P\overline{1}$)~\cite{Troyanov-1994}, and this same triclinic structure was also later reported for TiCl\thr\ \cite{Troyanov-2000}.

All three of the layered chromium trihalides are known to undergo temperature induced crystallographic phase transitions between the AlCl\thr\ and BiI\thr\ structure types \cite{CrCl3-str, CrI3-str}. At high temperatures all three adopt the AlCl\thr\ structure and transition to the BiI\thr\ structure upon cooling. This happens near 240, 420, and 210\,K in the chloride, bromide, and iodide, respectively. The phase transition is first order, displaying thermal hysteresis and a temperature range over which both phases coexist. Interestingly, it is the lower symmetry monoclinic phase that is preferred at higher temperatures. The transition must be driven by interlayer interactions, since the layers themselves are changed little between the two phases. As expected, twinning and stacking faults develops during the transition upon cooling as the layers rearrange themselves into the BiI\thr\ stacking, which can complicate interpretation of diffraction data \cite{CrI3-str}.

Multiple structure types have been assigned to the layered form of RuCl\thr, known as $\alpha$-RuCl\thr. Early reports assigned the trigonal space group $P3_112$ \cite{RuCl3-Stroganov1957} (known as the CrCl$_3$ structure type, although it has been shown that CrCl$_3$ does not actually adopt it) and the AlCl\thr\ type \cite{RuCl3-IrCl3-Brodersen1965}, and a tendency to form stacking defects has been noted \cite{IrBr3-IrI3-RuCl3-str}. The Ti$_3$O type was also reported \cite{Fletcher-1967}. More recently an X-ray and neutron diffraction study reported the monoclinic AlCl\thr\ structure for small single crystals at and below room temperature, and a phase transition in large single crystals from a trigonal structure at room temperature to the monoclinic AlCl\thr\ structure type below about 155\,K \cite{RuCl3-Cao-2016}. A recent report finds high quality crystals undergo a crystallographic phase transition upon cooling from the AlCl\thr-type at room temperature to the BiI\thr-type below about 60\,K \cite{Park-RuCl3}, the same transition described above for Cr$X_3$. Note that even in the monoclinic form the honeycomb net of Ru has little or no distortion (Table \ref{tab:MX3}).

Finally, layered IrCl\thr\ has the AlCl\thr\ structure with a nearly regular honeycomb net (Table \ref{tab:MX3}), but it is also known to adopt a less stable orthorhombic polymorph ($Fddd$). The orthorhombic structure is made up of edge sharing octahedra like the layered structure, but the connectivity extends the structure in three dimensions \cite{IrCl3-ortho}. It is interesting to note that the structure of orthorhombic IrCl\thr\ is made up of fragments of honeycomb nets like those found in the layered structures shown in Figure \ref{fig:trihalide-PT}.

Clearly there are many variants on the stacking sequence in these layered materials due to the weak van der Waals interactions between layers that results in small energy differences between arrangements with different stacking sequences. This is apparent from the crystallographic results from the Ti, Cr, Fe, and Ru trichlorides discussed above. This has been demonstrated using first principles calculations for RuCl\thr\, where multiple structures are found to be very close in energy, and the ground state can depend on the fine details of spin-orbit coupling and electron correlations \cite{RuCl3-Kim2016}. The possibility of mechanically separating these materials into thin specimens or even monolayers is of great interest from the point of view of low dimensional magnetism and potential applications and is greatly facilitated by the weakness of the interlayer interactions. The cleavability of several of these compounds has been studied with first principles calculations, using density functionals that incorporate the weak interlayer dispersion forces that are missing from many conventional functionals. For the Ti, V, and Cr trihalides, cleavage energies are reported to be near 0.3\,J/m$^2$, which is smaller than that of graphite \cite{CrI3-str, Zhang-2015, Zhou-TiCl3-VCl3}. Stable monolayer crystals of CrI\thr\ have recently been demonstrated experimentally \cite{Huang-2017}.

%%%%%%%%%%%%%%%%%%%%%%%%%%%%%%%%%%%%%%%%%%%%%%%%%%%%%%%%%%%%%%%%%%%%%%%%%%%%%%%%%%%%%%%%%%%%%%%%%%%%%%%%%%%%%%%%
%%%%%%%%%%%%%%%%%%%%%%%%%%%%%%%%%%%%%%%%%%%%%%%%%%%%%%%%%%%%%%%%%%%%%%%%%%%%%%%%%%%%%%%%%%%%%%%%%%%%%%%%%%%%%%%%
%%%%%%%%%%%%%%%%%%%%%%%%%%%%%%%%%%%%%%%%%%%%%%%%%%%%%%%%%%%%%%%%%%%%%%%%%%%%%%%%%%%%%%%%%%%%%%%%%%%%%%%%%%%%%%%%
%%%%%%%%%%%%%%%%%%%%%%%%%%%%%%%%%%%%%%%%%%%%%%%%%%%%%%%%%%%%%%%%%%%%%%%%%%%%%%%%%%%%%%%%%%%%%%%%%%%%%%%%%%%%%%%%

\section{Magnetic Structures of Layered, Binary, Transition Metal Halides}

The magnetic order in layered $MX_2$ and $MX_3$ compounds is described below, and some description of the high temperature paramagnetic behavior is given as well. Magnetic excitations and magnetic correlations that develop above the long range ordering temperature are not considered here. Magnetism in these insulating transition metal halide compounds arises from the angular momentum associated with partially filled \textit{d} orbitals. In octahedral coordination, interaction with the coordinating anions split the five \textit{d} orbitals into a set of three levels at lower energy, the $t_{2g}$ levels ($d_{xy}, d_{xz}, d_{yz}$), and two levels at higher energy, the $e_g$ levels ($d_{x^2-y^2}, d_{z^2}$). According to Hund's rules, the $d$ electrons first fill these states singly with their spins parallel, unless the energy cost of putting electrons in the higher energy $e_g$ states overcomes the cost of doubly occupying a single state. In addition to their spin, the electrons in these levels also have orbital angular momentum. In ideal octahedral coordination, the total orbital angular momentum can be shown to be zero for certain electronic configurations. This arises due to rotational symmetry of the system, and when this occurs the orbital angular momentum is said to be ``quenched''. For octahedral coordination the orbital angular momentum is quenched when there is exactly one electron in each of the $t_{2g}$ orbitals, and when there are two electrons in each of the $t_{2g}$ orbitals. Otherwise there is an orbital moment that must be considered. There is no orbital angular momentum associated with the $e_g$ orbitals. Of course, distortions of the octahedral environments can affect the details of the magnetism that are based on symmetry and degeneracy of electronic states. Despite the partially filled $d$-orbitals, the materials considered here are electrically insulating under ambient conditions. This can be attributed to a Mott-Hubbard type mechanism by which electron-electron interactions produce a band gap related to the Coulomb repulsion among the well-localized electrons (see for example \cite{Khomskii}). % delete? -the text "Ref." can be deleted before the reference number. (MAM)

Magnetism in a material is often first characterized by measurements of magnetization ($M$) as functions of applied magnetic field ($H$) and temperature ($T$). Considering magnetic interactions between localized magnetic moments, the temperature dependence of the magnetic susceptibility ($\chi = M/H$) can often be described by the Curie-Weiss formula, $\chi(T) = $C/$(T-\theta)$. The Curie constant (C) is a measure of the size of the magnetic moment and is given by $C = \frac{N_A}{3k_B}\mu_B^2g^2S(S+1)$, where~$N_A$ is Avogadro's number, $k_B$ is the Boltzmann constant, $\mu_B$ is the Bohr magneton, $S$ is the total spin, and $g = 2.00$ is the electron gyromagnetic ratio. The ``effective moment'' ($\mu_{eff}$) is also often quoted, $\mu_{eff} = g\sqrt{S(S+1)}\mu_B$. In cgs units, $\mu_{eff} \approx \sqrt{8C}$. Fully ordered magnetic moments are expected to be equal to $gS$ in units of Bohr magnetons. When both orbital and spin moments are present, the total angular momentum and associated g-factor must be used. The Weiss temperature ($\theta$) is a measure of the strength of the magnetic interactions. Considering magnetic interactions between nearest neighbors of the form $H_{ij} = -J \overrightarrow{S_i} \cdot \overrightarrow{S_j}$, it can be shown that the Weiss temperature depends on the spin $S$, the magnetic exchange interaction strength $J$, and the number of nearest neighbors $z$ according to $\theta = \frac{2zJ}{3k_B}S(S+1)$. Positive values of $\theta$ indicate positive values of $J$, which indicate ferromagnetic interactions. Negative values of $\theta$ indicate antiferromagnetic interactions. In a simple mean field model, the Weiss temperature corresponds to the ordering temperature ($T_{C,N}$\,=\,$|\theta|$). Note that the presence of multiple types of interactions, for example FM intralayer interactions and AFM interlayer interactions, complicates the interpretation of Weiss temperatures.

In the materials considered here, the in-plane magnetic interactions between transition metal cations are expected to arise mainly from superexchange through shared coordinating halogen anions. The sign of the superexchange interaction depends upon many factors, including the orbital occupations and the $M-X-M$ angle (see, for example, the discussion in \cite{Khomskii}). % delete? -the text "Ref." can be deleted before the reference number. (MAM)
It is often AFM and strong when the angle is 180$^{\circ}$. When this angle is 90$^{\circ}$, as it is in the edge sharing octahedral coordination found in layered \di\ and \tri\ compounds, superexchange can be either FM or AFM. There are also direct $M-M$ exchange interactions, which tend to be AFM, but this is expected to be relatively weak in these materials due to the relatively large $M-M$ distances. The in-plane magnetic order in most of the compounds described below either is ferromagnetic, contains ferromagnetic stripes, or has a helimagnetic arrangement. The later two scenarios are expected to arise from competing magnetic interactions. The exceptions are V$X_2$ in which the interactions are predominantly AFM \cite{Niel-1977}, and perhaps TiCl\two.

Note that in the figures below showing the magnetic structures of \di\ and \tri\ compounds, only the $M$ sublattices are shown. The magnetic moments directions are indicated by red arrows. In addition, to make the magnetic structures easier to view, different colored balls are used to represent atoms with moments along different directions, except for in the more complex helimagnetic structures.

\subsection{\textit{MX}$_2$ Compounds}

\subsubsection{Ti\textit{X}$_2$ and Zr\textit{X}$_2$}

These compounds have Ti and Zr in the unusual formal oxidation state of 2+. However, as noted above, metal-metal bonds are present in ZrI\two, so this simple electron counting is invalid in this case. Divalent~Ti and Zr in Ti$X_2$ and ZrCl\two\ have electron configurations of $3d^2$ and $4d^2$, respectively, with an expected spin of $S = 1$. There have been very few magnetic studies of these materials, likely due in part to their instability and reactivity. Magnetic susceptibility measurements on TiCl\two\ down to 80\,K have revealed a cusp near 85\,K \cite{Lewis-1962}. The authors suggest that this may indicate antiferromagnetic ordering at this temperature, although they note that previous measurements showed smoothly increasing susceptibility upon cooling from 300 to 20\,K \cite{Starr-1940}, but this was based on only six temperature points and significant features could have been overlooked. Magnetic susceptibility versus temperature curves have somewhat unusual shapes, and effective moments of 1.1 and 2.0\,$\mu_B$ per % is the \mu italics necessary? -no it is not. that's just how \mu comes out in math mode. (MAM)
Ti have been reported~\cite{Lewis-1962, Starr-1940}. A Weiss temperature of $-$702\,K was determined by Starr et al. \cite{Starr-1940}, which would indicate strong antiferromagnetic interactions. Frustration of these interactions by the triangular Ti lattice may be responsible for the relatively low ordering temperature of 85\,K proposed in Ref. \cite{Lewis-1962}.

ZrCl\two\ is reported to have a reduced magnetic moment at room temperature \cite{Lewis-1962}, but no temperature dependent data were reported. The authors suggest that this may indicate strong antiferromagnetic interactions between Zr magnetic moments. No magnetic structure determinations for TiCl\two\ were located in the literature and no magnetic information was found for TiBr\two\ or TiI\two.

\subsubsection{V\textit{X}$_2$}

These materials contain divalent V with an electronic configuration $3d^3$, $S = 3/2$. An early report on VCl\two\ found it to be paramagnetic with a large negative Weiss temperature ($-$565\,K) indicating strong antiferromagnetic interactions \cite{Starr-1940}. Niel et al. later reported Weiss temperatures of $-$437, $-$335, and $-$143\,K for VCl\two, VBr\two, and VI\two, respectively, with effective moments close to the expected value of 3.9\,$\mu_B$, and % is the \mu italics necessary? -no it is not. that's just how \mu comes out in math mode. (MAM)
explained their behavior in terms of a 2D Heisenberg model \cite{Niel-1977}.

\begin{figure}
\begin{center}
\includegraphics[width=3.0in]{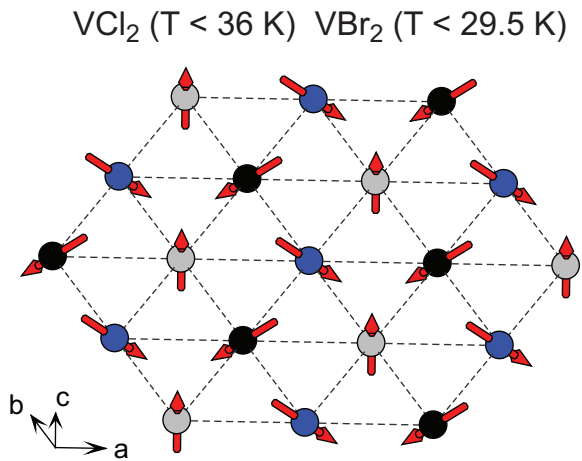}
\caption{\label{fig:VCl2-VBr2-mag-str}
The 120$^{\circ}$ magnetic structure determined for VCl\two\ and VBr\two. The image shows a single triangular net of V atoms lying in the \textit{ab} plane. Moments on the grey atoms are along the \textit{c} axis. Moments on the black and blue atoms lie in the \textit{ac} plane and are rotated by an angle of $\pm$120$^{\circ}$ from the \textit{c} axis.
}
\end{center}
\end{figure}

A neutron powder diffraction study showed that all three of the vanadium dihalides order antiferromagnetically with N\'{e}el temperatures of 36.0\,K for VCl\two, 29.5\,K for VBr\two\ and 16.3\,K for VI$_2$~\cite{Hirakawa-1983}. The strong suppression of these ordering temperatures relative to the Weiss temperatures is a result of geometrical frustration. Both temperatures trend to lower values as the halogen is changed from Cl to Br to I. Further neutron scattering experiments revealed that the magnetic order in VCl\two\ develops in two steps, with phase transition temperatures separated by about 0.1\,K, and found the magnetic structure at low temperature to be a 120$^{\circ}$ N\'{e}el state shown in Figure \ref{fig:VCl2-VBr2-mag-str}, where each moment in the triangular lattice is rotated by this angle with respect to its neighbors, with moments in the $ac$-plane \cite{Kadowaki-1987}. The ordered moment corresponded to a spin of 1.2. In that study, three types of critical behavior were observed, corresponding to 2D Heisenberg, 3D Heisenberg, and 3D Ising models. A similar magnetic structure was found for VBr\two\ with moments of about 83\% of the expected value \cite{Kadowaki-1985}. The magnetic order in VI\two\ develops in two steps with $T_{N1} = 16.3$\,K and $T_{N2}$ near 15\,K, but the low temperature magnetic structure of this compound was not resolved with any certainty \cite{Hirakawa-1983}.

Recently Abdul Wasey \textit{et al.} proposed V$X_2$ materials as promising candidates for extending 2D materials beyond graphene and dichalcogenides \cite{Abdul-Wasey-2013}. They report results of first principles calculations of the magnetic order in these systems in both bulk and monolayer forms. In the bulk crystal the experimental spin structure was reproduced. A similar structure is predicted for the monolayer, and the authors suggest that magnetic order in the monolayer may occur at much higher temperature than in the bulk.

\subsubsection{Mn\textit{X}$_2$}

Divalent Mn has a $3d^5$ electronic configuration, with $S=5/2$. Magnetization measurements for MnCl\two\ indicate weak antiferromagnetic interactions ($\theta$\,=\,$-$3.3\,K) and an effective moment of 5.7\,$\mu_B$, close to the expected value of 5.9\,$\mu_B$ \cite{Starr-1940}. Heat capacity measurements indicate magnetic phase transitions at 1.96 and 1.81\,K \cite{Murray-1962}. The magnetic structures of MnCl\two\ below these two transitions have not been completely determined. Neutron diffraction from single crystals were analyzed assuming a collinear structure and complex orderings with stripes of ferromagnetically aligned spins in the plane were proposed \cite{Wilkinson-MnCl2-1958}. A more recent investigation of MnCl\two-graphite intercalation compounds found that the magnetic order within isolated MnCl\two\ layers could be described by an incommensurate helimagnetic arrangement, and it was suggested that this may also hold for the magnetic structure of the bulk crystal~\cite{Wiesler-1995}.

\begin{figure}
\begin{center}
\includegraphics[width=2.4in]{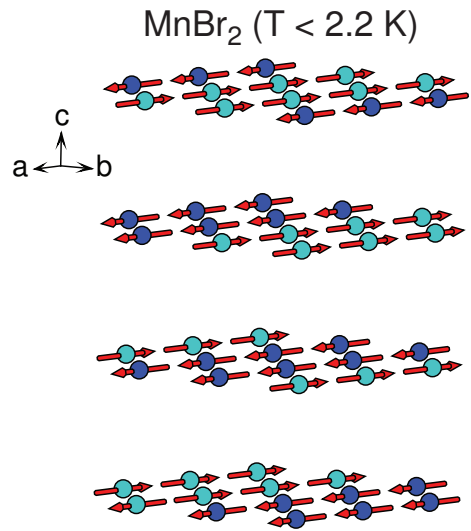}
\caption{\label{fig:MnBr2-mag-str}
The lowest temperature, commensurate magnetic structure of MnBr\two. The coordinate system refers to the underlying hexagonal crystal structure. Dark blue is used for Mn atoms with moments along the hexagonal \textit{a} direction, and light blue for those with moments along -\textit{a}. % change to minus? -minus is correct (MAM)
}
\end{center}
\end{figure}

A heat capacity anomaly was reported at 2.16\,K in MnBr\two, and neutron diffraction showed that antiferromagnetic order is present below this temperature \cite{MnBr2-str}. The magnetic structure has ferromagnetic stripes within the layers with antiferromagnetic coupling between neighboring stripes, as depicted in Figure \ref{fig:MnBr2-mag-str}. The moments are along the \textit{a} axis of the hexagonal cell of the crystal structure. There is antiferromagnetic order between the layers. Later, an incommensurate magnetic phase was identified between this phase and the paramagnetic state, persisting up to about 2.3\,K \cite{Sato-1995}.

MnI\two\ adopts a complicated helical magnetic structure below 3.4\,K \cite{Cable-1962}. The moments lie in the (307) planes, and are ferromagnetically aligned within each of these planes. The variation of the moment direction upon moving between (307) planes was originally reported to be a rotation by 2$\pi$/16~\cite{Cable-1962}. Further  measurements resolved multiple phase transitions as the magnetic order develops and find the helical ordering to be incommensurate, but with a wave vector close to that reported in the earlier work \cite{Sato-1995}.

It was recently noted that a ferroelectric polarization develops in the magnetically ordered state of MnI\two, spurring interest in this compound as a multiferroic material \cite{Kurumaji-2011, Wu-2012}. Density functional theory calculations suggest that spin-orbit coupling on the iodine ions is the main source of the ferroelectric polarization in MnI\two\ \cite{Wu-2012}, which has been measured to exceed 120\,$\upmu$C/m$^2$ \cite{Kurumaji-2011}. While spin-orbit coupling is required to accurately describe the polarization, is was found to have little influence on the magnetic interactions determined by fitting density functional theory results to a Heisenberg model \cite{Wu-2012}. In that study it was found that the observed helimagnetic order arises from competing magnetic interactions on the triangular Mn lattice, that electronic correlations, which weaken AFM superexchange, must be considered to accurately reproduce the experimental magnetic structures, and that the details of the spiral structure are sensitive to relatively strong interplane magnetic interactions. The ferroelectric polarization responds to applied magnetic fields in multiferroic MnI\two. Magnetic fields affect the polarization by modifying the helimagnetic domain structure at low fields, and by changing the magnetic order at higher fields \cite{Kurumaji-2011}. Ferroelectric distortions onsetting at the magnetic ordering temperatures and associated multiferroictiy is a common occurrence in $MX_2$ compounds that adopt non-collinear magnetic structures (see CoI\two, NiBr\two, and NiI\two\ below); however, the details of the coupling between the spin and electric polarization in these and related triangular lattice multiferroics is not well understood \cite{Kurumaji-2013}.

\subsubsection{Fe\textit{X}$_2$}

The divalent iron, $3d^6$, in these compounds is expected to be in the high spin state with $S=2$. The partially filled $t_{2g}$ levels means that orbital angular momentum is not quenched, and an orbital moment may be expected, as discussed in [\citenum{Katsumata-2010}] % delete -the text "Ref." can be deleted (MAM)
and references therein. Significant anisotropy is observed in the paramagnetic state in all three of the iron dihalides, with a larger effective moment measured along the $c$ axis \cite{Bertrand-1974}. Moments in the magnetically ordered states are also along this direction. Weiss temperatures determined from measurements with the field in the plane ($||$) and out of the plane ($\perp$) are  9\,K ($||$) and 21\,K ($\perp$) for FeCl\two, $-$3.0\,K ($||$) and 3.5\,K ($\perp$) for FeBr\two, and 24\,K ($||$) and 21.5\,K ($\perp$) for FeI\two\ \cite{Bertrand-1974}.

\begin{figure*}
\begin{center}
\includegraphics[width=5.4in]{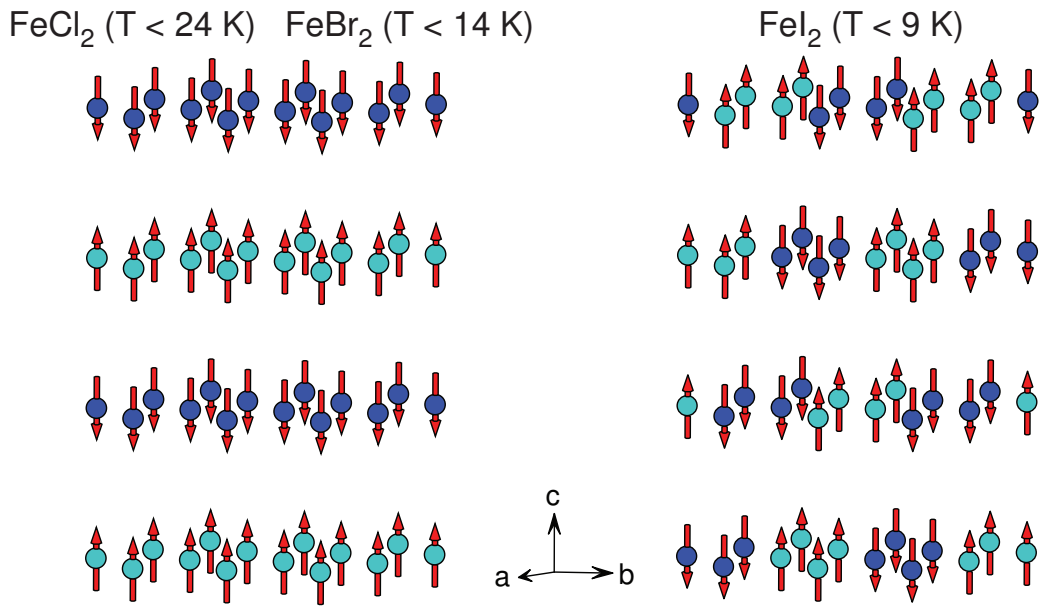}
\caption{\label{fig:FeX2-mag-str}
The magnetic structures of Fe$X_2$ with the hexagonal coordinate system of the underlying crystal lattice shown. All moments are along either the +\textit{c} or -\textit{c} direction.
}
\end{center}
\end{figure*}

Although the crystallographic structures of FeCl\two\ and FeBr\two\ differ (Table \ref{tab:MX2}), they have the same ordered arrangement of spins at low temperature. This magnetic structure is shown in Figure \ref{fig:FeX2-mag-str}, and contains ferromagnetic intralayer order and antiferromagnetic stacking. The chloride orders below~24\,K and has an ordered moment of 4.5\,$\mu_B$ \cite{Wilkinson-1959}, and the bromide orders below 14\,K and has an ordered moment of 3.9\,$\mu_B$ \cite{Wilkinson-1959, Fert-1973-FeBr2}. The iodide adopts a different low temperature structure below its N\'{e}el temperature of 9\,K, with two-atom-wide ferromagnetic stipes in the plane (ordered moment of 3.7\,$\mu_B$) that are aligned antiferromagnetically with neighboring stripes \cite{FeI2-str}. The moment arrangement is shifted from layer to layer so that the magnetic unit cell contains four layers. This is similar to the magnetic structure of MnBr\two\ shown in Figure \ref{fig:MnBr2-mag-str}, but the stripes run in different directions in the plane. There is no apparent correlation between the Weiss temperatures and magnetic ordering temperatures in the Fe$X_2$ series. This is likely related to the presence of both FM and AFM interaction in these materials, which complicates interpretation of the fitted Weiss temperatures.

As noted above, FeCl\two\ undergoes a transition from the CdCl\two\ structure to the CdI\two\ structure at a pressure of 0.6\,GPa. At higher pressures two additional phase transitions occur, with pronounced effects on the magnetic behavior. Above 32\,GPa the orbital moment is quenched and the magnetic moments cant away from the $c$ axis. A further increase in pressure results in the collapse of the magnetization and an insulator-metal transition that is attributed to delocalization of the Fe $d$ electrons~\cite{Xu-2002, FeCl2-pressure}. Similar behavior is reported for FeI\two\ \cite{Pasternak-2001}. In both materials the N\'{e}el temperature in increased with applied pressure, and reaches room temperature before collapsing into the non-magnetic~state.

In the antiferromagnetic state, magnetic field induced phase transitions, or metamagnetic transitions, occur in FeCl\two, FeBr\two, and FeI\two\ at applied fields near 11, 29, and 46\,kOe, \scalebox{0.95}[0.95]{respectively \cite{Jacobs-1967-FeCl2, Fert-1973-FeBr2, Fert-1973-FeI2}}. This arises from stronger ferromagnetic coupling within the layers compared to the weak antiferromagnetic coupling between them, and led to much of the early interest in these materials, as summarized in [\citenum{Lines-1963}] % delete? -the text "Ref." can be deleted (MAM)
and references therein. The most complex behavior is seen in FeI$_2$~\cite{Fert-1973-FeI2}. From magnetization and heat capacity measurements, Katsumata et al. identified~four different field induced phases, in addition to the antiferromagnetic ground state, and proposed ferrimagnetic structures for them \cite{Katsumata-2010}. In addition, Binek et al. have proposed the emergence of a Griffith's phase in FeCl$_2$~\cite{Binek-1994, Binek-1996}, and neutron diffraction has been used to construct the temperature-field magnetic phase diagram of FeBr\two\ \cite{Katsumata-1997}.

\subsubsection{Co\textit{X}$_2$}

Cobalt dihalides have cobalt in electronic configuration $3d^7$, which can have a high ($S = 3/2$) or low ($S = 1/2$) spin state. Orbital magnetic moments may be expected in either state. It is apparent from neutron diffraction results that the high spin state is preferred, at least for CoCl\two\ and CoBr\two. The ordered moment on Co in CoCl\two\, which orders below 25\,K \cite{Chisholm-1962}, is 3.0\,$\mu_B$ \cite{Wilkinson-1959}, and it is 2.8\,$\mu_B$ in CoBr$_2$~\cite{Wilkinson-1959}, which orders at 19\,K \cite{Yoshizawa-1980}. These are close to the expected value of $gS$ for $S = 3/2$ for high-spin only. However, magnetization measurements on CoCl$_2$~\cite{Starr-1940} indicate an enhanced effective moment in the paramagnetic state (5.3\,$\mu_B$), which suggests an orbital contribution, and a Weiss temperature of~38\,K.

\begin{figure*}
\begin{center}
\includegraphics[width=5.4in]{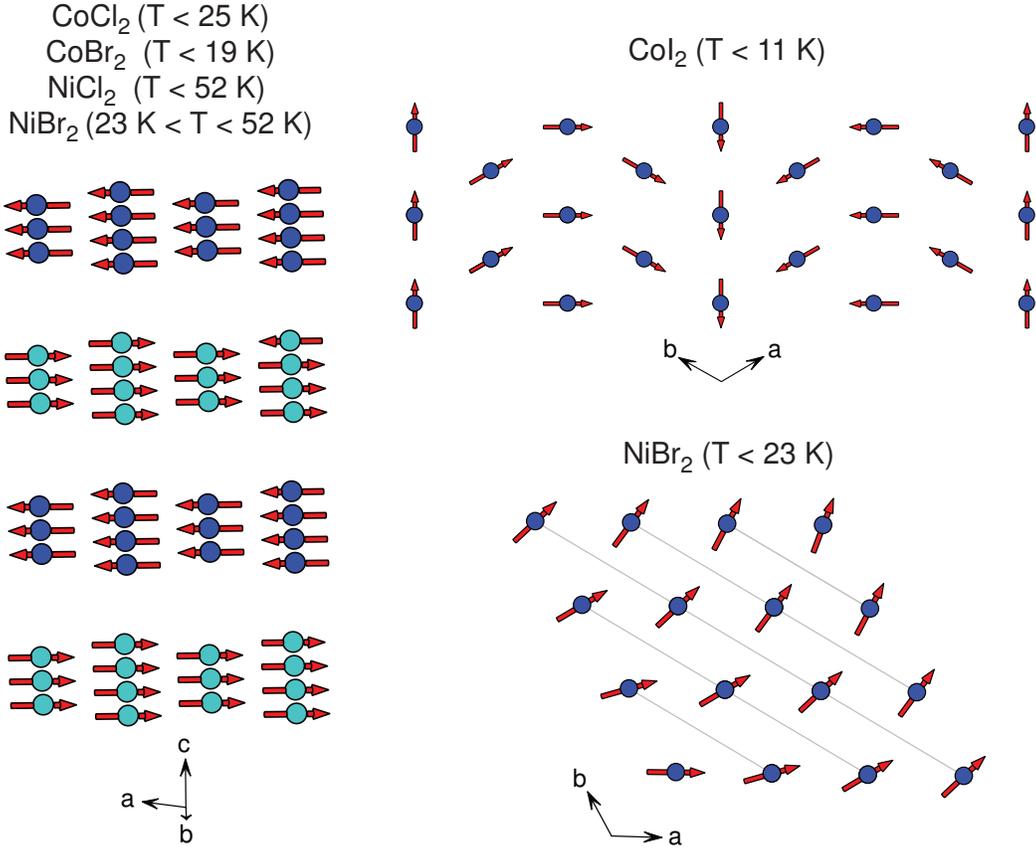}
\caption{\label{fig:NiCoClBr-mag-str}
Magnetic structures of Co$X_2$ and Ni$X_2$ for $X$\,=\,Cl and Br, and CoI\two.  The moments lie in the plane in each case, and are known to be parallel and antiparallel to the [210] direction in the commensurate structures of CoCl\two\ and Ni$X_2$. A layer of the commensurate, cycloidal, helimagnetic structure of CoI\two\ reported in \cite{Kuindersma-1981} is shown. % delete? -the text "Ref." can be deleted (MAM)
A more complex cycloidal structure has also been reported~\cite{Mekata-1992}. Only a small portion of one layer of the lower temperature (T\,$<$\,23\,K) long period helimagnetic structure of NiBr\two\ is shown. Coordinate systems of the hexagonal crystallographic unit cells shown for reference.
}
\end{center}
\end{figure*}

Below their ordering temperatures, both of these compounds adopt the magnetic structure shown in Figure \ref{fig:NiCoClBr-mag-str}, with ferromagnetic alignment within each layer and antiferromagnetic stacking. The~moments are known to be parallel or antiparallel to the hexagonal [210] direction for CoCl\two\ \cite{Wilkinson-1959}, as shown in the Figure. The moments in CoBr\two\ are only known to lie within the \textit{ab} plane \cite{Wilkinson-1959}.

The magnetic behavior in CoI\two\ is more complex. CoI\two\ is a helimagnet with a spiral spin structure, and anisotropic magnetic susceptibility in the paramagnetic state arising from spin-orbit coupling~\cite{Mekata-1992}. Powder neutron diffraction analysis indicated a cycloidal structure with moments in the plane and planes stacked antiferromagnetically \cite{Kuindersma-1981}, which is supported by M\"{o}ssbauer spectroscopy \cite{Friedt-1976}. The~corresponding in-plane spin arrangement is shown Figure \ref{fig:NiCoClBr-mag-str}.  Mekata et al. used single crystal neutron diffraction to examine the magnetic order in CoI\two\ and found evidence of a more complicated magnetic structure that requires an additional propagation vector to describe. The same study identified a first order magnetic phase transition at 9.4\,K, just below the magnetic ordering transition at 11.0\,K, and suggested that these successive transitions may arise due to in-plane magnetic frustration, but no change in the magnetic structure was observed at 9.4\,K \cite{Mekata-1992}.

An electric polarization of about 10\,$\mu$C/m$^2$ that varies with applied magnetic field is induced below the magnetic ordering transition in CoI\two\ indicating multiferroic behavior \cite{Kurumaji-2013} (see MnI\two\ above, NiBr\two, NiI\two\ below).

\subsubsection{Ni\textit{X}$_2$}

The octahedrally coordinated, divalent nickel in these compounds has a $3d^8$ electronic configuration, with filled $t_{2g}$ and half-filled $e_g$ orbitals. Magnetic moments are expected to be spin only, as orbital angular momentum is quenched in this configuration. Magnetization data for NiCl\two\ indicate an effective moment of 3.3\,$\mu_B$, somewhat larger than the spin only value of 2.8\,$\mu_B$ expected for $S = 1$, and Weiss temperature of 68\,K, suggesting predominantly ferromagnetic interactions~\cite{Starr-1940}. The N\'{e}el temperatures of NiCl\two\ and NiBr\two\ are quite similar; upon cooling, both develop long range antiferromagnetic order below 52\,K \cite{Busey-1952, Adam-1980}. Their fully ordered moments are 2.11 and 2.0\,$\mu_B$, respectively \cite{DeGunzbourg-1971, Adam-1980}, as expected for $S = 1$. The resulting magnetic structure is shown in Figure~\ref{fig:NiCoClBr-mag-str}a. The moments lie within the $ab$ plane and are ferromagnetically aligned within each layer, with antiferromagnetic stacking. The moment directions were determined from M\"{o}ssbauer spectroscopy to be parallel and antiparallel to the [210] direction, as depicted in the figure \cite{Pollard-1982}.

While the magnetic structure shown in Figure \ref{fig:NiCoClBr-mag-str} describes NiCl\two\ at all temperatures below $T_N$, NiBr\two\ undergoes a second phase transition, to a more complicated magnetic structure below 23\,K~\cite{Day-1976, Day-1980, Adam-1980}. Below this first order transition the magnetic moments adopts an incommensurate helimagnetic structure with a periodicity that varies with temperature. As described by Adam et~al., the magnetic moments still lie within the basal plane, but vary in direction at 4.2\,K by 9.72$^{\circ}$ from site to site along both the hexagonal $a$ and $b$ axes \cite{Adam-1980}, as depicted in Figure \ref{fig:NiCoClBr-mag-str}. This results in a periodicity of about 37 crystallographic unit cells along each in-plane direction. The stacking remains~antiferromagnetic.

Heat capacity data show that NiI\two\ undergoes two phase transitions upon cooling, at 75 and 60\,K~\cite{Billery-1977}. Helimagnetic order develops at 75\,K, and the phase transition at 60\,K is crystallographic~\cite{Kuindersma-1981}. The helimagnetic structure of NiI\two\ is incommensurate with the nuclear structure and the moments rotate in a plane that makes a 55$^{\circ}$ angle with the $c$ axis, as depicted in \cite{Kuindersma-1981}. %delete? The text "Ref." can be deleted (MAM)
The ordered moment at 4.2\,K was determined to be 1.6\,$\mu_B$.

Like helimagnetic MnI\two\ and CoI\two\ described above, NiBr\two\ and NiI\two\ also develop a ferroelectric polarization in their helimagnetic states \cite{Tokunaga-2011, Kurumaji-2013}. Polarizations of 20$-$25\,$\mu$C/m$^2$ are observed in the bromide, and polarizations exceeding 120\,$\mu$C/m$^2$ are reported for the iodide. As in MnI\two\ and CoI\two, the~polarization can be controlled by applied magnetic fields through their influence on the helimagnetic domain structure \cite{Tokunaga-2011, Kurumaji-2013}.

%%%%%%%%%%%%%%%%%%%%%%%%%%%%%%%%%%%%%%%%%%%%%%%%

\subsection{\textit{MX}$_3$ Compounds}

Several of the $MX_3$ compounds listed in Table \ref{tab:MX3} are not known to form magnetically ordered states. These include Ti$X_3$, MoCl\thr, TcCl\thr, Rh$X_3$, and Ir$X_3$. The later two materials have electron configuration $4d^6$, and are expected to have non-magnetic ground states with all electrons paired. A clue to the non-magnetic nature of MoCl\thr\ \cite{Schafer-1967} is found in the magnetic behavior of TiCl\thr. Although neutron diffraction shows no magnetic ordering in layered TiCl\thr\ at low temperature, magnetic susceptibility shows a dramatic and sharp decrease near 217\,K. This corresponds to the structural distortion noted above in the discussion of TiCl\thr\ and described in [\citenum{Troyanov-1991}] and [\citenum{Troyanov-2000}]. % delete? The text "Ref." can be deleted (MAM)
Ogawa had earlier observed a lattice response coincident with the magnetic anomaly, and proposed that the formation of covalently bonded Ti-Ti dimers that pair the $d$ electrons on each Ti as the reason for the collapse of the magnetic moment~\cite{Ogawa-1960}. Thus the strong dimerization in MoCl\thr\ (Table \ref{tab:MX3}) is expected to be responsible for its non-magnetic nature. Dimerized TcCl\thr\ is also expected to be non-magnetic \cite{Poineau-2013}.

\subsubsection{V\textit{X}$_3$}

Little information about magnetic order in VCl\thr\ or VBr\thr\ is available. These compounds are expected to be magnetic due to their electron configuration $3d^2$ ($S = 1$) and the undistorted honeycomb net of the BiI\thr\ structure type reported for these materials (Table \ref{tab:MX3}). Magnetic susceptibility data \cite{Starr-1940} for VCl\thr\ give an effective moment of 2.85\,$\mu_B$, close to the expected value for $S = 1$ (2.82\,$\mu_B$), and a Weiss temperature of $-$30\,K, indicating antiferromagnetic interactions. The maximum displayed near 20\,K in the temperature dependence of the susceptibility suggests antiferromagnetic order at lower temperatures. First principles calculations have been done to examine the electronic and magnetic properties of monolayers of VCl\thr\ and (hypothetical) VI\thr\ \cite{Zhou-TiCl3-VCl3, He-2016}. Both are predicted to be ferromagnetic.

\subsubsection{Cr\textit{X}$_3$}

\begin{figure}
\begin{center}
\includegraphics[width=3.5in]{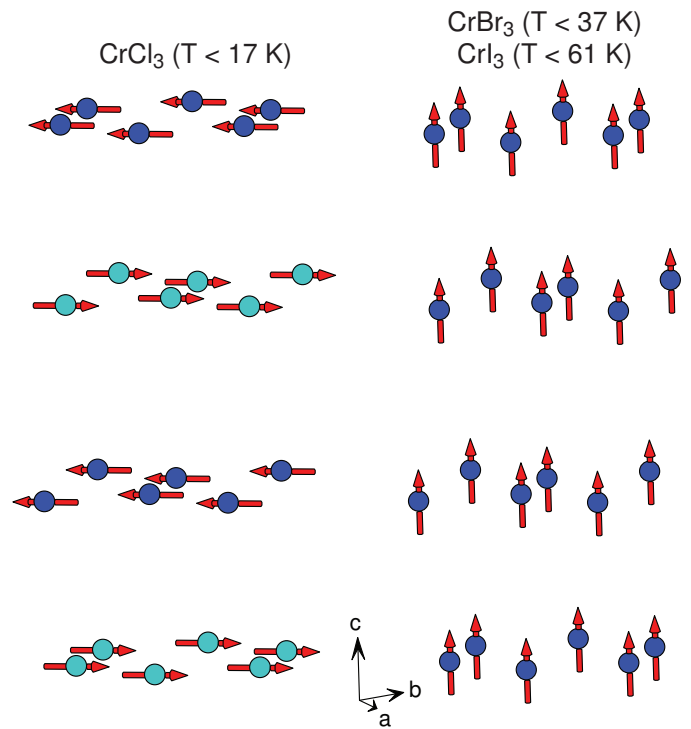}
\caption{\label{fig:CrX3-mag-str}
Magnetic structures of Cr$X_3$. The hexagonal coordinate system of the low temperature rhombohedral structure is shown. The moments in CrCl\thr\ are drawn along the [110] and % please confirm whether [100] and [110] are right. -[110] is correct (MAM)
 [$\overline{11}$0] directions here, but are only known to be in the \textit{ab} plane. Moments are along the \textit{c} axis in ferromagnetic CrBr\thr\ and CrI\thr.
}
\end{center}
\end{figure}

In these compounds Cr is expected to be in a $3d^3$ electronic configuration, and effective moments determined from high temperature magnetic susceptibility range from 3.7 to 3.9\,$\mu_B$ per Cr
% change to to? -I made the change (MAM)
 as expected for $S=3/2$ \cite{Starr-1940, Tsubokawa-1960-CrBr3, CrI3-str}. Weiss temperatures determined from these measurements are 27, 47, and 70\,K for CrCl\thr, CrBr\thr, and CrI\thr, respectively, indicating predominantly ferromagnetic interactions. In fact, among the layered \di\ and \tri\ materials, the chromium trihalide family contains the only compounds in which long-range, 3D ferromagnetic ground states are observed. The magnetic structures are shown in Figure \ref{fig:CrX3-mag-str}. Below 61\,K for CrI\thr\ and 37\,K for CrBr\thr, moments directed out of the plane order ferromagnetically \cite{Hansen-1959, Tsubokawa-1960-CrBr3, Dillon-1965, CrI3-str}. In CrCl\thr\ below about 17\,K, ferromagnetic order is also observed within the layers, but the layers stack antiferromagnetically \cite{Cable-1961-CrCl3, Kuhlow-1982-CrCl3}. Also unlike the tribromide and triiodide, the moments in CrCl\thr\ lie within the planes. In this series, the ordering temperatures scale nicely with the Weiss temperatures. The ordered moments determined by neutron diffraction and magnetic saturation are all close to 3$\mu_B$ as expected for the $3d^3$ electronic configuration of Cr$^{3+}$. Reported values are 2.7--3.2\,$\mu_B$  for CrCl\thr\ \cite{Cable-1961-CrCl3}, 3\,$\mu_B$ for CrBr\thr\ \cite{Tsubokawa-1960-CrBr3}, and 3.1\,$\mu_B$for CrI\thr\ \cite{CrI3-str, Dillon-1965}. Significant magnetic anisotropy is observed in the ferromagnetic state of CrI\thr; the anisotropy field, the field required to rotate the ordered moments away from the $c$-axis and into the $ab$-plane, is found to be near 30\,kOe near 2\,K \cite{CrI3-str, Dillon-1965}. Ferromagnetic CrBr\thr\ has a significantly lower anisotropy field of about 5\,kOe \cite{Hansen-1959, Tsubokawa-1960-CrBr3}.

 With moments in the plane and antiferromagnetic stacking of the layers CrCl\thr\ is unique among the chromium trihalides. It also has weak magnetic anisotropy. A magnetic field of only a few kOe is sufficient to overcome the antiferromagnetic order and fully polarize the magnetization in any direction \cite{Cable-1961-CrCl3, Kuhlow-1982-CrCl3}. Using the optical technique of Faraday rotation, Kuhlow followed closely the evolution of the magnetization in CrCl\thr\ with changing temperature and applied magnetic field \cite{Kuhlow-1982-CrCl3}. It was noted that the magnetic order appears to onset in two stages upon cooling, first developing ferromagnetic correlations and 16.7\,K with long range antiferromagnetic order as shown in Figure \ref{fig:CrX3-mag-str} below 15.5\,K.

The ferromagnetism in these Cr$X_3$ compounds makes them particulary interesting for incorporating magnetism into functional van der Waals heterostructures. Several relevant theoretical studies have been reported that suggest ferromagnetic order may persist into monolayer \scalebox{0.95}[0.95]{specimens \cite{Wang-2011, CrI3-str, Liu-2016, Zhang-2015}}. Recently ferromagnetic monolayers of CrI$_3$ were demonstrated experimentally \cite{Huang-2017}. Ferromagnetic CrI\thr\ was also recently incorporated into a van der Waals heterostructure in which an exchange field effect equivalent to a 13\,T applied magnetic field was observed in the electronic properties of monolayer WSe\two\ when the heterostructure was cooled through the Curie temperature of CrI\thr\ \cite{Zhong-2017}. Although CrCl\thr\ has an antiferromagnetic structure in the bulk, each layer is ferromagnetically ordered. If this proves to be independent of sample thickness then ferromagnetic monolayers may be realized in all three of the chromium trihalides, with a range of magnetic anisotropy that may allow easy tuning of the magnetization direction in the chloride or more robust moment orientation in the~iodide.

\subsubsection{Fe\textit{X}$_3$}

FeCl\thr\ and FeBr\thr\ have iron in the $3d^5$ configuration. There has been considerable study of the magnetism in the chloride, but very little for the bromide. Early magnetization measurements on FeCl\thr\ found an effective moment of 5.7\,$\mu_B$, close to the expected spin-only value of 5.9\,$\mu_B$, and a Weiss temperature of $-$11.5\,K, indicating antiferromagnetic interactions \cite{Starr-1940}. A neutron diffraction study found a helimagnetic structure for FeCl\thr\ below about 15\,K, with an ordered moment on 4.3\,$\mu_B$ per iron at 4.2\,K \cite{Cable-1962-FeCl3}. The reduction from the expected value of 5\,$\mu_B$ may be due to some disorder still present at 4.2\,K or could arise from a slight distortion from the periodic model used to describe the magnetic order. Later magnetization measurements place the N\'{e}el temperature at 9--10\,K \cite{Jones-1969, Johnson-1981}.

\begin{figure}
\begin{center}
\includegraphics[width=3in]{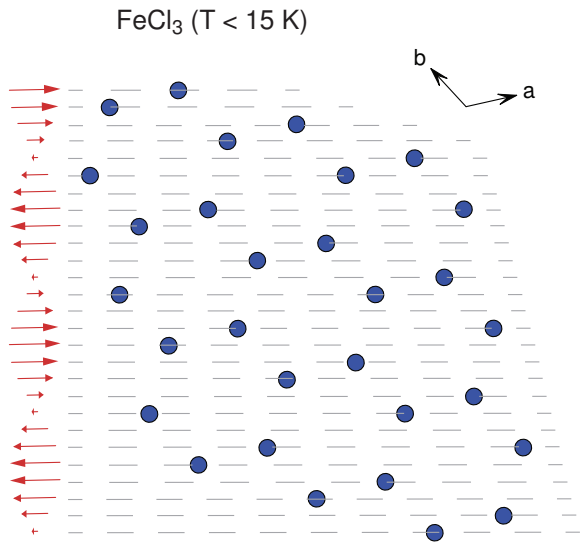}
\caption{\label{fig:FeCl3-mag-str}
The helimagnetic structure of FeCl\thr. A portion of one layer of Fe atoms is shown, along with the hexagonal coordinate system. The dotted lines indicate (140) planes. The moment of the atoms lying on these planes is indicated by the red arrows at the left of each line. Moments are in the (140) planes and have components in and out of the page.
}
\end{center}
\end{figure}

The magnetic structure of FeCl\thr\ is shown in Figure \ref{fig:FeCl3-mag-str}. The figure shows one layer of Fe atoms, with dashed lines denoting (140) planes. Sites in this layer on a common (140) plane have parallel moments with their orientation indicated at the left of the Figure. Note that the moments all have the same magnitude, but their projections on to the plane of the page vary as their orientations rotate about the [140] direction by 2$\pi$/15 between neighboring planes \cite{Cable-1962-FeCl3}. The layers stack antiferromagnetically. A field induced magnetic phase transition was noted in FeCl\thr\ by Stampfel et al. and Johnson et~al. with the magnetic structure evolving with field up to about 15\,kOe and experiencing a spin-flop \scalebox{0.95}[0.95]{near 40\,kOe \cite{Stampfel-1973, Johnson-1981}}. A M\"{o}ssbauer spectroscopy study of FeBr\thr\ found magnetic order below 15.7\,K, and the authors proposed the order below this temperature to be antiferromagnetic in analogy with FeCl\thr\ \cite{Oosterhuis-1975}.

\subsubsection{Ru\textit{X}$_3$}

Recent interest in RuCl\thr\ began with Plumb et al. identifying it as a spin-orbit assisted Mott insulator in which a small band gap arises from a combination of spin-orbit interactions and strong electron-electron correlations \cite{Plumb-2014}. In this compound, Ru has electron configuration $3d^5$. Considering spin only gives an option of a high spin configuration ($S=5/2$) with all $t_{2g}$ and $e_g$ levels half filled, or~a lower spin configurations with some levels doubly occupied. In this case the crystal field splitting is large enough so that all five of the $d$ electrons go into the lower, $t_{2g}$ set leaving only one unpaired spin ($S=1/2$). However, the orbital angular momentum is not quenched and cannot be neglected. In~addition, the spin orbit coupling interaction, which increases in strength as $Z^4$ where $Z$ is the atomic number of the nucleus,  must also be considered for this heavy, $4d$ transition metal. In RuCl\thr, spin orbit coupling, along with significant electron-electron correlations, splits the otherwise degenerate $t_{2g}$ states into states with effective angular momentum $j_{eff} = 3/2$ and $j_{eff} = 1/2$ \cite{Kim-2008, Kim-2015}. The $j_{eff} = 3/2$ states are lower in energy, and hold four of the five $d$ electrons, leaving one for the higher energy level, and giving Ru in this compound an angular momentum of $j_{eff} = 1/2$. Magnetization measurements in the paramagnetic state have been reported for both powder and single crystals. Powder measurements give effective moments of 2.2--2.3\,$\mu_B$ and Weiss temperatures of 23--40\,K \cite{Fletcher-1963, Fletcher-1967, Kobayashi-1992, Banerjee-2016}. Single crystal measurements show strong paramagnetic anisotropy, and give $\mu_{eff}$\,=\,2.1$\mu_B$ and $\theta$\,=\,37\,K with the field applied in the plane, and $\mu_{eff}$\,=\,2.7$\mu_B$ and $\theta$\,=\,$-$150\,K with the field applied perpendicular to the layers~\cite{Majumder-2015}.

\begin{figure*}
\begin{center}
\includegraphics[width=4.2in]{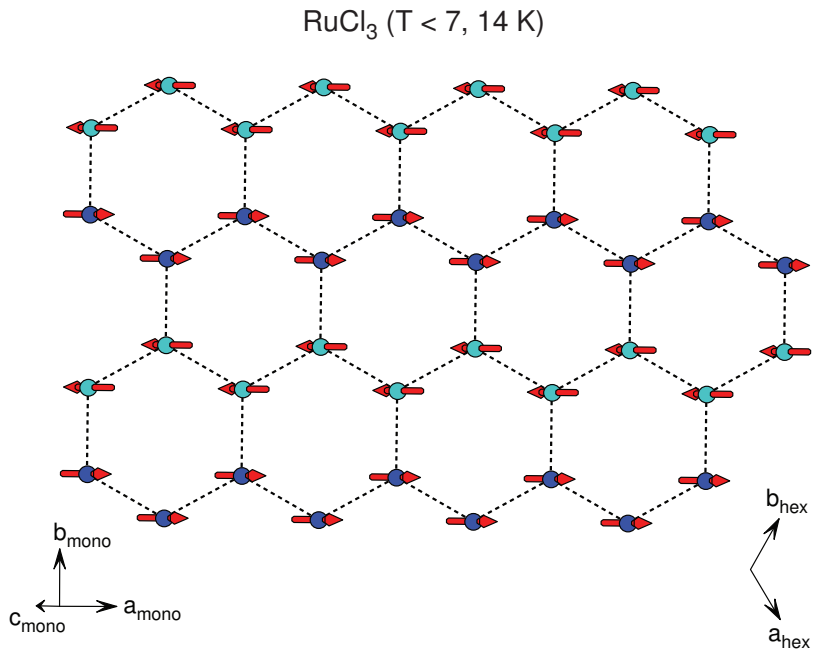}
\caption{\label{fig:RuCl3-mag-str}
The zig-zag in plane magnetic structure of RuCl\thr. Coordinate systems corresponding to the underlying monoclinic and hexagonal crystal lattices are shown. The moments lie in the monoclinic \textit{ac} plane, but the direction in this plane is not known well.
}
\end{center}
\end{figure*}

Two magnetic phase transitions have been observed in RuCl\thr, at 14\,K and 7\,K. It is believed that the difference depends upon the details of the stacking sequence of the RuCl\thr\ layers and the density of stacking faults \cite{Banerjee-2016}. Some crystals show only one transition or the other, while others samples show both. In crystals which undergo no crystallographic phase transition upon cooling (see above) and remain monoclinic at all temperatures magnetic order occurs below 14\,K, while crystals that undergo a structural transition upon cooling show only the 7\,K transition \cite{Banerjee-2016, Park-RuCl3}. Pristine crystals that have shown a phase transition at 7\,K can be transformed into crystals with only the 14\,K transition through mechanical deformation \cite{Banerjee-2016, RuCl3-Cao-2016}. Although all of the details of the magnetic structures of the two phases are have not been settled, there is consensus that the in-plane magnetic structures are of the so-called zig-zag type \cite{Sears-2015, Johnson-2015, Banerjee-2016, RuCl3-Cao-2016} shown in Figure \ref{fig:RuCl3-mag-str}. Determinations of the size of the ordered moment include $\leq$0.4\,$\mu_B$\cite{Banerjee-2016},  $\leq$0.45\,$\mu_B$ \cite{RuCl3-Cao-2016}, $\geq$0.64\,$\mu_B$ \cite{Johnson-2015}, and 0.73\,$\mu_B$\cite{Park-RuCl3}. %I reinserted \leq and \geq that had been deleted
The moment direction is reported to lie in the monoclinic $ac$-plane, with components both in and out of the plane of the RuCl\thr\ layers~\cite{Johnson-2015, RuCl3-Cao-2016, Park-RuCl3}. The layers stack antiferromagnetically with a different stacking sequence associated with the different transition temperatures. AB magnetic stacking is seen in crystals with a 14\,K transition, ABC stacking is seen in crystals with a 7\,K transition, and both types of stacking onsetting at the appropriate temperatures are seen in samples with both transitions \cite{Johnson-2015, Banerjee-2016, RuCl3-Cao-2016}.

With $j_{eff} = 1/2$ and strong spin orbit coupling on a honeycomb lattice, RuCl\thr\ is identified as a promising system for studying the Kitaev model \cite{Kitaev, Plumb-2014, Kim-2015}. In this model, anisotropic interactions result in a type of magnetic frustration. This can give rise to a quantum spin liquid ground state, in which fluctuations prevent magnetic order even at very low temperature, and in which particulary exotic magnetic excitations are predicted \cite{Baskaran-2007, Knolle-2014, Banerjee-2016}. This is, in fact, the motivation for much of the current interest in RuCl\thr.

%

%%%%%%%%%%%%%%%%%%%%%%%%%%%%%%%%%%%%%%%%%%%%%%%%

\section{Summary and Conclusions}
% change to Conclusions? -How about "Summary and Conclusions"? (MAM)
The binary transition-metal halides \di\ and \tri\ reviewed here have simple layered crystal structures containing triangular and honeycomb transition metal nets, yet they display a wide variety of interesting crystallographic and magnetic behaviors. Several compounds display polymorphism, with multiple layered and non-layered structures reported. Temperature and pressure induced crystallographic phase transitions are observed in some. Dimerization of transition metal cations results in a quenching of the magnetic moment in materials like Ti$X_3$ and MoCl\thr. All of the materials which maintain a local magnetic moment are observed to order magnetically, although the magnetic order in TiCl\two\ is not definitively confirmed. This compound and the vanadium dihalides clearly show evidence of geometrical frustration of strong antiferromagnetic interactions on their triangular lattices, with ordering temperatures an order of magnitude smaller than their Weiss temperatures. Effects~of a different kind of frustration, due to competing anisotropic exchange interactions, is observed in RuCl\thr, making it a promising candidate for the realization of a Kitaev spin liquid. It appears that the in-plane magnetic interactions are at least partly ferromagnetic in most of the other magnetic \di\ and \tri\ compounds, and field induced phase transitions that may arise from competing magnetic interactions and multiple low energy magnetic configurations are observed in several of cases. Several dihalides adopt helimagnetic structures and develop electric polarization at their magnetic ordering temperature, providing an interesting class of multiferroic materials. Finally, interest is growing in producing monolayer magnetic materials from several of these compounds, in particular the chromium trihalides, which will enable exciting advances in functional van der Waals heterostructures. Particularly interesting for this application is the wide variety of in-plane magnetic structures that occur in \di\ and \tri\ compounds. Although several of these materials have been studied for many decades, it is likely that layered, binary, transition-metal halides will continue to provide a fruitful playground for solid state chemists, physicists, and materials scientists seeking to further our understanding of low dimensional magnetism and to develop new functional materials.

\vspace{6pt}

\section*{Acknowledgments}{This work is supported by the US Department of Energy, Office of Science, Basic Energy Sciences, Materials Sciences and Engineering Division.}

%\bibliography{layered-halides}

\begin{thebibliography}{146}%
\makeatletter
\providecommand \@ifxundefined [1]{%
 \@ifx{#1\undefined}
}%
\providecommand \@ifnum [1]{%
 \ifnum #1\expandafter \@firstoftwo
 \else \expandafter \@secondoftwo
 \fi
}%
\providecommand \@ifx [1]{%
 \ifx #1\expandafter \@firstoftwo
 \else \expandafter \@secondoftwo
 \fi
}%
\providecommand \natexlab [1]{#1}%
\providecommand \enquote  [1]{``#1''}%
\providecommand \bibnamefont  [1]{#1}%
\providecommand \bibfnamefont [1]{#1}%
\providecommand \citenamefont [1]{#1}%
\providecommand \href@noop [0]{\@secondoftwo}%
\providecommand \href [0]{\begingroup \@sanitize@url \@href}%
\providecommand \@href[1]{\@@startlink{#1}\@@href}%
\providecommand \@@href[1]{\endgroup#1\@@endlink}%
\providecommand \@sanitize@url [0]{\catcode `\\12\catcode `\$12\catcode
  `\&12\catcode `\#12\catcode `\^12\catcode `\_12\catcode `\%12\relax}%
\providecommand \@@startlink[1]{}%
\providecommand \@@endlink[0]{}%
\providecommand \url  [0]{\begingroup\@sanitize@url \@url }%
\providecommand \@url [1]{\endgroup\@href {#1}{\urlprefix }}%
\providecommand \urlprefix  [0]{URL }%
\providecommand \Eprint [0]{\href }%
\providecommand \doibase [0]{http://dx.doi.org/}%
\providecommand \selectlanguage [0]{\@gobble}%
\providecommand \bibinfo  [0]{\@secondoftwo}%
\providecommand \bibfield  [0]{\@secondoftwo}%
\providecommand \translation [1]{[#1]}%
\providecommand \BibitemOpen [0]{}%
\providecommand \bibitemStop [0]{}%
\providecommand \bibitemNoStop [0]{.\EOS\space}%
\providecommand \EOS [0]{\spacefactor3000\relax}%
\providecommand \BibitemShut  [1]{\csname bibitem#1\endcsname}%
\let\auto@bib@innerbib\@empty
%</preamble>
\bibitem [{\citenamefont {{de Jongh}}(1990)}]{layered-TM-book}%
  \BibitemOpen
  \bibfield  {author} {\bibinfo {author} {\bibfnamefont {L.~J.}\ \bibnamefont
  {{de Jongh}}},\ }\href@noop {} {\emph {\bibinfo {title} {Magnetic Properties
  of Layered Transition Metal Compounds}}}\ (\bibinfo  {publisher} {Kluwer
  Academic Press: Dordrecht, The Netherlands},\ \bibinfo {year}
  {1990})\BibitemShut {NoStop}%
\bibitem [{\citenamefont {Kadowaki}\ \emph {et~al.}(1987)\citenamefont
  {Kadowaki}, \citenamefont {Ubukoshi}, \citenamefont {Hirakawa}, \citenamefont
  {Martinez},\ and\ \citenamefont {Shirane}}]{Kadowaki-1987}%
  \BibitemOpen
  \bibfield  {author} {\bibinfo {author} {\bibfnamefont {H.}~\bibnamefont
  {Kadowaki}}, \bibinfo {author} {\bibfnamefont {K.}~\bibnamefont {Ubukoshi}},
  \bibinfo {author} {\bibfnamefont {K.}~\bibnamefont {Hirakawa}}, \bibinfo
  {author} {\bibfnamefont {J.~L.}\ \bibnamefont {Martinez}}, \ and\ \bibinfo
  {author} {\bibfnamefont {G.}~\bibnamefont {Shirane}},\ }\href@noop {}
  {\bibfield  {journal} {\bibinfo  {journal} {J. Phys. Soc. Japan}\ }\textbf
  {\bibinfo {volume} {56}},\ \bibinfo {pages} {4027} (\bibinfo {year}
  {1987})}\BibitemShut {NoStop}%
\bibitem [{\citenamefont {Ramirez}(1994)}]{Ramirez-1994}%
  \BibitemOpen
  \bibfield  {author} {\bibinfo {author} {\bibfnamefont {A.~P.}\ \bibnamefont
  {Ramirez}},\ }\href@noop {} {\bibfield  {journal} {\bibinfo  {journal} {Annu.
  Rev. Mater. Sci.}\ }\textbf {\bibinfo {volume} {24}},\ \bibinfo {pages} {453}
  (\bibinfo {year} {1994})}\BibitemShut {NoStop}%
\bibitem [{\citenamefont {Collins}\ and\ \citenamefont
  {Petrenko}(1997)}]{TriangularAFM}%
  \BibitemOpen
  \bibfield  {author} {\bibinfo {author} {\bibfnamefont {M.~F.}\ \bibnamefont
  {Collins}}\ and\ \bibinfo {author} {\bibfnamefont {O.~A.}\ \bibnamefont
  {Petrenko}},\ }\href@noop {} {\bibfield  {journal} {\bibinfo  {journal} {Can.
  J. Phys.}\ }\textbf {\bibinfo {volume} {75}},\ \bibinfo {pages} {605}
  (\bibinfo {year} {1997})}\BibitemShut {NoStop}%
\bibitem [{\citenamefont {Tokunaga}\ \emph {et~al.}(2011)\citenamefont
  {Tokunaga}, \citenamefont {Okuyama}, \citenamefont {Kurumaji}, \citenamefont
  {Arima}, \citenamefont {Nakao}, \citenamefont {Murakami}, \citenamefont
  {Taguchi},\ and\ \citenamefont {Tokura}}]{Tokunaga-2011}%
  \BibitemOpen
  \bibfield  {author} {\bibinfo {author} {\bibfnamefont {Y.}~\bibnamefont
  {Tokunaga}}, \bibinfo {author} {\bibfnamefont {D.}~\bibnamefont {Okuyama}},
  \bibinfo {author} {\bibfnamefont {T.}~\bibnamefont {Kurumaji}}, \bibinfo
  {author} {\bibfnamefont {T.}~\bibnamefont {Arima}}, \bibinfo {author}
  {\bibfnamefont {H.}~\bibnamefont {Nakao}}, \bibinfo {author} {\bibfnamefont
  {Y.}~\bibnamefont {Murakami}}, \bibinfo {author} {\bibfnamefont
  {Y.}~\bibnamefont {Taguchi}}, \ and\ \bibinfo {author} {\bibfnamefont
  {Y.}~\bibnamefont {Tokura}},\ }\href@noop {} {\bibfield  {journal} {\bibinfo
  {journal} {Phys. Rev. B}\ }\textbf {\bibinfo {volume} {84}},\ \bibinfo
  {pages} {060406(R)} (\bibinfo {year} {2011})}\BibitemShut {NoStop}%
\bibitem [{\citenamefont {Kurumaji}\ \emph {et~al.}(2011)\citenamefont
  {Kurumaji}, \citenamefont {Seki}, \citenamefont {Ishiwata}, \citenamefont
  {Murakawa}, \citenamefont {Tokunaga}, \citenamefont {Kaneko},\ and\
  \citenamefont {Tokura}}]{Kurumaji-2011}%
  \BibitemOpen
  \bibfield  {author} {\bibinfo {author} {\bibfnamefont {T.}~\bibnamefont
  {Kurumaji}}, \bibinfo {author} {\bibfnamefont {S.}~\bibnamefont {Seki}},
  \bibinfo {author} {\bibfnamefont {S.}~\bibnamefont {Ishiwata}}, \bibinfo
  {author} {\bibfnamefont {H.}~\bibnamefont {Murakawa}}, \bibinfo {author}
  {\bibfnamefont {Y.}~\bibnamefont {Tokunaga}}, \bibinfo {author}
  {\bibfnamefont {Y.}~\bibnamefont {Kaneko}}, \ and\ \bibinfo {author}
  {\bibfnamefont {Y.}~\bibnamefont {Tokura}},\ }\href@noop {} {\bibfield
  {journal} {\bibinfo  {journal} {Phys. Rev. Lett.}\ }\textbf {\bibinfo
  {volume} {106}},\ \bibinfo {pages} {167206} (\bibinfo {year}
  {2011})}\BibitemShut {NoStop}%
\bibitem [{\citenamefont {Wu}\ \emph {et~al.}(2012)\citenamefont {Wu},
  \citenamefont {Cai}, \citenamefont {Xie}, \citenamefont {Weng}, \citenamefont
  {Fan},\ and\ \citenamefont {Hu}}]{Wu-2012}%
  \BibitemOpen
  \bibfield  {author} {\bibinfo {author} {\bibfnamefont {X.}~\bibnamefont
  {Wu}}, \bibinfo {author} {\bibfnamefont {Y.}~\bibnamefont {Cai}}, \bibinfo
  {author} {\bibfnamefont {Q.}~\bibnamefont {Xie}}, \bibinfo {author}
  {\bibfnamefont {H.}~\bibnamefont {Weng}}, \bibinfo {author} {\bibfnamefont
  {H.}~\bibnamefont {Fan}}, \ and\ \bibinfo {author} {\bibfnamefont
  {J.}~\bibnamefont {Hu}},\ }\href@noop {} {\bibfield  {journal} {\bibinfo
  {journal} {Phys. Rev. B}\ }\textbf {\bibinfo {volume} {86}},\ \bibinfo
  {pages} {134413} (\bibinfo {year} {2012})}\BibitemShut {NoStop}%
\bibitem [{\citenamefont {Kurumaji}\ \emph {et~al.}(2013)\citenamefont
  {Kurumaji}, \citenamefont {Seki}, \citenamefont {Ishiwata}, \citenamefont
  {Murakawa}, \citenamefont {Kaneko},\ and\ \citenamefont
  {Tokura}}]{Kurumaji-2013}%
  \BibitemOpen
  \bibfield  {author} {\bibinfo {author} {\bibfnamefont {T.}~\bibnamefont
  {Kurumaji}}, \bibinfo {author} {\bibfnamefont {S.}~\bibnamefont {Seki}},
  \bibinfo {author} {\bibfnamefont {S.}~\bibnamefont {Ishiwata}}, \bibinfo
  {author} {\bibfnamefont {H.}~\bibnamefont {Murakawa}}, \bibinfo {author}
  {\bibfnamefont {Y.}~\bibnamefont {Kaneko}}, \ and\ \bibinfo {author}
  {\bibfnamefont {Y.}~\bibnamefont {Tokura}},\ }\href@noop {} {\bibfield
  {journal} {\bibinfo  {journal} {Phys. Rev. B}\ }\textbf {\bibinfo {volume}
  {87}},\ \bibinfo {pages} {014429} (\bibinfo {year} {2013})}\BibitemShut
  {NoStop}%
\bibitem [{\citenamefont {Plumb}\ \emph {et~al.}(2014)\citenamefont {Plumb},
  \citenamefont {Clancy}, \citenamefont {Sandilands}, \citenamefont {Shankar},
  \citenamefont {Hu}, \citenamefont {Burch}, \citenamefont {Kee},\ and\
  \citenamefont {Kim}}]{Plumb-2014}%
  \BibitemOpen
  \bibfield  {author} {\bibinfo {author} {\bibfnamefont {K.~W.}\ \bibnamefont
  {Plumb}}, \bibinfo {author} {\bibfnamefont {J.~P.}\ \bibnamefont {Clancy}},
  \bibinfo {author} {\bibfnamefont {L.~J.}\ \bibnamefont {Sandilands}},
  \bibinfo {author} {\bibfnamefont {V.~V.}\ \bibnamefont {Shankar}}, \bibinfo
  {author} {\bibfnamefont {Y.~F.}\ \bibnamefont {Hu}}, \bibinfo {author}
  {\bibfnamefont {K.~S.}\ \bibnamefont {Burch}}, \bibinfo {author}
  {\bibfnamefont {H.-Y.}\ \bibnamefont {Kee}}, \ and\ \bibinfo {author}
  {\bibfnamefont {Y.-J.}\ \bibnamefont {Kim}},\ }\href {\doibase
  10.1103/PhysRevB.90.041112} {\bibfield  {journal} {\bibinfo  {journal} {Phys.
  Rev. B}\ }\textbf {\bibinfo {volume} {90}},\ \bibinfo {pages} {041112}
  (\bibinfo {year} {2014})}\BibitemShut {NoStop}%
\bibitem [{\citenamefont {Kim}\ \emph {et~al.}(2015)\citenamefont {Kim},
  \citenamefont {{Shankar V.}}, \citenamefont {Catuneanu},\ and\ \citenamefont
  {Kee}}]{Kim-2015}%
  \BibitemOpen
  \bibfield  {author} {\bibinfo {author} {\bibfnamefont {H.}~\bibnamefont
  {Kim}}, \bibinfo {author} {\bibfnamefont {V.}~\bibnamefont {{Shankar V.}}},
  \bibinfo {author} {\bibfnamefont {A.}~\bibnamefont {Catuneanu}}, \ and\
  \bibinfo {author} {\bibfnamefont {H.}~\bibnamefont {Kee}},\ }\href@noop {}
  {\bibfield  {journal} {\bibinfo  {journal} {Phys. Rev. B}\ }\textbf {\bibinfo
  {volume} {91}},\ \bibinfo {pages} {241110(R)} (\bibinfo {year}
  {2015})}\BibitemShut {NoStop}%
\bibitem [{\citenamefont {Sears}\ \emph {et~al.}(2015)\citenamefont {Sears},
  \citenamefont {Songvilay}, \citenamefont {Plumb}, \citenamefont {Clancy},
  \citenamefont {Qiu}, \citenamefont {Zhao}, \citenamefont {Parshall},\ and\
  \citenamefont {Kim}}]{Sears-2015}%
  \BibitemOpen
  \bibfield  {author} {\bibinfo {author} {\bibfnamefont {J.~A.}\ \bibnamefont
  {Sears}}, \bibinfo {author} {\bibfnamefont {M.}~\bibnamefont {Songvilay}},
  \bibinfo {author} {\bibfnamefont {K.~W.}\ \bibnamefont {Plumb}}, \bibinfo
  {author} {\bibfnamefont {J.~P.}\ \bibnamefont {Clancy}}, \bibinfo {author}
  {\bibfnamefont {Y.}~\bibnamefont {Qiu}}, \bibinfo {author} {\bibfnamefont
  {Y.}~\bibnamefont {Zhao}}, \bibinfo {author} {\bibfnamefont {D.}~\bibnamefont
  {Parshall}}, \ and\ \bibinfo {author} {\bibfnamefont {Y.~J.}\ \bibnamefont
  {Kim}},\ }\href@noop {} {\bibfield  {journal} {\bibinfo  {journal} {Phys.
  Rev. B}\ }\textbf {\bibinfo {volume} {91}},\ \bibinfo {pages} {144420}
  (\bibinfo {year} {2015})}\BibitemShut {NoStop}%
\bibitem [{\citenamefont {Johnson}\ \emph {et~al.}(2015)\citenamefont
  {Johnson}, \citenamefont {Williams}, \citenamefont {Haghighirad},
  \citenamefont {Singleton}, \citenamefont {Zapf}, \citenamefont {Manuel},
  \citenamefont {Mazin}, \citenamefont {Li}, \citenamefont {Jeschke},
  \citenamefont {Valenti},\ and\ \citenamefont {Coldea}}]{Johnson-2015}%
  \BibitemOpen
  \bibfield  {author} {\bibinfo {author} {\bibfnamefont {R.~D.}\ \bibnamefont
  {Johnson}}, \bibinfo {author} {\bibfnamefont {S.~C.}\ \bibnamefont
  {Williams}}, \bibinfo {author} {\bibfnamefont {A.~A.}\ \bibnamefont
  {Haghighirad}}, \bibinfo {author} {\bibfnamefont {J.}~\bibnamefont
  {Singleton}}, \bibinfo {author} {\bibfnamefont {V.}~\bibnamefont {Zapf}},
  \bibinfo {author} {\bibfnamefont {P.}~\bibnamefont {Manuel}}, \bibinfo
  {author} {\bibfnamefont {I.~I.}\ \bibnamefont {Mazin}}, \bibinfo {author}
  {\bibfnamefont {Y.}~\bibnamefont {Li}}, \bibinfo {author} {\bibfnamefont
  {H.~O.}\ \bibnamefont {Jeschke}}, \bibinfo {author} {\bibfnamefont
  {R.}~\bibnamefont {Valenti}}, \ and\ \bibinfo {author} {\bibfnamefont
  {R.}~\bibnamefont {Coldea}},\ }\href@noop {} {\bibfield  {journal} {\bibinfo
  {journal} {Phys. Rev. B}\ }\textbf {\bibinfo {volume} {92}},\ \bibinfo
  {pages} {235119} (\bibinfo {year} {2015})}\BibitemShut {NoStop}%
\bibitem [{\citenamefont {Banerjee}\ \emph {et~al.}(2016)\citenamefont
  {Banerjee}, \citenamefont {Bridges}, \citenamefont {Yan}, \citenamefont
  {Aczel}, \citenamefont {Li}, \citenamefont {Stone}, \citenamefont {Granroth},
  \citenamefont {Lumsden}, \citenamefont {Yiu}, \citenamefont {Knolle},
  \citenamefont {Bhattacharjee}, \citenamefont {Kovrizhin}, \citenamefont
  {Moessner}, \citenamefont {Tennant}, \citenamefont {Mandrus},\ and\
  \citenamefont {Nagler}}]{Banerjee-2016}%
  \BibitemOpen
  \bibfield  {author} {\bibinfo {author} {\bibfnamefont {A.}~\bibnamefont
  {Banerjee}}, \bibinfo {author} {\bibfnamefont {C.~A.}\ \bibnamefont
  {Bridges}}, \bibinfo {author} {\bibfnamefont {J.~Q.}\ \bibnamefont {Yan}},
  \bibinfo {author} {\bibfnamefont {A.~A.}\ \bibnamefont {Aczel}}, \bibinfo
  {author} {\bibfnamefont {L.}~\bibnamefont {Li}}, \bibinfo {author}
  {\bibfnamefont {M.~B.}\ \bibnamefont {Stone}}, \bibinfo {author}
  {\bibfnamefont {G.~E.}\ \bibnamefont {Granroth}}, \bibinfo {author}
  {\bibfnamefont {M.~D.}\ \bibnamefont {Lumsden}}, \bibinfo {author}
  {\bibfnamefont {Y.}~\bibnamefont {Yiu}}, \bibinfo {author} {\bibfnamefont
  {J.}~\bibnamefont {Knolle}}, \bibinfo {author} {\bibfnamefont
  {S.}~\bibnamefont {Bhattacharjee}}, \bibinfo {author} {\bibfnamefont {D.~L.}\
  \bibnamefont {Kovrizhin}}, \bibinfo {author} {\bibfnamefont {R.}~\bibnamefont
  {Moessner}}, \bibinfo {author} {\bibfnamefont {D.~A.}\ \bibnamefont
  {Tennant}}, \bibinfo {author} {\bibfnamefont {D.~G.}\ \bibnamefont
  {Mandrus}}, \ and\ \bibinfo {author} {\bibfnamefont {S.~E.}\ \bibnamefont
  {Nagler}},\ }\href@noop {} {\bibfield  {journal} {\bibinfo  {journal} {Nature
  Materials}\ }\textbf {\bibinfo {volume} {15}},\ \bibinfo {pages} {733}
  (\bibinfo {year} {2016})}\BibitemShut {NoStop}%
\bibitem [{\citenamefont {Geim}\ and\ \citenamefont
  {Geigorieva}(2013)}]{Geim-2013}%
  \BibitemOpen
  \bibfield  {author} {\bibinfo {author} {\bibfnamefont {A.~K.}\ \bibnamefont
  {Geim}}\ and\ \bibinfo {author} {\bibfnamefont {I.~V.}\ \bibnamefont
  {Geigorieva}},\ }\href@noop {} {\bibfield  {journal} {\bibinfo  {journal}
  {Nature}\ }\textbf {\bibinfo {volume} {499}},\ \bibinfo {pages} {419}
  (\bibinfo {year} {2013})}\BibitemShut {NoStop}%
\bibitem [{\citenamefont {Wang}\ \emph {et~al.}(2011)\citenamefont {Wang},
  \citenamefont {Eyert},\ and\ \citenamefont
  {Schwingenschl{\"{o}}gl}}]{Wang-2011}%
  \BibitemOpen
  \bibfield  {author} {\bibinfo {author} {\bibfnamefont {H.}~\bibnamefont
  {Wang}}, \bibinfo {author} {\bibfnamefont {V.}~\bibnamefont {Eyert}}, \ and\
  \bibinfo {author} {\bibfnamefont {U.}~\bibnamefont
  {Schwingenschl{\"{o}}gl}},\ }\href@noop {} {\bibfield  {journal} {\bibinfo
  {journal} {J. Phys. Condens. Matter}\ }\textbf {\bibinfo {volume} {23}},\
  \bibinfo {pages} {116003} (\bibinfo {year} {2011})}\BibitemShut {NoStop}%
\bibitem [{\citenamefont {McGuire}\ \emph {et~al.}(2015)\citenamefont
  {McGuire}, \citenamefont {Dixit}, \citenamefont {Cooper},\ and\ \citenamefont
  {Sales}}]{CrI3-str}%
  \BibitemOpen
  \bibfield  {author} {\bibinfo {author} {\bibfnamefont {M.~A.}\ \bibnamefont
  {McGuire}}, \bibinfo {author} {\bibfnamefont {H.}~\bibnamefont {Dixit}},
  \bibinfo {author} {\bibfnamefont {V.~R.}\ \bibnamefont {Cooper}}, \ and\
  \bibinfo {author} {\bibfnamefont {B.~C.}\ \bibnamefont {Sales}},\ }\href@noop
  {} {\bibfield  {journal} {\bibinfo  {journal} {Chem. Mater.}\ }\textbf
  {\bibinfo {volume} {27}},\ \bibinfo {pages} {612} (\bibinfo {year}
  {2015})}\BibitemShut {NoStop}%
\bibitem [{\citenamefont {Zhang}\ \emph {et~al.}(2015)\citenamefont {Zhang},
  \citenamefont {Qu}, \citenamefont {Zhu},\ and\ \citenamefont
  {Lam}}]{Zhang-2015}%
  \BibitemOpen
  \bibfield  {author} {\bibinfo {author} {\bibfnamefont {W.-B.}\ \bibnamefont
  {Zhang}}, \bibinfo {author} {\bibfnamefont {Q.}~\bibnamefont {Qu}}, \bibinfo
  {author} {\bibfnamefont {P.}~\bibnamefont {Zhu}}, \ and\ \bibinfo {author}
  {\bibfnamefont {C.-H.}\ \bibnamefont {Lam}},\ }\href@noop {} {\bibfield
  {journal} {\bibinfo  {journal} {J. Mater. Chem. C}\ }\textbf {\bibinfo
  {volume} {3}},\ \bibinfo {pages} {12457} (\bibinfo {year}
  {2015})}\BibitemShut {NoStop}%
\bibitem [{\citenamefont {Liu}\ \emph {et~al.}(2016)\citenamefont {Liu},
  \citenamefont {Sun}, \citenamefont {Kawazoe},\ and\ \citenamefont
  {Jena}}]{Liu-2016}%
  \BibitemOpen
  \bibfield  {author} {\bibinfo {author} {\bibfnamefont {J.}~\bibnamefont
  {Liu}}, \bibinfo {author} {\bibfnamefont {Q.}~\bibnamefont {Sun}}, \bibinfo
  {author} {\bibfnamefont {Y.}~\bibnamefont {Kawazoe}}, \ and\ \bibinfo
  {author} {\bibfnamefont {P.}~\bibnamefont {Jena}},\ }\href@noop {} {\bibfield
   {journal} {\bibinfo  {journal} {Plys. Chem. Chem. Phys.}\ }\textbf {\bibinfo
  {volume} {18}},\ \bibinfo {pages} {8777} (\bibinfo {year}
  {2016})}\BibitemShut {NoStop}%
\bibitem [{\citenamefont {Wang}\ \emph {et~al.}(2016)\citenamefont {Wang},
  \citenamefont {Fan}, \citenamefont {Zhu},\ and\ \citenamefont
  {Wu}}]{Wang-2016}%
  \BibitemOpen
  \bibfield  {author} {\bibinfo {author} {\bibfnamefont {H.}~\bibnamefont
  {Wang}}, \bibinfo {author} {\bibfnamefont {F.}~\bibnamefont {Fan}}, \bibinfo
  {author} {\bibfnamefont {S.}~\bibnamefont {Zhu}}, \ and\ \bibinfo {author}
  {\bibfnamefont {H.}~\bibnamefont {Wu}},\ }\href@noop {} {\bibfield  {journal}
  {\bibinfo  {journal} {EPL}\ }\textbf {\bibinfo {volume} {114}},\ \bibinfo
  {pages} {47001} (\bibinfo {year} {2016})}\BibitemShut {NoStop}%
\bibitem [{\citenamefont {Zhong}\ \emph {et~al.}(2017)\citenamefont {Zhong},
  \citenamefont {Seyler}, \citenamefont {Linpeng}, \citenamefont {Cheng},
  \citenamefont {Sivadas}, \citenamefont {Schmidgall}, \citenamefont
  {Taniguchi}, \citenamefont {Watanabe}, \citenamefont {McGuire}, \citenamefont
  {Yao}, \citenamefont {Xiao}, \citenamefont {Fu},\ and\ \citenamefont
  {Xu}}]{Zhong-2017}%
  \BibitemOpen
  \bibfield  {author} {\bibinfo {author} {\bibfnamefont {D.}~\bibnamefont
  {Zhong}}, \bibinfo {author} {\bibfnamefont {K.~L.}\ \bibnamefont {Seyler}},
  \bibinfo {author} {\bibfnamefont {X.}~\bibnamefont {Linpeng}}, \bibinfo
  {author} {\bibfnamefont {R.}~\bibnamefont {Cheng}}, \bibinfo {author}
  {\bibfnamefont {N.}~\bibnamefont {Sivadas}}, \bibinfo {author} {\bibfnamefont
  {B.~H.~E.}\ \bibnamefont {Schmidgall}}, \bibinfo {author} {\bibfnamefont
  {T.}~\bibnamefont {Taniguchi}}, \bibinfo {author} {\bibfnamefont
  {K.}~\bibnamefont {Watanabe}}, \bibinfo {author} {\bibfnamefont {M.~A.}\
  \bibnamefont {McGuire}}, \bibinfo {author} {\bibfnamefont {W.}~\bibnamefont
  {Yao}}, \bibinfo {author} {\bibfnamefont {D.}~\bibnamefont {Xiao}}, \bibinfo
  {author} {\bibfnamefont {K.-M.~C.}\ \bibnamefont {Fu}}, \ and\ \bibinfo
  {author} {\bibfnamefont {X.}~\bibnamefont {Xu}},\ }\href@noop {} {\bibfield
  {journal} {\bibinfo  {journal} {arXiv:1704.00841}\ } (\bibinfo {year}
  {2017})}\BibitemShut {NoStop}%
\bibitem [{\citenamefont {Huang}\ \emph {et~al.}(2017)\citenamefont {Huang},
  \citenamefont {Clark}, \citenamefont {Navarro-Moratalla}, \citenamefont
  {Klein}, \citenamefont {Cheng}, \citenamefont {Seyler}, \citenamefont
  {Zhong}, \citenamefont {Schmidgall}, \citenamefont {McGuire}, \citenamefont
  {Cobden}, \citenamefont {Yao}, \citenamefont {Xiao}, \citenamefont
  {Jarillo-Herrero},\ and\ \citenamefont {Xu}}]{Huang-2017}%
  \BibitemOpen
  \bibfield  {author} {\bibinfo {author} {\bibfnamefont {B.}~\bibnamefont
  {Huang}}, \bibinfo {author} {\bibfnamefont {G.}~\bibnamefont {Clark}},
  \bibinfo {author} {\bibfnamefont {E.}~\bibnamefont {Navarro-Moratalla}},
  \bibinfo {author} {\bibfnamefont {D.~R.}\ \bibnamefont {Klein}}, \bibinfo
  {author} {\bibfnamefont {R.}~\bibnamefont {Cheng}}, \bibinfo {author}
  {\bibfnamefont {K.~L.}\ \bibnamefont {Seyler}}, \bibinfo {author}
  {\bibfnamefont {D.}~\bibnamefont {Zhong}}, \bibinfo {author} {\bibfnamefont
  {E.}~\bibnamefont {Schmidgall}}, \bibinfo {author} {\bibfnamefont {M.~A.}\
  \bibnamefont {McGuire}}, \bibinfo {author} {\bibfnamefont {D.~H.}\
  \bibnamefont {Cobden}}, \bibinfo {author} {\bibfnamefont {W.}~\bibnamefont
  {Yao}}, \bibinfo {author} {\bibfnamefont {D.}~\bibnamefont {Xiao}}, \bibinfo
  {author} {\bibfnamefont {P.}~\bibnamefont {Jarillo-Herrero}}, \ and\ \bibinfo
  {author} {\bibfnamefont {X.}~\bibnamefont {Xu}},\ }\href@noop {} {\bibfield
  {journal} {\bibinfo  {journal} {arXiv:1703.05892}\ } (\bibinfo {year}
  {2017})}\BibitemShut {NoStop}%
\bibitem [{\citenamefont {Leb{\`{e}}gue}\ \emph {et~al.}(2013)\citenamefont
  {Leb{\`{e}}gue}, \citenamefont {Bj{\"{o}}rkman}, \citenamefont {Klintenberg},
  \citenamefont {Nieminen},\ and\ \citenamefont {Eriksson}}]{Lebegue-2013}%
  \BibitemOpen
  \bibfield  {author} {\bibinfo {author} {\bibfnamefont {S.}~\bibnamefont
  {Leb{\`{e}}gue}}, \bibinfo {author} {\bibfnamefont {T.}~\bibnamefont
  {Bj{\"{o}}rkman}}, \bibinfo {author} {\bibfnamefont {M.}~\bibnamefont
  {Klintenberg}}, \bibinfo {author} {\bibfnamefont {R.~M.}\ \bibnamefont
  {Nieminen}}, \ and\ \bibinfo {author} {\bibfnamefont {O.}~\bibnamefont
  {Eriksson}},\ }\href@noop {} {\bibfield  {journal} {\bibinfo  {journal}
  {Phys. Rev. X}\ }\textbf {\bibinfo {volume} {3}},\ \bibinfo {pages} {031002}
  (\bibinfo {year} {2013})}\BibitemShut {NoStop}%
\bibitem [{\citenamefont {Ajayan}\ \emph {et~al.}(2016)\citenamefont {Ajayan},
  \citenamefont {Kim},\ and\ \citenamefont {Banerjee}}]{PhysToday}%
  \BibitemOpen
  \bibfield  {author} {\bibinfo {author} {\bibfnamefont {P.}~\bibnamefont
  {Ajayan}}, \bibinfo {author} {\bibfnamefont {P.}~\bibnamefont {Kim}}, \ and\
  \bibinfo {author} {\bibfnamefont {K.}~\bibnamefont {Banerjee}},\ }\href@noop
  {} {\bibfield  {journal} {\bibinfo  {journal} {Phys. Today}\ }\textbf
  {\bibinfo {volume} {69}},\ \bibinfo {pages} {38} (\bibinfo {year}
  {2016})}\BibitemShut {NoStop}%
\bibitem [{\citenamefont {Park}(2016)}]{Park-2016}%
  \BibitemOpen
  \bibfield  {author} {\bibinfo {author} {\bibfnamefont {J.}~\bibnamefont
  {Park}},\ }\href@noop {} {\bibfield  {journal} {\bibinfo  {journal} {J.
  Phys.: Condens. Matter}\ }\textbf {\bibinfo {volume} {28}},\ \bibinfo {pages}
  {301001} (\bibinfo {year} {2016})}\BibitemShut {NoStop}%
\bibitem [{\citenamefont {MacDonald}\ and\ \citenamefont
  {Tsoi}(2011)}]{MacDonald-2011}%
  \BibitemOpen
  \bibfield  {author} {\bibinfo {author} {\bibfnamefont {A.~H.}\ \bibnamefont
  {MacDonald}}\ and\ \bibinfo {author} {\bibfnamefont {M.}~\bibnamefont
  {Tsoi}},\ }\href@noop {} {\bibfield  {journal} {\bibinfo  {journal} {Phil.
  Trans. R. Soc. A}\ }\textbf {\bibinfo {volume} {369}},\ \bibinfo {pages}
  {3098} (\bibinfo {year} {2011})}\BibitemShut {NoStop}%
\bibitem [{\citenamefont {Gomonay}\ and\ \citenamefont
  {Loktev}(2014)}]{Gomonay-2014}%
  \BibitemOpen
  \bibfield  {author} {\bibinfo {author} {\bibfnamefont {E.~V.}\ \bibnamefont
  {Gomonay}}\ and\ \bibinfo {author} {\bibfnamefont {V.~M.}\ \bibnamefont
  {Loktev}},\ }\href@noop {} {\bibfield  {journal} {\bibinfo  {journal} {Low
  Temp. Phys.}\ }\textbf {\bibinfo {volume} {40}},\ \bibinfo {pages} {17}
  (\bibinfo {year} {2014})}\BibitemShut {NoStop}%
\bibitem [{\citenamefont {Jungwirth}\ \emph {et~al.}(2016)\citenamefont
  {Jungwirth}, \citenamefont {Marti}, \citenamefont {Wadley},\ and\
  \citenamefont {Wunderlich}}]{Jungwirth-2016}%
  \BibitemOpen
  \bibfield  {author} {\bibinfo {author} {\bibfnamefont {T.}~\bibnamefont
  {Jungwirth}}, \bibinfo {author} {\bibfnamefont {X.}~\bibnamefont {Marti}},
  \bibinfo {author} {\bibfnamefont {P.}~\bibnamefont {Wadley}}, \ and\ \bibinfo
  {author} {\bibfnamefont {J.}~\bibnamefont {Wunderlich}},\ }\href@noop {}
  {\bibfield  {journal} {\bibinfo  {journal} {Nature Nanotechnology}\ }\textbf
  {\bibinfo {volume} {11}},\ \bibinfo {pages} {231} (\bibinfo {year}
  {2016})}\BibitemShut {NoStop}%
\bibitem [{\citenamefont {Klemm}\ and\ \citenamefont
  {Krose}(1947)}]{TiCl3-VCl3-str-BiI3}%
  \BibitemOpen
  \bibfield  {author} {\bibinfo {author} {\bibfnamefont {W.}~\bibnamefont
  {Klemm}}\ and\ \bibinfo {author} {\bibfnamefont {E.}~\bibnamefont {Krose}},\
  }\href@noop {} {\bibfield  {journal} {\bibinfo  {journal} {Z. Anorg. Chem.}\
  }\textbf {\bibinfo {volume} {253}},\ \bibinfo {pages} {218} (\bibinfo {year}
  {1947})}\BibitemShut {NoStop}%
\bibitem [{\citenamefont {Natta}\ \emph {et~al.}(1961)\citenamefont {Natta},
  \citenamefont {Corradini},\ and\ \citenamefont {Allegra}}]{TiCl3-1D}%
  \BibitemOpen
  \bibfield  {author} {\bibinfo {author} {\bibfnamefont {G.}~\bibnamefont
  {Natta}}, \bibinfo {author} {\bibfnamefont {P.}~\bibnamefont {Corradini}}, \
  and\ \bibinfo {author} {\bibfnamefont {G.}~\bibnamefont {Allegra}},\
  }\href@noop {} {\bibfield  {journal} {\bibinfo  {journal} {J. Polymer Sci.}\
  }\textbf {\bibinfo {volume} {51}},\ \bibinfo {pages} {399} (\bibinfo {year}
  {1961})}\BibitemShut {NoStop}%
\bibitem [{\citenamefont {{von Schnering}}\ \emph {et~al.}(1961)\citenamefont
  {{von Schnering}}, \citenamefont {W{\"{o}}hrle},\ and\ \citenamefont
  {Sch{\"{a}}fer}}]{von-Schnering-1961}%
  \BibitemOpen
  \bibfield  {author} {\bibinfo {author} {\bibfnamefont {H.~G.}\ \bibnamefont
  {{von Schnering}}}, \bibinfo {author} {\bibfnamefont {H.}~\bibnamefont
  {W{\"{o}}hrle}}, \ and\ \bibinfo {author} {\bibfnamefont {H.}~\bibnamefont
  {Sch{\"{a}}fer}},\ }\href@noop {} {\bibfield  {journal} {\bibinfo  {journal}
  {Naturwissensch.}\ }\textbf {\bibinfo {volume} {48}},\ \bibinfo {pages} {159}
  (\bibinfo {year} {1961})}\BibitemShut {NoStop}%
\bibitem [{\citenamefont {Sheckelton}\ \emph {et~al.}(2017)\citenamefont
  {Sheckelton}, \citenamefont {Plumb}, \citenamefont {Trump}, \citenamefont
  {Broholm},\ and\ \citenamefont {McQueen}}]{Sheckelton-2017}%
  \BibitemOpen
  \bibfield  {author} {\bibinfo {author} {\bibfnamefont {J.~P.}\ \bibnamefont
  {Sheckelton}}, \bibinfo {author} {\bibfnamefont {K.~W.}\ \bibnamefont
  {Plumb}}, \bibinfo {author} {\bibfnamefont {B.~A.}\ \bibnamefont {Trump}},
  \bibinfo {author} {\bibfnamefont {C.~L.}\ \bibnamefont {Broholm}}, \ and\
  \bibinfo {author} {\bibfnamefont {T.~M.}\ \bibnamefont {McQueen}},\
  }\href@noop {} {\bibfield  {journal} {\bibinfo  {journal} {Inorg. Chem.
  Front.}\ }\textbf {\bibinfo {volume} {4}},\ \bibinfo {pages} {481} (\bibinfo
  {year} {2017})}\BibitemShut {NoStop}%
\bibitem [{\citenamefont {Jiang}\ \emph {et~al.}(2017)\citenamefont {Jiang},
  \citenamefont {Liang}, \citenamefont {Meng}, \citenamefont {Yang},
  \citenamefont {Tan}, \citenamefont {Sun},\ and\ \citenamefont
  {Chen}}]{Jiang-2017}%
  \BibitemOpen
  \bibfield  {author} {\bibinfo {author} {\bibfnamefont {J.}~\bibnamefont
  {Jiang}}, \bibinfo {author} {\bibfnamefont {Q.}~\bibnamefont {Liang}},
  \bibinfo {author} {\bibfnamefont {R.}~\bibnamefont {Meng}}, \bibinfo {author}
  {\bibfnamefont {Q.}~\bibnamefont {Yang}}, \bibinfo {author} {\bibfnamefont
  {C.}~\bibnamefont {Tan}}, \bibinfo {author} {\bibfnamefont {X.}~\bibnamefont
  {Sun}}, \ and\ \bibinfo {author} {\bibfnamefont {X.}~\bibnamefont {Chen}},\
  }\href@noop {} {\bibfield  {journal} {\bibinfo  {journal} {Nanoscale}\
  }\textbf {\bibinfo {volume} {9}},\ \bibinfo {pages} {2992} (\bibinfo {year}
  {2017})}\BibitemShut {NoStop}%
\bibitem [{\citenamefont {Baenziger}\ and\ \citenamefont
  {Rundle}(1948)}]{TiCl2-str}%
  \BibitemOpen
  \bibfield  {author} {\bibinfo {author} {\bibfnamefont {N.~C.}\ \bibnamefont
  {Baenziger}}\ and\ \bibinfo {author} {\bibfnamefont {R.~E.}\ \bibnamefont
  {Rundle}},\ }\href@noop {} {\bibfield  {journal} {\bibinfo  {journal} {Acta
  Crystallogr.}\ }\textbf {\bibinfo {volume} {1}},\ \bibinfo {pages} {274}
  (\bibinfo {year} {1948})}\BibitemShut {NoStop}%
\bibitem [{\citenamefont {Ehrlich}\ \emph {et~al.}(1961)\citenamefont
  {Ehrlich}, \citenamefont {Gutsche},\ and\ \citenamefont
  {Seifert}}]{TiBr2-str}%
  \BibitemOpen
  \bibfield  {author} {\bibinfo {author} {\bibfnamefont {P.}~\bibnamefont
  {Ehrlich}}, \bibinfo {author} {\bibfnamefont {W.}~\bibnamefont {Gutsche}}, \
  and\ \bibinfo {author} {\bibfnamefont {H.~J.}\ \bibnamefont {Seifert}},\
  }\href@noop {} {\bibfield  {journal} {\bibinfo  {journal} {Z. Anorg. Allg.
  Chem.}\ }\textbf {\bibinfo {volume} {312}},\ \bibinfo {pages} {80} (\bibinfo
  {year} {1961})}\BibitemShut {NoStop}%
\bibitem [{\citenamefont {Klemm}\ and\ \citenamefont
  {Grimm}(1942)}]{TiI2-VBr2-str}%
  \BibitemOpen
  \bibfield  {author} {\bibinfo {author} {\bibfnamefont {W.}~\bibnamefont
  {Klemm}}\ and\ \bibinfo {author} {\bibfnamefont {L.}~\bibnamefont {Grimm}},\
  }\href@noop {} {\bibfield  {journal} {\bibinfo  {journal} {Z. Anorg. Allg.
  Chem.}\ }\textbf {\bibinfo {volume} {249}},\ \bibinfo {pages} {198} (\bibinfo
  {year} {1942})}\BibitemShut {NoStop}%
\bibitem [{\citenamefont {Villadsen}(1959)}]{VCl2-str}%
  \BibitemOpen
  \bibfield  {author} {\bibinfo {author} {\bibfnamefont {J.}~\bibnamefont
  {Villadsen}},\ }\href@noop {} {\bibfield  {journal} {\bibinfo  {journal}
  {Acta Chem. Scand.}\ }\textbf {\bibinfo {volume} {13}},\ \bibinfo {pages}
  {2146} (\bibinfo {year} {1959})}\BibitemShut {NoStop}%
\bibitem [{\citenamefont {Kuindersma}\ \emph {et~al.}(1979)\citenamefont
  {Kuindersma}, \citenamefont {Hass}, \citenamefont {Sanchez},\ and\
  \citenamefont {Al}}]{VI2-str}%
  \BibitemOpen
  \bibfield  {author} {\bibinfo {author} {\bibfnamefont {S.~R.}\ \bibnamefont
  {Kuindersma}}, \bibinfo {author} {\bibfnamefont {C.}~\bibnamefont {Hass}},
  \bibinfo {author} {\bibfnamefont {J.~P.}\ \bibnamefont {Sanchez}}, \ and\
  \bibinfo {author} {\bibfnamefont {R.}~\bibnamefont {Al}},\ }\href@noop {}
  {\bibfield  {journal} {\bibinfo  {journal} {Solid State Commun.}\ }\textbf
  {\bibinfo {volume} {30}},\ \bibinfo {pages} {403} (\bibinfo {year}
  {1979})}\BibitemShut {NoStop}%
\bibitem [{\citenamefont {Tornero}\ and\ \citenamefont
  {Fayos}(1990)}]{MnCl2-str}%
  \BibitemOpen
  \bibfield  {author} {\bibinfo {author} {\bibfnamefont {J.~D.}\ \bibnamefont
  {Tornero}}\ and\ \bibinfo {author} {\bibfnamefont {J.}~\bibnamefont
  {Fayos}},\ }\href@noop {} {\bibfield  {journal} {\bibinfo  {journal} {Z.
  Kristallogr.}\ }\textbf {\bibinfo {volume} {192}},\ \bibinfo {pages} {147}
  (\bibinfo {year} {1990})}\BibitemShut {NoStop}%
\bibitem [{\citenamefont {Wollan}\ \emph {et~al.}(1958)\citenamefont {Wollan},
  \citenamefont {Koehler},\ and\ \citenamefont {Wilkinson}}]{MnBr2-str}%
  \BibitemOpen
  \bibfield  {author} {\bibinfo {author} {\bibfnamefont {E.~O.}\ \bibnamefont
  {Wollan}}, \bibinfo {author} {\bibfnamefont {W.~C.}\ \bibnamefont {Koehler}},
  \ and\ \bibinfo {author} {\bibfnamefont {M.~K.}\ \bibnamefont {Wilkinson}},\
  }\href@noop {} {\bibfield  {journal} {\bibinfo  {journal} {Phys. Rev.}\
  }\textbf {\bibinfo {volume} {110}},\ \bibinfo {pages} {638} (\bibinfo {year}
  {1958})}\BibitemShut {NoStop}%
\bibitem [{\citenamefont {Ferrari}\ and\ \citenamefont
  {Giorgi}(1929{\natexlab{a}})}]{MnI2-CoI2-str}%
  \BibitemOpen
  \bibfield  {author} {\bibinfo {author} {\bibfnamefont {A.}~\bibnamefont
  {Ferrari}}\ and\ \bibinfo {author} {\bibfnamefont {F.}~\bibnamefont
  {Giorgi}},\ }\href@noop {} {\bibfield  {journal} {\bibinfo  {journal} {Atti
  Accad. Naz. Lincei, Cl. Sci. Fis., Mat. Nat., Rend.}\ }\textbf {\bibinfo
  {volume} {10}},\ \bibinfo {pages} {522} (\bibinfo {year}
  {1929}{\natexlab{a}})}\BibitemShut {NoStop}%
\bibitem [{\citenamefont {Vettier}\ and\ \citenamefont
  {Yellon}(1975)}]{FeCl2-str}%
  \BibitemOpen
  \bibfield  {author} {\bibinfo {author} {\bibfnamefont {C.}~\bibnamefont
  {Vettier}}\ and\ \bibinfo {author} {\bibfnamefont {W.~B.}\ \bibnamefont
  {Yellon}},\ }\href@noop {} {\bibfield  {journal} {\bibinfo  {journal} {J.
  Phys. Chem. Solids}\ }\textbf {\bibinfo {volume} {36}},\ \bibinfo {pages}
  {401} (\bibinfo {year} {1975})}\BibitemShut {NoStop}%
\bibitem [{\citenamefont {Haberecht}\ \emph {et~al.}(2001)\citenamefont
  {Haberecht}, \citenamefont {Borrmann},\ and\ \citenamefont
  {Kniep}}]{FeBr2-str}%
  \BibitemOpen
  \bibfield  {author} {\bibinfo {author} {\bibfnamefont {J.}~\bibnamefont
  {Haberecht}}, \bibinfo {author} {\bibfnamefont {H.}~\bibnamefont {Borrmann}},
  \ and\ \bibinfo {author} {\bibfnamefont {R.}~\bibnamefont {Kniep}},\
  }\href@noop {} {\bibfield  {journal} {\bibinfo  {journal} {Z. Kristallogr. -
  New Cryst. Struct.}\ }\textbf {\bibinfo {volume} {216}},\ \bibinfo {pages}
  {210} (\bibinfo {year} {2001})}\BibitemShut {NoStop}%
\bibitem [{\citenamefont {Gelard}\ \emph {et~al.}(1974)\citenamefont {Gelard},
  \citenamefont {Fert}, \citenamefont {M\'{e}riel},\ and\ \citenamefont
  {Allain}}]{FeI2-str}%
  \BibitemOpen
  \bibfield  {author} {\bibinfo {author} {\bibfnamefont {J.}~\bibnamefont
  {Gelard}}, \bibinfo {author} {\bibfnamefont {A.~R.}\ \bibnamefont {Fert}},
  \bibinfo {author} {\bibfnamefont {P.}~\bibnamefont {M\'{e}riel}}, \ and\
  \bibinfo {author} {\bibfnamefont {Y.}~\bibnamefont {Allain}},\ }\href@noop {}
  {\bibfield  {journal} {\bibinfo  {journal} {Solid State Commun.}\ }\textbf
  {\bibinfo {volume} {14}},\ \bibinfo {pages} {187} (\bibinfo {year}
  {1974})}\BibitemShut {NoStop}%
\bibitem [{\citenamefont {Grimme}\ and\ \citenamefont
  {Santos}(1934)}]{CoCl2-str}%
  \BibitemOpen
  \bibfield  {author} {\bibinfo {author} {\bibfnamefont {H.}~\bibnamefont
  {Grimme}}\ and\ \bibinfo {author} {\bibfnamefont {J.~A.}\ \bibnamefont
  {Santos}},\ }\href@noop {} {\bibfield  {journal} {\bibinfo  {journal} {Z.
  Kristallogr.}\ }\textbf {\bibinfo {volume} {88}},\ \bibinfo {pages} {136}
  (\bibinfo {year} {1934})}\BibitemShut {NoStop}%
\bibitem [{\citenamefont {Ferrari}\ and\ \citenamefont
  {Giorgi}(1929{\natexlab{b}})}]{CoBr2-str}%
  \BibitemOpen
  \bibfield  {author} {\bibinfo {author} {\bibfnamefont {A.}~\bibnamefont
  {Ferrari}}\ and\ \bibinfo {author} {\bibfnamefont {F.}~\bibnamefont
  {Giorgi}},\ }\href@noop {} {\bibfield  {journal} {\bibinfo  {journal} {Atti
  Accad. Naz. Lincei, Cl. Sci. Fis., Mat. Nat., Rend.}\ }\textbf {\bibinfo
  {volume} {9}},\ \bibinfo {pages} {1134} (\bibinfo {year}
  {1929}{\natexlab{b}})}\BibitemShut {NoStop}%
\bibitem [{\citenamefont {Ferrari}\ \emph {et~al.}(1963)\citenamefont
  {Ferrari}, \citenamefont {Braibanti},\ and\ \citenamefont
  {Bigliardi}}]{NiCl2-str}%
  \BibitemOpen
  \bibfield  {author} {\bibinfo {author} {\bibfnamefont {A.}~\bibnamefont
  {Ferrari}}, \bibinfo {author} {\bibfnamefont {A.}~\bibnamefont {Braibanti}},
  \ and\ \bibinfo {author} {\bibfnamefont {G.}~\bibnamefont {Bigliardi}},\
  }\href@noop {} {\bibfield  {journal} {\bibinfo  {journal} {Acta
  Crystallogr.}\ }\textbf {\bibinfo {volume} {16}},\ \bibinfo {pages} {846}
  (\bibinfo {year} {1963})}\BibitemShut {NoStop}%
\bibitem [{\citenamefont {Nasser}\ \emph {et~al.}(1992)\citenamefont {Nasser},
  \citenamefont {Kiat},\ and\ \citenamefont {Gabilly}}]{NiBr2-str}%
  \BibitemOpen
  \bibfield  {author} {\bibinfo {author} {\bibfnamefont {J.~A.}\ \bibnamefont
  {Nasser}}, \bibinfo {author} {\bibfnamefont {J.~M.}\ \bibnamefont {Kiat}}, \
  and\ \bibinfo {author} {\bibfnamefont {R.}~\bibnamefont {Gabilly}},\
  }\href@noop {} {\bibfield  {journal} {\bibinfo  {journal} {Solid State
  Commun.}\ }\textbf {\bibinfo {volume} {82}},\ \bibinfo {pages} {49} (\bibinfo
  {year} {1992})}\BibitemShut {NoStop}%
\bibitem [{\citenamefont {Ketalaar}(1934)}]{NiI2-str}%
  \BibitemOpen
  \bibfield  {author} {\bibinfo {author} {\bibfnamefont {J.~A.~A.}\
  \bibnamefont {Ketalaar}},\ }\href@noop {} {\bibfield  {journal} {\bibinfo
  {journal} {Z. Kristallogr.}\ }\textbf {\bibinfo {volume} {88}},\ \bibinfo
  {pages} {26} (\bibinfo {year} {1934})}\BibitemShut {NoStop}%
\bibitem [{\citenamefont {Cisar}\ \emph {et~al.}(1979)\citenamefont {Cisar},
  \citenamefont {Corbett},\ and\ \citenamefont {Daake}}]{ZrCl2-str}%
  \BibitemOpen
  \bibfield  {author} {\bibinfo {author} {\bibfnamefont {A.}~\bibnamefont
  {Cisar}}, \bibinfo {author} {\bibfnamefont {J.~D.}\ \bibnamefont {Corbett}},
  \ and\ \bibinfo {author} {\bibfnamefont {R.~L.}\ \bibnamefont {Daake}},\
  }\href@noop {} {\bibfield  {journal} {\bibinfo  {journal} {Inorg. Chem.}\
  }\textbf {\bibinfo {volume} {18}},\ \bibinfo {pages} {836} (\bibinfo {year}
  {1979})}\BibitemShut {NoStop}%
\bibitem [{\citenamefont {Guthrie}\ and\ \citenamefont
  {Corbett}(1981)}]{ZrI2-str-MoTe2}%
  \BibitemOpen
  \bibfield  {author} {\bibinfo {author} {\bibfnamefont {D.~H.}\ \bibnamefont
  {Guthrie}}\ and\ \bibinfo {author} {\bibfnamefont {J.~D.}\ \bibnamefont
  {Corbett}},\ }\href@noop {} {\bibfield  {journal} {\bibinfo  {journal} {J.
  Solid State Chem.}\ }\textbf {\bibinfo {volume} {37}},\ \bibinfo {pages}
  {256} (\bibinfo {year} {1981})}\BibitemShut {NoStop}%
\bibitem [{\citenamefont {Corbett}\ and\ \citenamefont
  {Guthrie}(1982)}]{ZrI2-str-WTe2}%
  \BibitemOpen
  \bibfield  {author} {\bibinfo {author} {\bibfnamefont {J.~D.}\ \bibnamefont
  {Corbett}}\ and\ \bibinfo {author} {\bibfnamefont {D.~H.}\ \bibnamefont
  {Guthrie}},\ }\href@noop {} {\bibfield  {journal} {\bibinfo  {journal}
  {Inorg. Chem.}\ }\textbf {\bibinfo {volume} {21}},\ \bibinfo {pages} {1747}
  (\bibinfo {year} {1982})}\BibitemShut {NoStop}%
\bibitem [{\citenamefont {Troyanov}\ \emph {et~al.}(1991)\citenamefont
  {Troyanov}, \citenamefont {Snigireva},\ and\ \citenamefont
  {Rybakov}}]{Troyanov-1991}%
  \BibitemOpen
  \bibfield  {author} {\bibinfo {author} {\bibfnamefont {S.~I.}\ \bibnamefont
  {Troyanov}}, \bibinfo {author} {\bibfnamefont {E.~M.}\ \bibnamefont
  {Snigireva}}, \ and\ \bibinfo {author} {\bibfnamefont {V.~B.}\ \bibnamefont
  {Rybakov}},\ }\href@noop {} {\bibfield  {journal} {\bibinfo  {journal} {Russ.
  J. Inorg. Chem.}\ }\textbf {\bibinfo {volume} {36}},\ \bibinfo {pages} {634}
  (\bibinfo {year} {1991})}\BibitemShut {NoStop}%
\bibitem [{\citenamefont {Troyanov}\ \emph {et~al.}(1990)\citenamefont
  {Troyanov}, \citenamefont {Rybakov},\ and\ \citenamefont
  {Ionov}}]{TiBr3-str}%
  \BibitemOpen
  \bibfield  {author} {\bibinfo {author} {\bibfnamefont {S.~I.}\ \bibnamefont
  {Troyanov}}, \bibinfo {author} {\bibfnamefont {V.~B.}\ \bibnamefont
  {Rybakov}}, \ and\ \bibinfo {author} {\bibfnamefont {V.~M.}\ \bibnamefont
  {Ionov}},\ }\href@noop {} {\bibfield  {journal} {\bibinfo  {journal} {Russ.
  J. Inorg. Chem.}\ }\textbf {\bibinfo {volume} {35}},\ \bibinfo {pages} {494}
  (\bibinfo {year} {1990})}\BibitemShut {NoStop}%
\bibitem [{\citenamefont {McCarley}\ \emph {et~al.}(1964)\citenamefont
  {McCarley}, \citenamefont {Roddy},\ and\ \citenamefont {Berry}}]{VBr3-str}%
  \BibitemOpen
  \bibfield  {author} {\bibinfo {author} {\bibfnamefont {R.~E.}\ \bibnamefont
  {McCarley}}, \bibinfo {author} {\bibfnamefont {J.~W.}\ \bibnamefont {Roddy}},
  \ and\ \bibinfo {author} {\bibfnamefont {K.~O.}\ \bibnamefont {Berry}},\
  }\href@noop {} {\bibfield  {journal} {\bibinfo  {journal} {Inorg. Chem.}\
  }\textbf {\bibinfo {volume} {3}},\ \bibinfo {pages} {50} (\bibinfo {year}
  {1964})}\BibitemShut {NoStop}%
\bibitem [{\citenamefont {Morosin}\ and\ \citenamefont
  {Narath}(1964)}]{CrCl3-str}%
  \BibitemOpen
  \bibfield  {author} {\bibinfo {author} {\bibfnamefont {B.}~\bibnamefont
  {Morosin}}\ and\ \bibinfo {author} {\bibfnamefont {A.}~\bibnamefont
  {Narath}},\ }\href@noop {} {\bibfield  {journal} {\bibinfo  {journal} {J.
  Chem. Phys.}\ }\textbf {\bibinfo {volume} {40}},\ \bibinfo {pages} {1958}
  (\bibinfo {year} {1964})}\BibitemShut {NoStop}%
\bibitem [{\citenamefont {Handy}\ and\ \citenamefont
  {Gregory}(1952)}]{CrBr3-str}%
  \BibitemOpen
  \bibfield  {author} {\bibinfo {author} {\bibfnamefont {L.~L.}\ \bibnamefont
  {Handy}}\ and\ \bibinfo {author} {\bibfnamefont {N.~W.}\ \bibnamefont
  {Gregory}},\ }\href@noop {} {\bibfield  {journal} {\bibinfo  {journal} {J.
  Am. Chem. Soc.}\ }\textbf {\bibinfo {volume} {74}},\ \bibinfo {pages} {891}
  (\bibinfo {year} {1952})}\BibitemShut {NoStop}%
\bibitem [{\citenamefont {Hashimoto}\ \emph {et~al.}(1989)\citenamefont
  {Hashimoto}, \citenamefont {Forster},\ and\ \citenamefont
  {Moss}}]{FeCl3-str}%
  \BibitemOpen
  \bibfield  {author} {\bibinfo {author} {\bibfnamefont {S.}~\bibnamefont
  {Hashimoto}}, \bibinfo {author} {\bibfnamefont {K.}~\bibnamefont {Forster}},
  \ and\ \bibinfo {author} {\bibfnamefont {S.~C.}\ \bibnamefont {Moss}},\
  }\href@noop {} {\bibfield  {journal} {\bibinfo  {journal} {J. Appl.
  Crystallogr.}\ }\textbf {\bibinfo {volume} {22}},\ \bibinfo {pages} {173}
  (\bibinfo {year} {1989})}\BibitemShut {NoStop}%
\bibitem [{\citenamefont {Troyanov}(1993)}]{FeCl3-str-polymorphs}%
  \BibitemOpen
  \bibfield  {author} {\bibinfo {author} {\bibfnamefont {S.}~\bibnamefont
  {Troyanov}},\ }\href@noop {} {\bibfield  {journal} {\bibinfo  {journal}
  {Russ. J. Inorg. Chem.}\ }\textbf {\bibinfo {volume} {38}},\ \bibinfo {pages}
  {1821} (\bibinfo {year} {1993})}\BibitemShut {NoStop}%
\bibitem [{\citenamefont {Armbr{\"{u}}ster}\ \emph {et~al.}(2000)\citenamefont
  {Armbr{\"{u}}ster}, \citenamefont {Ludwig}, \citenamefont {Rotter},
  \citenamefont {Thiele},\ and\ \citenamefont {Oppermann}}]{FeBr3-str}%
  \BibitemOpen
  \bibfield  {author} {\bibinfo {author} {\bibfnamefont {M.}~\bibnamefont
  {Armbr{\"{u}}ster}}, \bibinfo {author} {\bibfnamefont {T.}~\bibnamefont
  {Ludwig}}, \bibinfo {author} {\bibfnamefont {H.~W.}\ \bibnamefont {Rotter}},
  \bibinfo {author} {\bibfnamefont {G.}~\bibnamefont {Thiele}}, \ and\ \bibinfo
  {author} {\bibfnamefont {H.}~\bibnamefont {Oppermann}},\ }\href@noop {}
  {\bibfield  {journal} {\bibinfo  {journal} {Z. Anorg. Allg. Chem.}\ }\textbf
  {\bibinfo {volume} {626}},\ \bibinfo {pages} {187} (\bibinfo {year}
  {2000})}\BibitemShut {NoStop}%
\bibitem [{\citenamefont {Sch{\"{a}}fer}\ \emph {et~al.}(1967)\citenamefont
  {Sch{\"{a}}fer}, \citenamefont {{von Schnering}}, \citenamefont {Tillack},
  \citenamefont {Kuhnen}, \citenamefont {W{\"{o}}rle},\ and\ \citenamefont
  {Baumann}}]{Schafer-1967}%
  \BibitemOpen
  \bibfield  {author} {\bibinfo {author} {\bibfnamefont {H.}~\bibnamefont
  {Sch{\"{a}}fer}}, \bibinfo {author} {\bibfnamefont {H.~G.}\ \bibnamefont
  {{von Schnering}}}, \bibinfo {author} {\bibfnamefont {J.~V.}\ \bibnamefont
  {Tillack}}, \bibinfo {author} {\bibfnamefont {F.}~\bibnamefont {Kuhnen}},
  \bibinfo {author} {\bibfnamefont {H.}~\bibnamefont {W{\"{o}}rle}}, \ and\
  \bibinfo {author} {\bibfnamefont {H.}~\bibnamefont {Baumann}},\ }\href@noop
  {} {\bibfield  {journal} {\bibinfo  {journal} {Z. Anorg. Allgem. Chem.}\
  }\textbf {\bibinfo {volume} {353}},\ \bibinfo {pages} {281} (\bibinfo {year}
  {1967})}\BibitemShut {NoStop}%
\bibitem [{\citenamefont {Poineau}\ \emph {et~al.}(2012)\citenamefont
  {Poineau}, \citenamefont {Johnstone}, \citenamefont {Weck}, \citenamefont
  {Forster}, \citenamefont {Kim}, \citenamefont {Czerwinski},\ and\
  \citenamefont {Sattelberger}}]{TcCl3-str}%
  \BibitemOpen
  \bibfield  {author} {\bibinfo {author} {\bibfnamefont {F.}~\bibnamefont
  {Poineau}}, \bibinfo {author} {\bibfnamefont {E.~V.}\ \bibnamefont
  {Johnstone}}, \bibinfo {author} {\bibfnamefont {P.~F.}\ \bibnamefont {Weck}},
  \bibinfo {author} {\bibfnamefont {P.~M.}\ \bibnamefont {Forster}}, \bibinfo
  {author} {\bibfnamefont {E.}~\bibnamefont {Kim}}, \bibinfo {author}
  {\bibfnamefont {K.~R.}\ \bibnamefont {Czerwinski}}, \ and\ \bibinfo {author}
  {\bibfnamefont {A.~P.}\ \bibnamefont {Sattelberger}},\ }\href@noop {}
  {\bibfield  {journal} {\bibinfo  {journal} {Inorg. Chem.}\ }\textbf {\bibinfo
  {volume} {51}},\ \bibinfo {pages} {4915} (\bibinfo {year}
  {2012})}\BibitemShut {NoStop}%
\bibitem [{\citenamefont {Cao}\ \emph {et~al.}(2016)\citenamefont {Cao},
  \citenamefont {Banerjee}, \citenamefont {Yan}, \citenamefont {Bridges},
  \citenamefont {Lumsden}, \citenamefont {Mandrus}, \citenamefont {Tennant},
  \citenamefont {Chakoumakos},\ and\ \citenamefont {Nagler}}]{RuCl3-Cao-2016}%
  \BibitemOpen
  \bibfield  {author} {\bibinfo {author} {\bibfnamefont {H.~B.}\ \bibnamefont
  {Cao}}, \bibinfo {author} {\bibfnamefont {A.}~\bibnamefont {Banerjee}},
  \bibinfo {author} {\bibfnamefont {J.}~\bibnamefont {Yan}}, \bibinfo {author}
  {\bibfnamefont {C.~A.}\ \bibnamefont {Bridges}}, \bibinfo {author}
  {\bibfnamefont {M.~D.}\ \bibnamefont {Lumsden}}, \bibinfo {author}
  {\bibfnamefont {D.~G.}\ \bibnamefont {Mandrus}}, \bibinfo {author}
  {\bibfnamefont {D.~A.}\ \bibnamefont {Tennant}}, \bibinfo {author}
  {\bibfnamefont {B.~C.}\ \bibnamefont {Chakoumakos}}, \ and\ \bibinfo {author}
  {\bibfnamefont {S.~E.}\ \bibnamefont {Nagler}},\ }\href@noop {} {\bibfield
  {journal} {\bibinfo  {journal} {Phys. Rev. B}\ }\textbf {\bibinfo {volume}
  {93}},\ \bibinfo {pages} {134423} (\bibinfo {year} {2016})}\BibitemShut
  {NoStop}%
\bibitem [{\citenamefont {Fletcher}\ \emph {et~al.}(1967)\citenamefont
  {Fletcher}, \citenamefont {Gardner}, \citenamefont {Fox},\ and\ \citenamefont
  {Topping}}]{Fletcher-1967}%
  \BibitemOpen
  \bibfield  {author} {\bibinfo {author} {\bibfnamefont {J.~M.}\ \bibnamefont
  {Fletcher}}, \bibinfo {author} {\bibfnamefont {W.~E.}\ \bibnamefont
  {Gardner}}, \bibinfo {author} {\bibfnamefont {A.~C.}\ \bibnamefont {Fox}}, \
  and\ \bibinfo {author} {\bibfnamefont {G.}~\bibnamefont {Topping}},\
  }\href@noop {} {\bibfield  {journal} {\bibinfo  {journal} {J. Chem. Soc. A}\
  ,\ \bibinfo {pages} {1038}} (\bibinfo {year} {1967})}\BibitemShut {NoStop}%
\bibitem [{\citenamefont {Stroganov}\ and\ \citenamefont
  {Ovchinnikov}(1957)}]{RuCl3-Stroganov1957}%
  \BibitemOpen
  \bibfield  {author} {\bibinfo {author} {\bibfnamefont {E.~V.}\ \bibnamefont
  {Stroganov}}\ and\ \bibinfo {author} {\bibfnamefont {K.~V.}\ \bibnamefont
  {Ovchinnikov}},\ }\href@noop {} {\bibfield  {journal} {\bibinfo  {journal}
  {Vestn. Leningr. Univ., Ser. 4}\ }\textbf {\bibinfo {volume} {22}},\ \bibinfo
  {pages} {152} (\bibinfo {year} {1957})}\BibitemShut {NoStop}%
\bibitem [{\citenamefont {B{\"{a}}rnighausen}\ and\ \citenamefont
  {Handa}(1964)}]{RhCl3-str}%
  \BibitemOpen
  \bibfield  {author} {\bibinfo {author} {\bibfnamefont {H.}~\bibnamefont
  {B{\"{a}}rnighausen}}\ and\ \bibinfo {author} {\bibfnamefont {B.~K.}\
  \bibnamefont {Handa}},\ }\href@noop {} {\bibfield  {journal} {\bibinfo
  {journal} {J. Less-Common Met.}\ }\textbf {\bibinfo {volume} {6}},\ \bibinfo
  {pages} {226} (\bibinfo {year} {1964})}\BibitemShut {NoStop}%
\bibitem [{\citenamefont {Brodersen}\ \emph
  {et~al.}(1968{\natexlab{a}})\citenamefont {Brodersen}, \citenamefont
  {Thiele},\ and\ \citenamefont {Recke}}]{RhBr3-RhI3-str}%
  \BibitemOpen
  \bibfield  {author} {\bibinfo {author} {\bibfnamefont {K.}~\bibnamefont
  {Brodersen}}, \bibinfo {author} {\bibfnamefont {G.}~\bibnamefont {Thiele}}, \
  and\ \bibinfo {author} {\bibfnamefont {I.}~\bibnamefont {Recke}},\
  }\href@noop {} {\bibfield  {journal} {\bibinfo  {journal} {J. Less-Common
  Met.}\ }\textbf {\bibinfo {volume} {14}},\ \bibinfo {pages} {151} (\bibinfo
  {year} {1968}{\natexlab{a}})}\BibitemShut {NoStop}%
\bibitem [{\citenamefont {Brodersen}\ \emph {et~al.}(1965)\citenamefont
  {Brodersen}, \citenamefont {Moers},\ and\ \citenamefont {{von
  Schnering}}}]{RuCl3-IrCl3-Brodersen1965}%
  \BibitemOpen
  \bibfield  {author} {\bibinfo {author} {\bibfnamefont {K.}~\bibnamefont
  {Brodersen}}, \bibinfo {author} {\bibfnamefont {F.}~\bibnamefont {Moers}}, \
  and\ \bibinfo {author} {\bibfnamefont {H.~G.}\ \bibnamefont {{von
  Schnering}}},\ }\href@noop {} {\bibfield  {journal} {\bibinfo  {journal}
  {Naturwissensch.}\ }\textbf {\bibinfo {volume} {52}},\ \bibinfo {pages} {205}
  (\bibinfo {year} {1965})}\BibitemShut {NoStop}%
\bibitem [{\citenamefont {Brodersen}\ \emph
  {et~al.}(1968{\natexlab{b}})\citenamefont {Brodersen}, \citenamefont
  {Thiele}, \citenamefont {Ohnsorge}, \citenamefont {Recke},\ and\
  \citenamefont {Moers}}]{IrBr3-IrI3-RuCl3-str}%
  \BibitemOpen
  \bibfield  {author} {\bibinfo {author} {\bibfnamefont {K.}~\bibnamefont
  {Brodersen}}, \bibinfo {author} {\bibfnamefont {G.}~\bibnamefont {Thiele}},
  \bibinfo {author} {\bibfnamefont {H.}~\bibnamefont {Ohnsorge}}, \bibinfo
  {author} {\bibfnamefont {I.}~\bibnamefont {Recke}}, \ and\ \bibinfo {author}
  {\bibfnamefont {F.}~\bibnamefont {Moers}},\ }\href@noop {} {\bibfield
  {journal} {\bibinfo  {journal} {J. Less-Common Met.}\ }\textbf {\bibinfo
  {volume} {15}},\ \bibinfo {pages} {347} (\bibinfo {year}
  {1968}{\natexlab{b}})}\BibitemShut {NoStop}%
\bibitem [{\citenamefont {Brehzer}(1959)}]{ZnCl2}%
  \BibitemOpen
  \bibfield  {author} {\bibinfo {author} {\bibfnamefont {B.}~\bibnamefont
  {Brehzer}},\ }\href@noop {} {\bibfield  {journal} {\bibinfo  {journal}
  {Naturwissenschaften}\ }\textbf {\bibinfo {volume} {46}},\ \bibinfo {pages}
  {106b} (\bibinfo {year} {1959})}\BibitemShut {NoStop}%
\bibitem [{\citenamefont {Yamaguchi}(1942{\natexlab{a}})}]{ZnBr2}%
  \BibitemOpen
  \bibfield  {author} {\bibinfo {author} {\bibfnamefont {S.}~\bibnamefont
  {Yamaguchi}},\ }\href@noop {} {\bibfield  {journal} {\bibinfo  {journal}
  {Sci. Pap. Inst. Phys. Chem. Res. (Jpn.)}\ }\textbf {\bibinfo {volume}
  {39}},\ \bibinfo {pages} {291} (\bibinfo {year}
  {1942}{\natexlab{a}})}\BibitemShut {NoStop}%
\bibitem [{\citenamefont {Yamaguchi}(1942{\natexlab{b}})}]{ZnI2}%
  \BibitemOpen
  \bibfield  {author} {\bibinfo {author} {\bibfnamefont {S.}~\bibnamefont
  {Yamaguchi}},\ }\href@noop {} {\bibfield  {journal} {\bibinfo  {journal}
  {Sci. Pap. Inst. Phys. Chem. Res. (Jpn.)}\ }\textbf {\bibinfo {volume}
  {39}},\ \bibinfo {pages} {357} (\bibinfo {year}
  {1942}{\natexlab{b}})}\BibitemShut {NoStop}%
\bibitem [{\citenamefont {Pauling}(1929)}]{CdCl2}%
  \BibitemOpen
  \bibfield  {author} {\bibinfo {author} {\bibfnamefont {L.}~\bibnamefont
  {Pauling}},\ }\href@noop {} {\bibfield  {journal} {\bibinfo  {journal} {Proc.
  Natl. Acad. Sci. U.S.A.}\ }\textbf {\bibinfo {volume} {15}},\ \bibinfo
  {pages} {709} (\bibinfo {year} {1929})}\BibitemShut {NoStop}%
\bibitem [{\citenamefont {Mitchell}(1962)}]{CdBr2}%
  \BibitemOpen
  \bibfield  {author} {\bibinfo {author} {\bibfnamefont {R.~S.}\ \bibnamefont
  {Mitchell}},\ }\href@noop {} {\bibfield  {journal} {\bibinfo  {journal} {Z.
  Kristallogr.}\ }\textbf {\bibinfo {volume} {117}},\ \bibinfo {pages} {309}
  (\bibinfo {year} {1962})}\BibitemShut {NoStop}%
\bibitem [{\citenamefont {Bozorth}(1922)}]{CdI2}%
  \BibitemOpen
  \bibfield  {author} {\bibinfo {author} {\bibfnamefont {R.~M.}\ \bibnamefont
  {Bozorth}},\ }\href@noop {} {\bibfield  {journal} {\bibinfo  {journal} {J.
  Am. Chem. Soc.}\ }\textbf {\bibinfo {volume} {44}},\ \bibinfo {pages} {2232}
  (\bibinfo {year} {1922})}\BibitemShut {NoStop}%
\bibitem [{\citenamefont {Bijvoet}\ \emph {et~al.}(1926)\citenamefont
  {Bijvoet}, \citenamefont {Claassen},\ and\ \citenamefont {Karssen}}]{HgI2}%
  \BibitemOpen
  \bibfield  {author} {\bibinfo {author} {\bibfnamefont {J.~M.}\ \bibnamefont
  {Bijvoet}}, \bibinfo {author} {\bibfnamefont {A.}~\bibnamefont {Claassen}}, \
  and\ \bibinfo {author} {\bibfnamefont {A.}~\bibnamefont {Karssen}},\
  }\href@noop {} {\bibfield  {journal} {\bibinfo  {journal} {Proc. K. Ned.
  Akad. Wet.}\ }\textbf {\bibinfo {volume} {29}},\ \bibinfo {pages} {529}
  (\bibinfo {year} {1926})}\BibitemShut {NoStop}%
\bibitem [{\citenamefont {Schneider}\ \emph {et~al.}(1992)\citenamefont
  {Schneider}, \citenamefont {Kuske},\ and\ \citenamefont
  {Lutz}}]{MnBr2-CdCl2}%
  \BibitemOpen
  \bibfield  {author} {\bibinfo {author} {\bibfnamefont {M.}~\bibnamefont
  {Schneider}}, \bibinfo {author} {\bibfnamefont {P.}~\bibnamefont {Kuske}}, \
  and\ \bibinfo {author} {\bibfnamefont {H.~D.}\ \bibnamefont {Lutz}},\
  }\href@noop {} {\bibfield  {journal} {\bibinfo  {journal} {Acta Crystallogr.
  B}\ }\textbf {\bibinfo {volume} {48}},\ \bibinfo {pages} {761} (\bibinfo
  {year} {1992})}\BibitemShut {NoStop}%
\bibitem [{\citenamefont {Kuindersma}\ \emph {et~al.}(1981)\citenamefont
  {Kuindersma}, \citenamefont {Sanchez},\ and\ \citenamefont
  {Haas}}]{Kuindersma-1981}%
  \BibitemOpen
  \bibfield  {author} {\bibinfo {author} {\bibfnamefont {S.~R.}\ \bibnamefont
  {Kuindersma}}, \bibinfo {author} {\bibfnamefont {J.~P.}\ \bibnamefont
  {Sanchez}}, \ and\ \bibinfo {author} {\bibfnamefont {C.}~\bibnamefont
  {Haas}},\ }\href@noop {} {\bibfield  {journal} {\bibinfo  {journal} {Phys. B
  Condens. Matter}\ }\textbf {\bibinfo {volume} {111}},\ \bibinfo {pages} {231}
  (\bibinfo {year} {1981})}\BibitemShut {NoStop}%
\bibitem [{\citenamefont {Narath}\ and\ \citenamefont
  {Schirber}(1966)}]{Narath-1966}%
  \BibitemOpen
  \bibfield  {author} {\bibinfo {author} {\bibfnamefont {A.}~\bibnamefont
  {Narath}}\ and\ \bibinfo {author} {\bibfnamefont {J.~E.}\ \bibnamefont
  {Schirber}},\ }\href@noop {} {\bibfield  {journal} {\bibinfo  {journal} {J.
  Appl. Phys.}\ }\textbf {\bibinfo {volume} {37}},\ \bibinfo {pages} {1124}
  (\bibinfo {year} {1966})}\BibitemShut {NoStop}%
\bibitem [{\citenamefont {Rozenberg}\ \emph {et~al.}(2009)\citenamefont
  {Rozenberg}, \citenamefont {Pasternak}, \citenamefont {Gorodetsky},
  \citenamefont {Xu}, \citenamefont {Dubrovinsky}, \citenamefont {LeBihan},\
  and\ \citenamefont {Taylor}}]{FeCl2-pressure}%
  \BibitemOpen
  \bibfield  {author} {\bibinfo {author} {\bibfnamefont {G.~K.}\ \bibnamefont
  {Rozenberg}}, \bibinfo {author} {\bibfnamefont {M.~P.}\ \bibnamefont
  {Pasternak}}, \bibinfo {author} {\bibfnamefont {P.}~\bibnamefont
  {Gorodetsky}}, \bibinfo {author} {\bibfnamefont {W.~M.}\ \bibnamefont {Xu}},
  \bibinfo {author} {\bibfnamefont {L.~S.}\ \bibnamefont {Dubrovinsky}},
  \bibinfo {author} {\bibfnamefont {T.}~\bibnamefont {LeBihan}}, \ and\
  \bibinfo {author} {\bibfnamefont {R.~D.}\ \bibnamefont {Taylor}},\
  }\href@noop {} {\bibfield  {journal} {\bibinfo  {journal} {Phys. Rev. B}\
  }\textbf {\bibinfo {volume} {79}},\ \bibinfo {pages} {214105} (\bibinfo
  {year} {2009})}\BibitemShut {NoStop}%
\bibitem [{\citenamefont {Imoto}\ \emph {et~al.}(1981)\citenamefont {Imoto},
  \citenamefont {Corbett},\ and\ \citenamefont {Cisar}}]{ZrCl2-cluster}%
  \BibitemOpen
  \bibfield  {author} {\bibinfo {author} {\bibfnamefont {H.}~\bibnamefont
  {Imoto}}, \bibinfo {author} {\bibfnamefont {J.~D.}\ \bibnamefont {Corbett}},
  \ and\ \bibinfo {author} {\bibfnamefont {A.}~\bibnamefont {Cisar}},\
  }\href@noop {} {\bibfield  {journal} {\bibinfo  {journal} {Inorg. Chem.}\
  }\textbf {\bibinfo {volume} {20}},\ \bibinfo {pages} {145} (\bibinfo {year}
  {1981})}\BibitemShut {NoStop}%
\bibitem [{\citenamefont {{Men'kov}}\ and\ \citenamefont
  {Komissarova}(1964{\natexlab{a}})}]{ScBr3}%
  \BibitemOpen
  \bibfield  {author} {\bibinfo {author} {\bibfnamefont {A.~A.}\ \bibnamefont
  {{Men'kov}}}\ and\ \bibinfo {author} {\bibfnamefont {L.~N.}\ \bibnamefont
  {Komissarova}},\ }\href@noop {} {\bibfield  {journal} {\bibinfo  {journal}
  {Russ. J. Inorg. Chem.}\ }\textbf {\bibinfo {volume} {9}},\ \bibinfo {pages}
  {952} (\bibinfo {year} {1964}{\natexlab{a}})}\BibitemShut {NoStop}%
\bibitem [{\citenamefont {{Men'kov}}\ and\ \citenamefont
  {Komissarova}(1964{\natexlab{b}})}]{ScI3}%
  \BibitemOpen
  \bibfield  {author} {\bibinfo {author} {\bibfnamefont {A.~A.}\ \bibnamefont
  {{Men'kov}}}\ and\ \bibinfo {author} {\bibfnamefont {L.~N.}\ \bibnamefont
  {Komissarova}},\ }\href@noop {} {\bibfield  {journal} {\bibinfo  {journal}
  {Russ. J. Inorg. Chem.}\ }\textbf {\bibinfo {volume} {9}},\ \bibinfo {pages}
  {425} (\bibinfo {year} {1964}{\natexlab{b}})}\BibitemShut {NoStop}%
\bibitem [{\citenamefont {Templeton}\ and\ \citenamefont
  {Darter}(1954)}]{YCl3}%
  \BibitemOpen
  \bibfield  {author} {\bibinfo {author} {\bibfnamefont {D.~H.}\ \bibnamefont
  {Templeton}}\ and\ \bibinfo {author} {\bibfnamefont {G.~F.}\ \bibnamefont
  {Darter}},\ }\href@noop {} {\bibfield  {journal} {\bibinfo  {journal} {J.
  Phys. Chem.}\ }\textbf {\bibinfo {volume} {58}},\ \bibinfo {pages} {940}
  (\bibinfo {year} {1954})}\BibitemShut {NoStop}%
\bibitem [{\citenamefont {Brown}\ \emph {et~al.}(1968)\citenamefont {Brown},
  \citenamefont {Fletcher},\ and\ \citenamefont {Holah}}]{YBr3}%
  \BibitemOpen
  \bibfield  {author} {\bibinfo {author} {\bibfnamefont {D.}~\bibnamefont
  {Brown}}, \bibinfo {author} {\bibfnamefont {S.}~\bibnamefont {Fletcher}}, \
  and\ \bibinfo {author} {\bibfnamefont {D.~G.}\ \bibnamefont {Holah}},\
  }\href@noop {} {\bibfield  {journal} {\bibinfo  {journal} {J. Chem. Soc. A}\
  ,\ \bibinfo {pages} {1889}} (\bibinfo {year} {1968})}\BibitemShut {NoStop}%
\bibitem [{\citenamefont {Jongen}\ and\ \citenamefont {Meyer}(2005)}]{YI3}%
  \BibitemOpen
  \bibfield  {author} {\bibinfo {author} {\bibfnamefont {L.}~\bibnamefont
  {Jongen}}\ and\ \bibinfo {author} {\bibfnamefont {G.}~\bibnamefont {Meyer}},\
  }\href@noop {} {\bibfield  {journal} {\bibinfo  {journal} {Acta Crystallogr.
  E}\ }\textbf {\bibinfo {volume} {61}},\ \bibinfo {pages} {i151} (\bibinfo
  {year} {2005})}\BibitemShut {NoStop}%
\bibitem [{\citenamefont {Poineau}\ \emph {et~al.}(2013)\citenamefont
  {Poineau}, \citenamefont {Johnstone}, \citenamefont {Czerwinski},\ and\
  \citenamefont {Sattelberger}}]{Poineau-2013}%
  \BibitemOpen
  \bibfield  {author} {\bibinfo {author} {\bibfnamefont {F.}~\bibnamefont
  {Poineau}}, \bibinfo {author} {\bibfnamefont {E.~V.}\ \bibnamefont
  {Johnstone}}, \bibinfo {author} {\bibfnamefont {K.~R.}\ \bibnamefont
  {Czerwinski}}, \ and\ \bibinfo {author} {\bibfnamefont {A.~P.}\ \bibnamefont
  {Sattelberger}},\ }\href@noop {} {\bibfield  {journal} {\bibinfo  {journal}
  {Accounts of Chemical Research}\ }\textbf {\bibinfo {volume} {47}},\ \bibinfo
  {pages} {624} (\bibinfo {year} {2013})}\BibitemShut {NoStop}%
\bibitem [{\citenamefont {Ogawa}(1960)}]{Ogawa-1960}%
  \BibitemOpen
  \bibfield  {author} {\bibinfo {author} {\bibfnamefont {S.}~\bibnamefont
  {Ogawa}},\ }\href@noop {} {\bibfield  {journal} {\bibinfo  {journal} {J.
  Phys. Soc. Japan}\ }\textbf {\bibinfo {volume} {15}},\ \bibinfo {pages}
  {1901} (\bibinfo {year} {1960})}\BibitemShut {NoStop}%
\bibitem [{\citenamefont {Troyanov}\ \emph {et~al.}(1994)\citenamefont
  {Troyanov}, \citenamefont {Snigireva}, \citenamefont {Pirarevskii},
  \citenamefont {Yanovskii},\ and\ \citenamefont {Struchkov}}]{Troyanov-1994}%
  \BibitemOpen
  \bibfield  {author} {\bibinfo {author} {\bibfnamefont {S.~I.}\ \bibnamefont
  {Troyanov}}, \bibinfo {author} {\bibfnamefont {E.~M.}\ \bibnamefont
  {Snigireva}}, \bibinfo {author} {\bibfnamefont {A.~P.}\ \bibnamefont
  {Pirarevskii}}, \bibinfo {author} {\bibfnamefont {A.~I.}\ \bibnamefont
  {Yanovskii}}, \ and\ \bibinfo {author} {\bibfnamefont {Y.~T.}\ \bibnamefont
  {Struchkov}},\ }\href@noop {} {\bibfield  {journal} {\bibinfo  {journal}
  {Russ. J. Inorg. Chem.}\ }\textbf {\bibinfo {volume} {39}},\ \bibinfo {pages}
  {360} (\bibinfo {year} {1994})}\BibitemShut {NoStop}%
\bibitem [{\citenamefont {Troyanov}\ and\ \citenamefont
  {Snigireva}(2000)}]{Troyanov-2000}%
  \BibitemOpen
  \bibfield  {author} {\bibinfo {author} {\bibfnamefont {S.~I.}\ \bibnamefont
  {Troyanov}}\ and\ \bibinfo {author} {\bibfnamefont {E.~M.}\ \bibnamefont
  {Snigireva}},\ }\href@noop {} {\bibfield  {journal} {\bibinfo  {journal}
  {Russ. J. Inorg. Chem.}\ }\textbf {\bibinfo {volume} {45}},\ \bibinfo {pages}
  {580} (\bibinfo {year} {2000})}\BibitemShut {NoStop}%
\bibitem [{\citenamefont {Park}\ \emph {et~al.}(2016)\citenamefont {Park},
  \citenamefont {Do}, \citenamefont {Choi}, \citenamefont {Jang}, \citenamefont
  {Jang}, \citenamefont {Schefer}, \citenamefont {Wu}, \citenamefont {Gardner},
  \citenamefont {Park}, \citenamefont {Park},\ and\ \citenamefont
  {Ji}}]{Park-RuCl3}%
  \BibitemOpen
  \bibfield  {author} {\bibinfo {author} {\bibfnamefont {S.-Y.}\ \bibnamefont
  {Park}}, \bibinfo {author} {\bibfnamefont {S.-H.}\ \bibnamefont {Do}},
  \bibinfo {author} {\bibfnamefont {K.-Y.}\ \bibnamefont {Choi}}, \bibinfo
  {author} {\bibfnamefont {D.}~\bibnamefont {Jang}}, \bibinfo {author}
  {\bibfnamefont {T.-H.}\ \bibnamefont {Jang}}, \bibinfo {author}
  {\bibfnamefont {J.}~\bibnamefont {Schefer}}, \bibinfo {author} {\bibfnamefont
  {C.-M.}\ \bibnamefont {Wu}}, \bibinfo {author} {\bibfnamefont {J.~S.}\
  \bibnamefont {Gardner}}, \bibinfo {author} {\bibfnamefont {J.~M.~S.}\
  \bibnamefont {Park}}, \bibinfo {author} {\bibfnamefont {J.-H.}\ \bibnamefont
  {Park}}, \ and\ \bibinfo {author} {\bibfnamefont {S.}~\bibnamefont {Ji}},\
  }\href@noop {} {\bibfield  {journal} {\bibinfo  {journal} {arXiv:1609.05690}\
  } (\bibinfo {year} {2016})}\BibitemShut {NoStop}%
\bibitem [{\citenamefont {Babel}\ and\ \citenamefont
  {Deigner}(1965)}]{IrCl3-ortho}%
  \BibitemOpen
  \bibfield  {author} {\bibinfo {author} {\bibfnamefont {D.}~\bibnamefont
  {Babel}}\ and\ \bibinfo {author} {\bibfnamefont {P.}~\bibnamefont
  {Deigner}},\ }\href@noop {} {\bibfield  {journal} {\bibinfo  {journal} {Z.
  Anorg. Allg. Chem.}\ }\textbf {\bibinfo {volume} {339}},\ \bibinfo {pages}
  {57} (\bibinfo {year} {1965})}\BibitemShut {NoStop}%
\bibitem [{\citenamefont {Kim}\ and\ \citenamefont
  {Kee}(2016)}]{RuCl3-Kim2016}%
  \BibitemOpen
  \bibfield  {author} {\bibinfo {author} {\bibfnamefont {H.-S.}\ \bibnamefont
  {Kim}}\ and\ \bibinfo {author} {\bibfnamefont {H.-Y.}\ \bibnamefont {Kee}},\
  }\href@noop {} {\bibfield  {journal} {\bibinfo  {journal} {Phys. Rev. B}\
  }\textbf {\bibinfo {volume} {93}},\ \bibinfo {pages} {155143} (\bibinfo
  {year} {2016})}\BibitemShut {NoStop}%
\bibitem [{\citenamefont {Zhou}\ \emph {et~al.}(2016)\citenamefont {Zhou},
  \citenamefont {Lu}, \citenamefont {Zu},\ and\ \citenamefont
  {Gao}}]{Zhou-TiCl3-VCl3}%
  \BibitemOpen
  \bibfield  {author} {\bibinfo {author} {\bibfnamefont {Y.}~\bibnamefont
  {Zhou}}, \bibinfo {author} {\bibfnamefont {H.}~\bibnamefont {Lu}}, \bibinfo
  {author} {\bibfnamefont {X.}~\bibnamefont {Zu}}, \ and\ \bibinfo {author}
  {\bibfnamefont {F.}~\bibnamefont {Gao}},\ }\href@noop {} {\bibfield
  {journal} {\bibinfo  {journal} {Scientific Reports}\ }\textbf {\bibinfo
  {volume} {6}},\ \bibinfo {pages} {19407} (\bibinfo {year}
  {2016})}\BibitemShut {NoStop}%
\bibitem [{\citenamefont {Khomskii}(2014)}]{Khomskii}%
  \BibitemOpen
  \bibfield  {author} {\bibinfo {author} {\bibfnamefont {D.~I.}\ \bibnamefont
  {Khomskii}},\ }\href@noop {} {\emph {\bibinfo {title} {Transition Metal
  Compounds}}}\ (\bibinfo  {publisher} {Cambridge University Press},\ \bibinfo
  {year} {2014})\BibitemShut {NoStop}%
\bibitem [{\citenamefont {Niel}\ \emph {et~al.}(1977)\citenamefont {Niel},
  \citenamefont {Cros}, \citenamefont {{Le Flem}}, \citenamefont {Pouchard},\
  and\ \citenamefont {Hagenmuller}}]{Niel-1977}%
  \BibitemOpen
  \bibfield  {author} {\bibinfo {author} {\bibfnamefont {M.}~\bibnamefont
  {Niel}}, \bibinfo {author} {\bibfnamefont {C.}~\bibnamefont {Cros}}, \bibinfo
  {author} {\bibfnamefont {G.}~\bibnamefont {{Le Flem}}}, \bibinfo {author}
  {\bibfnamefont {M.}~\bibnamefont {Pouchard}}, \ and\ \bibinfo {author}
  {\bibfnamefont {P.}~\bibnamefont {Hagenmuller}},\ }\href@noop {} {\bibfield
  {journal} {\bibinfo  {journal} {Phys. B Condens. Matter}\ }\textbf {\bibinfo
  {volume} {86-88}},\ \bibinfo {pages} {702} (\bibinfo {year}
  {1977})}\BibitemShut {NoStop}%
\bibitem [{\citenamefont {Lewis}\ \emph {et~al.}(1962)\citenamefont {Lewis},
  \citenamefont {Machin}, \citenamefont {Newnham},\ and\ \citenamefont
  {Nyholm}}]{Lewis-1962}%
  \BibitemOpen
  \bibfield  {author} {\bibinfo {author} {\bibfnamefont {J.}~\bibnamefont
  {Lewis}}, \bibinfo {author} {\bibfnamefont {D.~J.}\ \bibnamefont {Machin}},
  \bibinfo {author} {\bibfnamefont {I.~E.}\ \bibnamefont {Newnham}}, \ and\
  \bibinfo {author} {\bibfnamefont {R.~S.}\ \bibnamefont {Nyholm}},\
  }\href@noop {} {\bibfield  {journal} {\bibinfo  {journal} {J. Chem Soc.}\ ,\
  \bibinfo {pages} {2036}} (\bibinfo {year} {1962})}\BibitemShut {NoStop}%
\bibitem [{\citenamefont {Starr}\ \emph {et~al.}(1940)\citenamefont {Starr},
  \citenamefont {Bitter},\ and\ \citenamefont {Kaufmann}}]{Starr-1940}%
  \BibitemOpen
  \bibfield  {author} {\bibinfo {author} {\bibfnamefont {C.}~\bibnamefont
  {Starr}}, \bibinfo {author} {\bibfnamefont {R.}~\bibnamefont {Bitter}}, \
  and\ \bibinfo {author} {\bibfnamefont {A.~R.}\ \bibnamefont {Kaufmann}},\
  }\href@noop {} {\bibfield  {journal} {\bibinfo  {journal} {Phys. Rev.}\
  }\textbf {\bibinfo {volume} {58}},\ \bibinfo {pages} {977} (\bibinfo {year}
  {1940})}\BibitemShut {NoStop}%
\bibitem [{\citenamefont {Hirakawa}\ \emph {et~al.}(1983)\citenamefont
  {Hirakawa}, \citenamefont {Kadowaki},\ and\ \citenamefont
  {Ubukoshi}}]{Hirakawa-1983}%
  \BibitemOpen
  \bibfield  {author} {\bibinfo {author} {\bibfnamefont {K.}~\bibnamefont
  {Hirakawa}}, \bibinfo {author} {\bibfnamefont {H.}~\bibnamefont {Kadowaki}},
  \ and\ \bibinfo {author} {\bibfnamefont {K.}~\bibnamefont {Ubukoshi}},\
  }\href@noop {} {\bibfield  {journal} {\bibinfo  {journal} {J. Phys. Soc.
  Japan}\ }\textbf {\bibinfo {volume} {52}},\ \bibinfo {pages} {1814} (\bibinfo
  {year} {1983})}\BibitemShut {NoStop}%
\bibitem [{\citenamefont {Kadowaki}\ \emph {et~al.}(1985)\citenamefont
  {Kadowaki}, \citenamefont {Ubukoshi},\ and\ \citenamefont
  {Hirakawa}}]{Kadowaki-1985}%
  \BibitemOpen
  \bibfield  {author} {\bibinfo {author} {\bibfnamefont {H.}~\bibnamefont
  {Kadowaki}}, \bibinfo {author} {\bibfnamefont {K.}~\bibnamefont {Ubukoshi}},
  \ and\ \bibinfo {author} {\bibfnamefont {K.}~\bibnamefont {Hirakawa}},\
  }\href@noop {} {\bibfield  {journal} {\bibinfo  {journal} {J. Phys. Soc.
  Japan}\ }\textbf {\bibinfo {volume} {54}},\ \bibinfo {pages} {363} (\bibinfo
  {year} {1985})}\BibitemShut {NoStop}%
\bibitem [{\citenamefont {{Abdul Wasey}}\ \emph {et~al.}(2013)\citenamefont
  {{Abdul Wasey}}, \citenamefont {Karmakar},\ and\ \citenamefont
  {Das}}]{Abdul-Wasey-2013}%
  \BibitemOpen
  \bibfield  {author} {\bibinfo {author} {\bibfnamefont {A.~H.~M.}\
  \bibnamefont {{Abdul Wasey}}}, \bibinfo {author} {\bibfnamefont
  {D.}~\bibnamefont {Karmakar}}, \ and\ \bibinfo {author} {\bibfnamefont
  {G.~P.}\ \bibnamefont {Das}},\ }\href@noop {} {\bibfield  {journal} {\bibinfo
   {journal} {J. Phys.: Condens. Matter}\ }\textbf {\bibinfo {volume} {25}},\
  \bibinfo {pages} {476001} (\bibinfo {year} {2013})}\BibitemShut {NoStop}%
\bibitem [{\citenamefont {Murray}(1962)}]{Murray-1962}%
  \BibitemOpen
  \bibfield  {author} {\bibinfo {author} {\bibfnamefont {R.~B.}\ \bibnamefont
  {Murray}},\ }\href@noop {} {\bibfield  {journal} {\bibinfo  {journal} {Phys.
  Rev.}\ }\textbf {\bibinfo {volume} {128}},\ \bibinfo {pages} {1570} (\bibinfo
  {year} {1962})}\BibitemShut {NoStop}%
\bibitem [{\citenamefont {Wilkinson}\ \emph {et~al.}(1958)\citenamefont
  {Wilkinson}, \citenamefont {Cable}, \citenamefont {Wollan},\ and\
  \citenamefont {Koehler}}]{Wilkinson-MnCl2-1958}%
  \BibitemOpen
  \bibfield  {author} {\bibinfo {author} {\bibfnamefont {M.~K.}\ \bibnamefont
  {Wilkinson}}, \bibinfo {author} {\bibfnamefont {J.~W.}\ \bibnamefont
  {Cable}}, \bibinfo {author} {\bibfnamefont {E.~O.}\ \bibnamefont {Wollan}}, \
  and\ \bibinfo {author} {\bibfnamefont {W.~C.}\ \bibnamefont {Koehler}},\
  }\href@noop {} {\bibfield  {journal} {\bibinfo  {journal} {Oak Ridge National
  Laboratory Report}\ }\textbf {\bibinfo {volume} {ORNL-2430}},\ \bibinfo
  {pages} {65} (\bibinfo {year} {1958})}\BibitemShut {NoStop}%
\bibitem [{\citenamefont {Wiesler}\ \emph {et~al.}(1995)\citenamefont
  {Wiesler}, \citenamefont {Suzuki}, \citenamefont {Suzuki},\ and\
  \citenamefont {Rosov}}]{Wiesler-1995}%
  \BibitemOpen
  \bibfield  {author} {\bibinfo {author} {\bibfnamefont {D.~G.}\ \bibnamefont
  {Wiesler}}, \bibinfo {author} {\bibfnamefont {M.}~\bibnamefont {Suzuki}},
  \bibinfo {author} {\bibfnamefont {I.~S.}\ \bibnamefont {Suzuki}}, \ and\
  \bibinfo {author} {\bibfnamefont {N.}~\bibnamefont {Rosov}},\ }\href@noop {}
  {\bibfield  {journal} {\bibinfo  {journal} {Phys. Rev. Lett.}\ }\textbf
  {\bibinfo {volume} {75}},\ \bibinfo {pages} {942} (\bibinfo {year}
  {1995})}\BibitemShut {NoStop}%
\bibitem [{\citenamefont {Sato}\ \emph {et~al.}(1995)\citenamefont {Sato},
  \citenamefont {Kadowaki},\ and\ \citenamefont {Iio}}]{Sato-1995}%
  \BibitemOpen
  \bibfield  {author} {\bibinfo {author} {\bibfnamefont {T.}~\bibnamefont
  {Sato}}, \bibinfo {author} {\bibfnamefont {H.}~\bibnamefont {Kadowaki}}, \
  and\ \bibinfo {author} {\bibfnamefont {K.}~\bibnamefont {Iio}},\ }\href@noop
  {} {\bibfield  {journal} {\bibinfo  {journal} {Phys. B Condens. Matter}\
  }\textbf {\bibinfo {volume} {213-214}},\ \bibinfo {pages} {224} (\bibinfo
  {year} {1995})}\BibitemShut {NoStop}%
\bibitem [{\citenamefont {Cable}\ \emph
  {et~al.}(1962{\natexlab{a}})\citenamefont {Cable}, \citenamefont {Wilkinson},
  \citenamefont {Wollan},\ and\ \citenamefont {Koehler}}]{Cable-1962}%
  \BibitemOpen
  \bibfield  {author} {\bibinfo {author} {\bibfnamefont {J.~W.}\ \bibnamefont
  {Cable}}, \bibinfo {author} {\bibfnamefont {M.~K.}\ \bibnamefont
  {Wilkinson}}, \bibinfo {author} {\bibfnamefont {E.~O.}\ \bibnamefont
  {Wollan}}, \ and\ \bibinfo {author} {\bibfnamefont {W.~C.}\ \bibnamefont
  {Koehler}},\ }\href@noop {} {\bibfield  {journal} {\bibinfo  {journal} {Phys.
  Rev.}\ }\textbf {\bibinfo {volume} {125}},\ \bibinfo {pages} {1860} (\bibinfo
  {year} {1962}{\natexlab{a}})}\BibitemShut {NoStop}%
\bibitem [{\citenamefont {Katsumata}\ \emph {et~al.}(2010)\citenamefont
  {Katsumata}, \citenamefont {{Aruga Katori}}, \citenamefont {Kimura},
  \citenamefont {Narumi}, \citenamefont {Hagiwara},\ and\ \citenamefont
  {Kindo}}]{Katsumata-2010}%
  \BibitemOpen
  \bibfield  {author} {\bibinfo {author} {\bibfnamefont {K.}~\bibnamefont
  {Katsumata}}, \bibinfo {author} {\bibfnamefont {H.}~\bibnamefont {{Aruga
  Katori}}}, \bibinfo {author} {\bibfnamefont {S.}~\bibnamefont {Kimura}},
  \bibinfo {author} {\bibfnamefont {Y.}~\bibnamefont {Narumi}}, \bibinfo
  {author} {\bibfnamefont {M.}~\bibnamefont {Hagiwara}}, \ and\ \bibinfo
  {author} {\bibfnamefont {K.}~\bibnamefont {Kindo}},\ }\href@noop {}
  {\bibfield  {journal} {\bibinfo  {journal} {Phys. Rev. B}\ }\textbf {\bibinfo
  {volume} {82}},\ \bibinfo {pages} {104402} (\bibinfo {year}
  {2010})}\BibitemShut {NoStop}%
\bibitem [{\citenamefont {Bertrand}\ \emph {et~al.}(1974)\citenamefont
  {Bertrand}, \citenamefont {Fert},\ and\ \citenamefont
  {G\'{e}lard}}]{Bertrand-1974}%
  \BibitemOpen
  \bibfield  {author} {\bibinfo {author} {\bibfnamefont {Y.}~\bibnamefont
  {Bertrand}}, \bibinfo {author} {\bibfnamefont {A.~R.}\ \bibnamefont {Fert}},
  \ and\ \bibinfo {author} {\bibfnamefont {J.}~\bibnamefont {G\'{e}lard}},\
  }\href@noop {} {\bibfield  {journal} {\bibinfo  {journal} {J. Physique}\
  }\textbf {\bibinfo {volume} {35}},\ \bibinfo {pages} {385} (\bibinfo {year}
  {1974})}\BibitemShut {NoStop}%
\bibitem [{\citenamefont {Wilkinson}\ \emph {et~al.}(1959)\citenamefont
  {Wilkinson}, \citenamefont {Cable}, \citenamefont {Wollan},\ and\
  \citenamefont {Koehler}}]{Wilkinson-1959}%
  \BibitemOpen
  \bibfield  {author} {\bibinfo {author} {\bibfnamefont {M.~K.}\ \bibnamefont
  {Wilkinson}}, \bibinfo {author} {\bibfnamefont {J.~W.}\ \bibnamefont
  {Cable}}, \bibinfo {author} {\bibfnamefont {E.~O.}\ \bibnamefont {Wollan}}, \
  and\ \bibinfo {author} {\bibfnamefont {W.~C.}\ \bibnamefont {Koehler}},\
  }\href@noop {} {\bibfield  {journal} {\bibinfo  {journal} {Phys. Rev.}\
  }\textbf {\bibinfo {volume} {113}},\ \bibinfo {pages} {497} (\bibinfo {year}
  {1959})}\BibitemShut {NoStop}%
\bibitem [{\citenamefont {Fert}\ \emph
  {et~al.}(1973{\natexlab{a}})\citenamefont {Fert}, \citenamefont {Carrara},
  \citenamefont {Lanusse}, \citenamefont {Mischler},\ and\ \citenamefont
  {Redoules}}]{Fert-1973-FeBr2}%
  \BibitemOpen
  \bibfield  {author} {\bibinfo {author} {\bibfnamefont {A.~R.}\ \bibnamefont
  {Fert}}, \bibinfo {author} {\bibfnamefont {P.}~\bibnamefont {Carrara}},
  \bibinfo {author} {\bibfnamefont {M.~C.}\ \bibnamefont {Lanusse}}, \bibinfo
  {author} {\bibfnamefont {G.}~\bibnamefont {Mischler}}, \ and\ \bibinfo
  {author} {\bibfnamefont {J.~P.}\ \bibnamefont {Redoules}},\ }\href@noop {}
  {\bibfield  {journal} {\bibinfo  {journal} {J. Phys. Chem. Solids}\ }\textbf
  {\bibinfo {volume} {34}},\ \bibinfo {pages} {223} (\bibinfo {year}
  {1973}{\natexlab{a}})}\BibitemShut {NoStop}%
\bibitem [{\citenamefont {Xu}\ and\ \citenamefont {Pasternak}(2002)}]{Xu-2002}%
  \BibitemOpen
  \bibfield  {author} {\bibinfo {author} {\bibfnamefont {W.~M.}\ \bibnamefont
  {Xu}}\ and\ \bibinfo {author} {\bibfnamefont {M.~P.}\ \bibnamefont
  {Pasternak}},\ }\href@noop {} {\bibfield  {journal} {\bibinfo  {journal}
  {Hyperfine Interactions}\ }\textbf {\bibinfo {volume} {144/145}},\ \bibinfo
  {pages} {175} (\bibinfo {year} {2002})}\BibitemShut {NoStop}%
\bibitem [{\citenamefont {Pasternak}\ \emph {et~al.}(2001)\citenamefont
  {Pasternak}, \citenamefont {Xu}, \citenamefont {Rozenberg}, \citenamefont
  {Taylor}, \citenamefont {Hearne},\ and\ \citenamefont
  {Sterer}}]{Pasternak-2001}%
  \BibitemOpen
  \bibfield  {author} {\bibinfo {author} {\bibfnamefont {M.~P.}\ \bibnamefont
  {Pasternak}}, \bibinfo {author} {\bibfnamefont {W.~M.}\ \bibnamefont {Xu}},
  \bibinfo {author} {\bibfnamefont {G.~K.}\ \bibnamefont {Rozenberg}}, \bibinfo
  {author} {\bibfnamefont {R.~D.}\ \bibnamefont {Taylor}}, \bibinfo {author}
  {\bibfnamefont {G.~R.}\ \bibnamefont {Hearne}}, \ and\ \bibinfo {author}
  {\bibfnamefont {E.}~\bibnamefont {Sterer}},\ }\href@noop {} {\bibfield
  {journal} {\bibinfo  {journal} {Phys. Rev. B}\ }\textbf {\bibinfo {volume}
  {65}},\ \bibinfo {pages} {035106} (\bibinfo {year} {2001})}\BibitemShut
  {NoStop}%
\bibitem [{\citenamefont {Jacobs}\ and\ \citenamefont
  {Lawrence}(1967)}]{Jacobs-1967-FeCl2}%
  \BibitemOpen
  \bibfield  {author} {\bibinfo {author} {\bibfnamefont {I.~S.}\ \bibnamefont
  {Jacobs}}\ and\ \bibinfo {author} {\bibfnamefont {P.~E.}\ \bibnamefont
  {Lawrence}},\ }\href@noop {} {\bibfield  {journal} {\bibinfo  {journal}
  {Phys. Rev.}\ }\textbf {\bibinfo {volume} {164}},\ \bibinfo {pages} {866}
  (\bibinfo {year} {1967})}\BibitemShut {NoStop}%
\bibitem [{\citenamefont {Fert}\ \emph
  {et~al.}(1973{\natexlab{b}})\citenamefont {Fert}, \citenamefont {Gelard},\
  and\ \citenamefont {Carrara}}]{Fert-1973-FeI2}%
  \BibitemOpen
  \bibfield  {author} {\bibinfo {author} {\bibfnamefont {A.~R.}\ \bibnamefont
  {Fert}}, \bibinfo {author} {\bibfnamefont {J.}~\bibnamefont {Gelard}}, \ and\
  \bibinfo {author} {\bibfnamefont {P.}~\bibnamefont {Carrara}},\ }\href@noop
  {} {\bibfield  {journal} {\bibinfo  {journal} {Solid State Commun.}\ }\textbf
  {\bibinfo {volume} {13}},\ \bibinfo {pages} {1219} (\bibinfo {year}
  {1973}{\natexlab{b}})}\BibitemShut {NoStop}%
\bibitem [{\citenamefont {Lines}(1963)}]{Lines-1963}%
  \BibitemOpen
  \bibfield  {author} {\bibinfo {author} {\bibfnamefont {M.~E.}\ \bibnamefont
  {Lines}},\ }\href@noop {} {\bibfield  {journal} {\bibinfo  {journal} {Phys.
  Rev.}\ }\textbf {\bibinfo {volume} {131}},\ \bibinfo {pages} {546} (\bibinfo
  {year} {1963})}\BibitemShut {NoStop}%
\bibitem [{\citenamefont {Binek}\ and\ \citenamefont
  {Kleemann}(1994)}]{Binek-1994}%
  \BibitemOpen
  \bibfield  {author} {\bibinfo {author} {\bibfnamefont {C.}~\bibnamefont
  {Binek}}\ and\ \bibinfo {author} {\bibfnamefont {W.}~\bibnamefont
  {Kleemann}},\ }\href@noop {} {\bibfield  {journal} {\bibinfo  {journal}
  {Phys. Rev. Lett.}\ }\textbf {\bibinfo {volume} {72}},\ \bibinfo {pages}
  {1287} (\bibinfo {year} {1994})}\BibitemShut {NoStop}%
\bibitem [{\citenamefont {Binek}\ \emph {et~al.}(1996)\citenamefont {Binek},
  \citenamefont {Bertrand}, \citenamefont {Regnault},\ and\ \citenamefont
  {Kleemann}}]{Binek-1996}%
  \BibitemOpen
  \bibfield  {author} {\bibinfo {author} {\bibfnamefont {C.}~\bibnamefont
  {Binek}}, \bibinfo {author} {\bibfnamefont {D.}~\bibnamefont {Bertrand}},
  \bibinfo {author} {\bibfnamefont {L.~P.}\ \bibnamefont {Regnault}}, \ and\
  \bibinfo {author} {\bibfnamefont {W.}~\bibnamefont {Kleemann}},\ }\href@noop
  {} {\bibfield  {journal} {\bibinfo  {journal} {Phys. Rev. B}\ }\textbf
  {\bibinfo {volume} {54}},\ \bibinfo {pages} {9015} (\bibinfo {year}
  {1996})}\BibitemShut {NoStop}%
\bibitem [{\citenamefont {Katsumata}\ \emph {et~al.}(1997)\citenamefont
  {Katsumata}, \citenamefont {{Aruga Katori}}, \citenamefont {Shapiro},\ and\
  \citenamefont {Shirane}}]{Katsumata-1997}%
  \BibitemOpen
  \bibfield  {author} {\bibinfo {author} {\bibfnamefont {K.}~\bibnamefont
  {Katsumata}}, \bibinfo {author} {\bibfnamefont {H.}~\bibnamefont {{Aruga
  Katori}}}, \bibinfo {author} {\bibfnamefont {S.~M.}\ \bibnamefont {Shapiro}},
  \ and\ \bibinfo {author} {\bibfnamefont {G.}~\bibnamefont {Shirane}},\
  }\href@noop {} {\bibfield  {journal} {\bibinfo  {journal} {Phys. Rev. B}\
  }\textbf {\bibinfo {volume} {55}},\ \bibinfo {pages} {11466} (\bibinfo {year}
  {1997})}\BibitemShut {NoStop}%
\bibitem [{\citenamefont {Chisholm}\ and\ \citenamefont
  {Stout}(1962)}]{Chisholm-1962}%
  \BibitemOpen
  \bibfield  {author} {\bibinfo {author} {\bibfnamefont {R.~C.}\ \bibnamefont
  {Chisholm}}\ and\ \bibinfo {author} {\bibfnamefont {J.~W.}\ \bibnamefont
  {Stout}},\ }\href@noop {} {\bibfield  {journal} {\bibinfo  {journal} {J.
  Chem. Phys.}\ }\textbf {\bibinfo {volume} {36}},\ \bibinfo {pages} {972}
  (\bibinfo {year} {1962})}\BibitemShut {NoStop}%
\bibitem [{\citenamefont {Yoshizawa}\ \emph {et~al.}(1980)\citenamefont
  {Yoshizawa}, \citenamefont {Ubukoshi},\ and\ \citenamefont
  {Hirakawa}}]{Yoshizawa-1980}%
  \BibitemOpen
  \bibfield  {author} {\bibinfo {author} {\bibfnamefont {H.}~\bibnamefont
  {Yoshizawa}}, \bibinfo {author} {\bibfnamefont {K.}~\bibnamefont {Ubukoshi}},
  \ and\ \bibinfo {author} {\bibfnamefont {K.}~\bibnamefont {Hirakawa}},\
  }\href@noop {} {\bibfield  {journal} {\bibinfo  {journal} {J. Phys. Soc.
  Japan}\ }\textbf {\bibinfo {volume} {48}},\ \bibinfo {pages} {42} (\bibinfo
  {year} {1980})}\BibitemShut {NoStop}%
\bibitem [{\citenamefont {Mekata}\ \emph {et~al.}(1992)\citenamefont {Mekata},
  \citenamefont {Kuriyama}, \citenamefont {Ajiro}, \citenamefont {Mitsuda},\
  and\ \citenamefont {Yoshizawa}}]{Mekata-1992}%
  \BibitemOpen
  \bibfield  {author} {\bibinfo {author} {\bibfnamefont {M.}~\bibnamefont
  {Mekata}}, \bibinfo {author} {\bibfnamefont {H.}~\bibnamefont {Kuriyama}},
  \bibinfo {author} {\bibfnamefont {Y.}~\bibnamefont {Ajiro}}, \bibinfo
  {author} {\bibfnamefont {S.}~\bibnamefont {Mitsuda}}, \ and\ \bibinfo
  {author} {\bibfnamefont {H.}~\bibnamefont {Yoshizawa}},\ }\href@noop {}
  {\bibfield  {journal} {\bibinfo  {journal} {J. Magn. Magn. Mater.}\ }\textbf
  {\bibinfo {volume} {104-107}},\ \bibinfo {pages} {859} (\bibinfo {year}
  {1992})}\BibitemShut {NoStop}%
\bibitem [{\citenamefont {Friedt}\ \emph {et~al.}(1976)\citenamefont {Friedt},
  \citenamefont {Sanchez},\ and\ \citenamefont {Shenoy}}]{Friedt-1976}%
  \BibitemOpen
  \bibfield  {author} {\bibinfo {author} {\bibfnamefont {J.~M.}\ \bibnamefont
  {Friedt}}, \bibinfo {author} {\bibfnamefont {J.~P.}\ \bibnamefont {Sanchez}},
  \ and\ \bibinfo {author} {\bibfnamefont {G.~K.}\ \bibnamefont {Shenoy}},\
  }\href@noop {} {\bibfield  {journal} {\bibinfo  {journal} {J. Chem. Phys.}\
  }\textbf {\bibinfo {volume} {65}},\ \bibinfo {pages} {5093} (\bibinfo {year}
  {1976})}\BibitemShut {NoStop}%
\bibitem [{\citenamefont {Busey}\ and\ \citenamefont
  {Giauque}(1952)}]{Busey-1952}%
  \BibitemOpen
  \bibfield  {author} {\bibinfo {author} {\bibfnamefont {R.~H.}\ \bibnamefont
  {Busey}}\ and\ \bibinfo {author} {\bibfnamefont {W.~F.}\ \bibnamefont
  {Giauque}},\ }\href@noop {} {\bibfield  {journal} {\bibinfo  {journal} {J.
  Am. Chem. Soc.}\ }\textbf {\bibinfo {volume} {74}},\ \bibinfo {pages} {4443}
  (\bibinfo {year} {1952})}\BibitemShut {NoStop}%
\bibitem [{\citenamefont {Adam}\ \emph {et~al.}(1980)\citenamefont {Adam},
  \citenamefont {Billery}, \citenamefont {Terrier}, \citenamefont {Mainard},
  \citenamefont {Regnault}, \citenamefont {Rossat-Mignod},\ and\ \citenamefont
  {M{\'{e}}riel}}]{Adam-1980}%
  \BibitemOpen
  \bibfield  {author} {\bibinfo {author} {\bibfnamefont {A.}~\bibnamefont
  {Adam}}, \bibinfo {author} {\bibfnamefont {D.}~\bibnamefont {Billery}},
  \bibinfo {author} {\bibfnamefont {C.}~\bibnamefont {Terrier}}, \bibinfo
  {author} {\bibfnamefont {R.}~\bibnamefont {Mainard}}, \bibinfo {author}
  {\bibfnamefont {L.~P.}\ \bibnamefont {Regnault}}, \bibinfo {author}
  {\bibfnamefont {J.}~\bibnamefont {Rossat-Mignod}}, \ and\ \bibinfo {author}
  {\bibfnamefont {P.}~\bibnamefont {M{\'{e}}riel}},\ }\href@noop {} {\bibfield
  {journal} {\bibinfo  {journal} {Solid State Commun.}\ }\textbf {\bibinfo
  {volume} {35}},\ \bibinfo {pages} {1} (\bibinfo {year} {1980})}\BibitemShut
  {NoStop}%
\bibitem [{\citenamefont {{De Gunzbourg}}\ \emph {et~al.}(1971)\citenamefont
  {{De Gunzbourg}}, \citenamefont {Papassimacopoulos}, \citenamefont
  {{Miedan-Gros}},\ and\ \citenamefont {Allain}}]{DeGunzbourg-1971}%
  \BibitemOpen
  \bibfield  {author} {\bibinfo {author} {\bibfnamefont {J.}~\bibnamefont {{De
  Gunzbourg}}}, \bibinfo {author} {\bibfnamefont {S.}~\bibnamefont
  {Papassimacopoulos}}, \bibinfo {author} {\bibfnamefont {A.}~\bibnamefont
  {{Miedan-Gros}}}, \ and\ \bibinfo {author} {\bibfnamefont {Y.}~\bibnamefont
  {Allain}},\ }\href@noop {} {\bibfield  {journal} {\bibinfo  {journal} {J.
  Phys. Colloq.}\ }\textbf {\bibinfo {volume} {32}},\ \bibinfo {pages} {C1}
  (\bibinfo {year} {1971})}\BibitemShut {NoStop}%
\bibitem [{\citenamefont {Pollard}\ \emph {et~al.}(1982)\citenamefont
  {Pollard}, \citenamefont {McCann},\ and\ \citenamefont
  {Ward}}]{Pollard-1982}%
  \BibitemOpen
  \bibfield  {author} {\bibinfo {author} {\bibfnamefont {R.~J.}\ \bibnamefont
  {Pollard}}, \bibinfo {author} {\bibfnamefont {V.~H.}\ \bibnamefont {McCann}},
  \ and\ \bibinfo {author} {\bibfnamefont {J.~B.}\ \bibnamefont {Ward}},\
  }\href@noop {} {\bibfield  {journal} {\bibinfo  {journal} {J. Phys. C: Solid
  State Phys.}\ }\textbf {\bibinfo {volume} {15}},\ \bibinfo {pages} {6807}
  (\bibinfo {year} {1982})}\BibitemShut {NoStop}%
\bibitem [{\citenamefont {Day}\ \emph {et~al.}(1976)\citenamefont {Day},
  \citenamefont {Dinsdale}, \citenamefont {Krausz},\ and\ \citenamefont
  {Robbins}}]{Day-1976}%
  \BibitemOpen
  \bibfield  {author} {\bibinfo {author} {\bibfnamefont {P.}~\bibnamefont
  {Day}}, \bibinfo {author} {\bibfnamefont {A.}~\bibnamefont {Dinsdale}},
  \bibinfo {author} {\bibfnamefont {E.~R.}\ \bibnamefont {Krausz}}, \ and\
  \bibinfo {author} {\bibfnamefont {D.~J.}\ \bibnamefont {Robbins}},\
  }\href@noop {} {\bibfield  {journal} {\bibinfo  {journal} {J. Phys. C: Solid
  State Phys.}\ }\textbf {\bibinfo {volume} {9}},\ \bibinfo {pages} {2481}
  (\bibinfo {year} {1976})}\BibitemShut {NoStop}%
\bibitem [{\citenamefont {Day}\ and\ \citenamefont {Ziebeck}(1980)}]{Day-1980}%
  \BibitemOpen
  \bibfield  {author} {\bibinfo {author} {\bibfnamefont {P.}~\bibnamefont
  {Day}}\ and\ \bibinfo {author} {\bibfnamefont {K.~R.~A.}\ \bibnamefont
  {Ziebeck}},\ }\href@noop {} {\bibfield  {journal} {\bibinfo  {journal} {J.
  Phys. C: Solid State Phys.}\ }\textbf {\bibinfo {volume} {13}},\ \bibinfo
  {pages} {L523} (\bibinfo {year} {1980})}\BibitemShut {NoStop}%
\bibitem [{\citenamefont {Billery}\ \emph {et~al.}(1977)\citenamefont
  {Billery}, \citenamefont {Terrier}, \citenamefont {Ciret},\ and\
  \citenamefont {Kleinclauss}}]{Billery-1977}%
  \BibitemOpen
  \bibfield  {author} {\bibinfo {author} {\bibfnamefont {D.}~\bibnamefont
  {Billery}}, \bibinfo {author} {\bibfnamefont {C.}~\bibnamefont {Terrier}},
  \bibinfo {author} {\bibfnamefont {N.}~\bibnamefont {Ciret}}, \ and\ \bibinfo
  {author} {\bibfnamefont {J.}~\bibnamefont {Kleinclauss}},\ }\href@noop {}
  {\bibfield  {journal} {\bibinfo  {journal} {Phys. Lett. A}\ }\textbf
  {\bibinfo {volume} {61}},\ \bibinfo {pages} {138} (\bibinfo {year}
  {1977})}\BibitemShut {NoStop}%
\bibitem [{\citenamefont {He}\ \emph {et~al.}(2016)\citenamefont {He},
  \citenamefont {Ma}, \citenamefont {Lyu},\ and\ \citenamefont
  {Nachtigall}}]{He-2016}%
  \BibitemOpen
  \bibfield  {author} {\bibinfo {author} {\bibfnamefont {J.}~\bibnamefont
  {He}}, \bibinfo {author} {\bibfnamefont {S.}~\bibnamefont {Ma}}, \bibinfo
  {author} {\bibfnamefont {P.}~\bibnamefont {Lyu}}, \ and\ \bibinfo {author}
  {\bibfnamefont {P.}~\bibnamefont {Nachtigall}},\ }\href@noop {} {\bibfield
  {journal} {\bibinfo  {journal} {J. Mater. Chem. C}\ }\textbf {\bibinfo
  {volume} {4}},\ \bibinfo {pages} {2518} (\bibinfo {year} {2016})}\BibitemShut
  {NoStop}%
\bibitem [{\citenamefont {Tsubokawa}(1960)}]{Tsubokawa-1960-CrBr3}%
  \BibitemOpen
  \bibfield  {author} {\bibinfo {author} {\bibfnamefont {I.}~\bibnamefont
  {Tsubokawa}},\ }\href@noop {} {\bibfield  {journal} {\bibinfo  {journal} {J.
  Phys. Soc. Japan}\ }\textbf {\bibinfo {volume} {15}},\ \bibinfo {pages}
  {1664} (\bibinfo {year} {1960})}\BibitemShut {NoStop}%
\bibitem [{\citenamefont {Hansen}(1959)}]{Hansen-1959}%
  \BibitemOpen
  \bibfield  {author} {\bibinfo {author} {\bibfnamefont {W.~N.}\ \bibnamefont
  {Hansen}},\ }\href@noop {} {\bibfield  {journal} {\bibinfo  {journal} {J.
  Appl. Phys.}\ }\textbf {\bibinfo {volume} {30}},\ \bibinfo {pages} {S304}
  (\bibinfo {year} {1959})}\BibitemShut {NoStop}%
\bibitem [{\citenamefont {{Dillon Jr.}}\ and\ \citenamefont
  {Olson}(1965)}]{Dillon-1965}%
  \BibitemOpen
  \bibfield  {author} {\bibinfo {author} {\bibfnamefont {J.~F.}\ \bibnamefont
  {{Dillon Jr.}}}\ and\ \bibinfo {author} {\bibfnamefont {C.~E.}\ \bibnamefont
  {Olson}},\ }\href@noop {} {\bibfield  {journal} {\bibinfo  {journal} {J.
  Appl. Phys.}\ }\textbf {\bibinfo {volume} {36}},\ \bibinfo {pages} {1259}
  (\bibinfo {year} {1965})}\BibitemShut {NoStop}%
\bibitem [{\citenamefont {Cable}\ \emph {et~al.}(1961)\citenamefont {Cable},
  \citenamefont {Wilkinson},\ and\ \citenamefont {Wollan}}]{Cable-1961-CrCl3}%
  \BibitemOpen
  \bibfield  {author} {\bibinfo {author} {\bibfnamefont {J.~W.}\ \bibnamefont
  {Cable}}, \bibinfo {author} {\bibfnamefont {M.~K.}\ \bibnamefont
  {Wilkinson}}, \ and\ \bibinfo {author} {\bibfnamefont {E.~O.}\ \bibnamefont
  {Wollan}},\ }\href@noop {} {\bibfield  {journal} {\bibinfo  {journal} {J.
  Phys. Chem. Solids}\ }\textbf {\bibinfo {volume} {19}},\ \bibinfo {pages}
  {29} (\bibinfo {year} {1961})}\BibitemShut {NoStop}%
\bibitem [{\citenamefont {Kuhlow}(1982)}]{Kuhlow-1982-CrCl3}%
  \BibitemOpen
  \bibfield  {author} {\bibinfo {author} {\bibfnamefont {B.}~\bibnamefont
  {Kuhlow}},\ }\href@noop {} {\bibfield  {journal} {\bibinfo  {journal} {Phys.
  Stat. Sol. A}\ }\textbf {\bibinfo {volume} {72}},\ \bibinfo {pages} {161}
  (\bibinfo {year} {1982})}\BibitemShut {NoStop}%
\bibitem [{\citenamefont {Cable}\ \emph
  {et~al.}(1962{\natexlab{b}})\citenamefont {Cable}, \citenamefont {Wilkinson},
  \citenamefont {Wollan},\ and\ \citenamefont {Koehler}}]{Cable-1962-FeCl3}%
  \BibitemOpen
  \bibfield  {author} {\bibinfo {author} {\bibfnamefont {J.~W.}\ \bibnamefont
  {Cable}}, \bibinfo {author} {\bibfnamefont {M.~K.}\ \bibnamefont
  {Wilkinson}}, \bibinfo {author} {\bibfnamefont {E.~O.}\ \bibnamefont
  {Wollan}}, \ and\ \bibinfo {author} {\bibfnamefont {W.~C.}\ \bibnamefont
  {Koehler}},\ }\href@noop {} {\bibfield  {journal} {\bibinfo  {journal} {Phys.
  Rev.}\ }\textbf {\bibinfo {volume} {127}},\ \bibinfo {pages} {714} (\bibinfo
  {year} {1962}{\natexlab{b}})}\BibitemShut {NoStop}%
\bibitem [{\citenamefont {{Jones Jr.}}\ \emph {et~al.}(1969)\citenamefont
  {{Jones Jr.}}, \citenamefont {Morton}, \citenamefont {Cathey}, \citenamefont
  {Auel},\ and\ \citenamefont {Amma}}]{Jones-1969}%
  \BibitemOpen
  \bibfield  {author} {\bibinfo {author} {\bibfnamefont {E.~R.}\ \bibnamefont
  {{Jones Jr.}}}, \bibinfo {author} {\bibfnamefont {O.~B.}\ \bibnamefont
  {Morton}}, \bibinfo {author} {\bibfnamefont {L.}~\bibnamefont {Cathey}},
  \bibinfo {author} {\bibfnamefont {T.}~\bibnamefont {Auel}}, \ and\ \bibinfo
  {author} {\bibfnamefont {E.~L.}\ \bibnamefont {Amma}},\ }\href@noop {}
  {\bibfield  {journal} {\bibinfo  {journal} {J. Chem. Phys.}\ }\textbf
  {\bibinfo {volume} {50}},\ \bibinfo {pages} {4755} (\bibinfo {year}
  {1969})}\BibitemShut {NoStop}%
\bibitem [{\citenamefont {Johnson}\ \emph {et~al.}(1981)\citenamefont
  {Johnson}, \citenamefont {Friedberg},\ and\ \citenamefont
  {Rayne}}]{Johnson-1981}%
  \BibitemOpen
  \bibfield  {author} {\bibinfo {author} {\bibfnamefont {P.~B.}\ \bibnamefont
  {Johnson}}, \bibinfo {author} {\bibfnamefont {S.~A.}\ \bibnamefont
  {Friedberg}}, \ and\ \bibinfo {author} {\bibfnamefont {J.~A.}\ \bibnamefont
  {Rayne}},\ }\href@noop {} {\bibfield  {journal} {\bibinfo  {journal} {J.
  Appl. Phys.}\ }\textbf {\bibinfo {volume} {52}},\ \bibinfo {pages} {1932}
  (\bibinfo {year} {1981})}\BibitemShut {NoStop}%
\bibitem [{\citenamefont {Stampfel}\ \emph {et~al.}(1973)\citenamefont
  {Stampfel}, \citenamefont {Oosterhuis}, \citenamefont {Window},\ and\
  \citenamefont {{Barros, F. deS}}}]{Stampfel-1973}%
  \BibitemOpen
  \bibfield  {author} {\bibinfo {author} {\bibfnamefont {J.~P.}\ \bibnamefont
  {Stampfel}}, \bibinfo {author} {\bibfnamefont {W.~T.}\ \bibnamefont
  {Oosterhuis}}, \bibinfo {author} {\bibfnamefont {B.}~\bibnamefont {Window}},
  \ and\ \bibinfo {author} {\bibnamefont {{Barros, F. deS}}},\ }\href@noop {}
  {\bibfield  {journal} {\bibinfo  {journal} {Phys. Rev. B}\ }\textbf {\bibinfo
  {volume} {8}},\ \bibinfo {pages} {4371} (\bibinfo {year} {1973})}\BibitemShut
  {NoStop}%
\bibitem [{\citenamefont {Oosterhuis}\ \emph {et~al.}(1975)\citenamefont
  {Oosterhuis}, \citenamefont {Window},\ and\ \citenamefont
  {Spartalian}}]{Oosterhuis-1975}%
  \BibitemOpen
  \bibfield  {author} {\bibinfo {author} {\bibfnamefont {W.~T.}\ \bibnamefont
  {Oosterhuis}}, \bibinfo {author} {\bibfnamefont {B.}~\bibnamefont {Window}},
  \ and\ \bibinfo {author} {\bibfnamefont {K.}~\bibnamefont {Spartalian}},\
  }\href@noop {} {\bibfield  {journal} {\bibinfo  {journal} {Phys. Rev. B}\
  }\textbf {\bibinfo {volume} {10}},\ \bibinfo {pages} {4616} (\bibinfo {year}
  {1975})}\BibitemShut {NoStop}%
\bibitem [{\citenamefont {Kim}\ \emph {et~al.}(2008)\citenamefont {Kim},
  \citenamefont {Jin}, \citenamefont {Moon}, \citenamefont {Kim}, \citenamefont
  {Park}, \citenamefont {Leem}, \citenamefont {Yu}, \citenamefont {Noh},
  \citenamefont {Kim}, \citenamefont {Oh}, \citenamefont {Park}, \citenamefont
  {Durairaj}, \citenamefont {Cao},\ and\ \citenamefont {Rotenberg}}]{Kim-2008}%
  \BibitemOpen
  \bibfield  {author} {\bibinfo {author} {\bibfnamefont {B.~J.}\ \bibnamefont
  {Kim}}, \bibinfo {author} {\bibfnamefont {H.}~\bibnamefont {Jin}}, \bibinfo
  {author} {\bibfnamefont {S.~J.}\ \bibnamefont {Moon}}, \bibinfo {author}
  {\bibfnamefont {J.~Y.}\ \bibnamefont {Kim}}, \bibinfo {author} {\bibfnamefont
  {B.~G.}\ \bibnamefont {Park}}, \bibinfo {author} {\bibfnamefont {C.~S.}\
  \bibnamefont {Leem}}, \bibinfo {author} {\bibfnamefont {J.}~\bibnamefont
  {Yu}}, \bibinfo {author} {\bibfnamefont {T.~W.}\ \bibnamefont {Noh}},
  \bibinfo {author} {\bibfnamefont {C.}~\bibnamefont {Kim}}, \bibinfo {author}
  {\bibfnamefont {S.~J.}\ \bibnamefont {Oh}}, \bibinfo {author} {\bibfnamefont
  {J.~H.}\ \bibnamefont {Park}}, \bibinfo {author} {\bibfnamefont
  {V.}~\bibnamefont {Durairaj}}, \bibinfo {author} {\bibfnamefont
  {G.}~\bibnamefont {Cao}}, \ and\ \bibinfo {author} {\bibfnamefont
  {E.}~\bibnamefont {Rotenberg}},\ }\href@noop {} {\bibfield  {journal}
  {\bibinfo  {journal} {Phys. Rev. Lett.}\ }\textbf {\bibinfo {volume} {101}},\
  \bibinfo {pages} {076402} (\bibinfo {year} {2008})}\BibitemShut {NoStop}%
\bibitem [{\citenamefont {Fletcher}\ \emph {et~al.}(1963)\citenamefont
  {Fletcher}, \citenamefont {Gardner}, \citenamefont {Hooper}, \citenamefont
  {Hyde}, \citenamefont {Moore},\ and\ \citenamefont
  {Woodhead}}]{Fletcher-1963}%
  \BibitemOpen
  \bibfield  {author} {\bibinfo {author} {\bibfnamefont {J.~M.}\ \bibnamefont
  {Fletcher}}, \bibinfo {author} {\bibfnamefont {W.~E.}\ \bibnamefont
  {Gardner}}, \bibinfo {author} {\bibfnamefont {E.~W.}\ \bibnamefont {Hooper}},
  \bibinfo {author} {\bibfnamefont {K.~R.}\ \bibnamefont {Hyde}}, \bibinfo
  {author} {\bibfnamefont {F.~H.}\ \bibnamefont {Moore}}, \ and\ \bibinfo
  {author} {\bibfnamefont {J.~L.}\ \bibnamefont {Woodhead}},\ }\href@noop {}
  {\bibfield  {journal} {\bibinfo  {journal} {Nature}\ }\textbf {\bibinfo
  {volume} {199}},\ \bibinfo {pages} {1089} (\bibinfo {year}
  {1963})}\BibitemShut {NoStop}%
\bibitem [{\citenamefont {Kobayashi}\ \emph {et~al.}(1992)\citenamefont
  {Kobayashi}, \citenamefont {Okada}, \citenamefont {Asai}, \citenamefont
  {Katada}, \citenamefont {Sano},\ and\ \citenamefont {Ambe}}]{Kobayashi-1992}%
  \BibitemOpen
  \bibfield  {author} {\bibinfo {author} {\bibfnamefont {Y.}~\bibnamefont
  {Kobayashi}}, \bibinfo {author} {\bibfnamefont {T.}~\bibnamefont {Okada}},
  \bibinfo {author} {\bibfnamefont {K.}~\bibnamefont {Asai}}, \bibinfo {author}
  {\bibfnamefont {M.}~\bibnamefont {Katada}}, \bibinfo {author} {\bibfnamefont
  {H.}~\bibnamefont {Sano}}, \ and\ \bibinfo {author} {\bibfnamefont
  {F.}~\bibnamefont {Ambe}},\ }\href@noop {} {\bibfield  {journal} {\bibinfo
  {journal} {Inorg. Chem.}\ }\textbf {\bibinfo {volume} {31}},\ \bibinfo
  {pages} {4570} (\bibinfo {year} {1992})}\BibitemShut {NoStop}%
\bibitem [{\citenamefont {Majumder}\ \emph {et~al.}(2015)\citenamefont
  {Majumder}, \citenamefont {Schmidt}, \citenamefont {Rosner}, \citenamefont
  {Tsirlin}, \citenamefont {Yasuoka},\ and\ \citenamefont
  {Baenitz}}]{Majumder-2015}%
  \BibitemOpen
  \bibfield  {author} {\bibinfo {author} {\bibfnamefont {M.}~\bibnamefont
  {Majumder}}, \bibinfo {author} {\bibfnamefont {M.}~\bibnamefont {Schmidt}},
  \bibinfo {author} {\bibfnamefont {H.}~\bibnamefont {Rosner}}, \bibinfo
  {author} {\bibfnamefont {A.~A.}\ \bibnamefont {Tsirlin}}, \bibinfo {author}
  {\bibfnamefont {H.}~\bibnamefont {Yasuoka}}, \ and\ \bibinfo {author}
  {\bibfnamefont {M.}~\bibnamefont {Baenitz}},\ }\href@noop {} {\bibfield
  {journal} {\bibinfo  {journal} {Phys. Rev. B}\ }\textbf {\bibinfo {volume}
  {91}},\ \bibinfo {pages} {180401} (\bibinfo {year} {2015})}\BibitemShut
  {NoStop}%
\bibitem [{\citenamefont {Kitaev}(2006)}]{Kitaev}%
  \BibitemOpen
  \bibfield  {author} {\bibinfo {author} {\bibfnamefont {A.}~\bibnamefont
  {Kitaev}},\ }\href@noop {} {\bibfield  {journal} {\bibinfo  {journal} {Ann.
  Phys.}\ }\textbf {\bibinfo {volume} {321}},\ \bibinfo {pages} {2} (\bibinfo
  {year} {2006})}\BibitemShut {NoStop}%
\bibitem [{\citenamefont {Baskaran}\ \emph {et~al.}(2007)\citenamefont
  {Baskaran}, \citenamefont {Mandal},\ and\ \citenamefont
  {Shankar}}]{Baskaran-2007}%
  \BibitemOpen
  \bibfield  {author} {\bibinfo {author} {\bibfnamefont {G.}~\bibnamefont
  {Baskaran}}, \bibinfo {author} {\bibfnamefont {S.}~\bibnamefont {Mandal}}, \
  and\ \bibinfo {author} {\bibfnamefont {R.}~\bibnamefont {Shankar}},\
  }\href@noop {} {\bibfield  {journal} {\bibinfo  {journal} {Phys. Rev. Lett.}\
  }\textbf {\bibinfo {volume} {98}},\ \bibinfo {pages} {247201} (\bibinfo
  {year} {2007})}\BibitemShut {NoStop}%
\bibitem [{\citenamefont {Knolle}\ \emph {et~al.}(2014)\citenamefont {Knolle},
  \citenamefont {Kovrizhin}, \citenamefont {Chalker},\ and\ \citenamefont
  {Moessner}}]{Knolle-2014}%
  \BibitemOpen
  \bibfield  {author} {\bibinfo {author} {\bibfnamefont {J.}~\bibnamefont
  {Knolle}}, \bibinfo {author} {\bibfnamefont {D.~L.}\ \bibnamefont
  {Kovrizhin}}, \bibinfo {author} {\bibfnamefont {J.~T.}\ \bibnamefont
  {Chalker}}, \ and\ \bibinfo {author} {\bibfnamefont {R.}~\bibnamefont
  {Moessner}},\ }\href@noop {} {\bibfield  {journal} {\bibinfo  {journal}
  {Phys. Rev. Lett.}\ }\textbf {\bibinfo {volume} {112}},\ \bibinfo {pages}
  {207203} (\bibinfo {year} {2014})}\BibitemShut {NoStop}%
\end{thebibliography}


\end{document}